\documentclass[fleqn,usenatbib]{mnras}
\usepackage{newtxtext,newtxmath}
\usepackage[T1]{fontenc}
\usepackage{subfigure}
\usepackage{orcidlink}
\usepackage{array}
\usepackage{multirow}
\DeclareRobustCommand{\VAN}[3]{#2}
\let\VANthebibliography\thebibliography
\def\thebibliography{\DeclareRobustCommand{\VAN}[3]{##3}\VANthebibliography}

\usepackage{graphicx}	
\usepackage{amsmath}	

\newcommand{\Msun}{\ensuremath{\mathrm{M}_\odot}}
\newcommand{\Mbh}{\ensuremath{M_\mathrm{BH}}}

\newcommand{\Mstar}{$M_{\star}$}

\newcommand{\Lsun}{\ensuremath{\mathrm{L}_\odot}}

\newcommand{\hst}{\emph{HST}}
\newcommand{\jwst}{\emph{JWST}}
\newcommand{\kms}{km~s$^{-1}$}

\newcommand{\cotwo}{$^{12}$CO(2$-$1)}

\mathchardef\mhyphen="2D

\title[Measuring SMBHs at $1\lesssim z\lesssim2$ with MICADO $\&$ HARMONI]{Extending the Frontier of Spatially-Resolved Supermassive Black Hole Mass Measurements to at $1\lesssim z\lesssim2$: Simulations with ELT/MICADO High-Resolution Mass Models and HARMONI Integral-Field Stellar Kinematics}

\author[D.\ D.\ Nguyen, M. Cappellari $\&$ Tinh Q.\ T.\ Le et al.]{
Dieu D.\ Nguyen$^{\orcidlink{0000-0002-5678-1008}1}$\thanks{E-mail: dieun@umich.edu},
Michele Cappellari$^{\orcidlink{0000-0002-1283-8420}2}$\thanks{E-mail: michele.cappellari@physics.ox.ac.uk},
Tinh Q.\ T.\ Le$^{\orcidlink{0009-0004-3689-8577}3}$,
Hai N.\ Ngo$^{\orcidlink{0009-0006-5852-4538}4}$, 
Elena Gallo$^{\orcidlink{0000-0001-5802-6041}1}$, \newauthor
Niranjan Thatte$^{\orcidlink{0000-0002-6694-5184}2}$, 
Fan Zou$^{\orcidlink{0000-0002-4436-6923}1}$, 
Tien H.\ T.\ Ho$^{\orcidlink{0009-0005-8845-9725}4}$,
Tuan N.\ Le$^{\orcidlink{0009-0009-0015-1208}5}$,
Huy G.\ Tong$^{\orcidlink{0009-0008-6050-5736}4}$, and \newauthor
Miguel Pereira-Santaella$^{\orcidlink{0000-0002-4005-9619}6}$ \\
$^{1}$Department of Astronomy, University of Michigan, 1085 South University Avenue, Ann Arbor, MI 48109, USA\\
$^{2}$Sub-Department of Astrophysics, Department of Physics, University of Oxford, Denys Wilkinson Building, Keble Road, Oxford, OX1 3RH, UK\\ 
$^{3}$Department of Physics, International University,Vietnam National University in Ho Chi Minh City, Vietnam\\
$^{4}$Faculty of Physics – Engineering Physics, University of Science, Vietnam National University in Ho Chi Minh City, Vietnam\\  
$^{5}$International Centre for Interdisciplinary Science and Education, 07 Science Avenue, Ghenh Rang, 55121 Quy Nhon, Vietnam\\
$^{6}$Instituto de F\'isica Fundamental, CSIC, Calle Serrano 123, 28006 Madrid, Spain 
}

\date{Accepted 2026 February 3. Received 2025 December 22; in original form 2025 November 7}

\pubyear{2025}

\begin{document}
\label{firstpage}
\pagerange{\pageref{firstpage}--\pageref{lastpage}}
\maketitle

\begin{abstract}
Current spatially resolved kinematic measurements of supermassive black hole (SMBH) masses are largely confined to the local Universe (distances $\lesssim100$ Mpc). We investigate the potential of the Extremely Large Telescope's (ELT) first-light instruments, MICADO and HARMONI, to extend these dynamical measurements to galaxies at redshift $1\lesssim z\lesssim2$. We select a sample of five bright, massive, quiescent galaxies at these redshifts, adopting their Sérsic profiles, from \hst\ photometry, as their intrinsic surface brightness distributions. Based on these intrinsic models, we generate mock MICADO images using \textsc{SimCADO} and mock HARMONI integral-field spectroscopic data cubes using \textsc{HSIM}. The HARMONI simulations utilize input stellar kinematics derived from Jeans Anisotropic Models (JAM). We then process these mock observations: the simulated MICADO images are fitted with Multi-Gaussian Expansion (MGE) to derive stellar mass models, and stellar kinematics are extracted from mock HARMONI cubes with \textsc{pPXF}. Finally, these derived stellar mass models and kinematics are used to constrain JAM dynamical models within a Bayesian framework. Our analysis demonstrates that SMBH masses can be recovered with an accuracy of $\sim$10\%. We find that MICADO can provide detailed stellar mass models with $\sim$1 hour of on-source exposure. HARMONI requires longer minimum integrations for reliable stellar kinematic measurements of SMBHs. The required on-source time scales with apparent brightness, ranging from 5–7.5~hours for galaxies at $z\approx1$ (F814W, 20–20.5~mag) to 5~hours for galaxies at $1<z\lesssim2$ (F160W, 20.8~mag). These findings highlight the ELT's capability to push the frontier of SMBH mass measurements to $z\approx2$, enabling crucial tests of SMBH-galaxy co-evolution at the top end of the galaxies mass function. 
\end{abstract}

\begin{keywords}
galaxies: general – galaxies: supermassive black holes – galaxies: nuclei – galaxies: kinematics and dynamics – galaxies: evolution – galaxies: formation
\end{keywords}


\section{Introduction}

\begin{figure}
    \centering
    \includegraphics[width=0.45\textwidth]{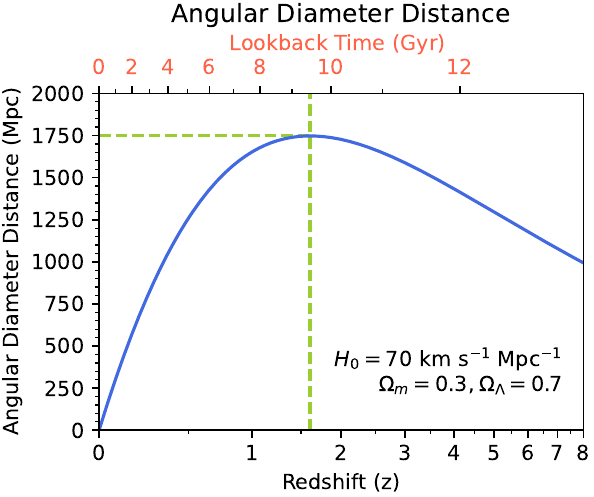}
    \caption{Angular-diameter distance ($D_A$) versus redshift ($z$) in a standard flat concordance $\Lambda$CDM cosmology, showing $D_A$ peaking at $\approx 1750$ Mpc around $z\approx1.6$.}
    \label{fig:cosmological_dist}
\end{figure}

\subsection{SMBH-Galaxy Co-evolution and Observational Challenges}

It is well-established that the mass of nearby supermassive black holes (SMBHs, \Mbh\ $\approx10^{6-9}$ \Msun) correlates with the macroscopic properties of massive galaxies \citep[$M_{\star}\gtrsim10^{10}$ \Msun; see review by][]{Kormendy13}. Notable correlations include the \Mbh--galaxy stellar mass \citep[$M_{\star}$;][]{Magorrian98}, the \Mbh--velocity dispersion ($\sigma_\star$) of the stars in a galaxy \citep{Ferrarese00,Gebhardt00}, and the \Mbh--luminosity \citep[$L_K$;][]{Kormendy95}. These correlations indicate the mutual growth of both black holes (BHs) and their host galaxies \citep[see reviews by][]{Fabian12,Kormendy13,Greene20}.

The discovery of these correlations has primarily relied on dynamical modelling of the stellar kinematics, for both massive galaxies \citep[e.g.,][]{Gebhardt2003, McConnell11, Cappellari2009, Rusli2013, Walsh2016, Thater22, Thater23} and lower mass ones \citep[e.g.,][]{Ahn18, Nguyen17conf, Nguyen14, Nguyen17, Nguyen18, Nguyen19, Nguyen2025a, Voggel18}. BH determination were also performed using gas kinematics from either ionized gas \citep[e.g.,][]{Barth2001, Shapiro2006, Walsh13} or molecular gas \citep[e.g.,][]{Davis20, Barth16, Onishi17, Nguyen19conf, Nguyen20, Nguyen21, Nguyen22, Ngo2025a}. These direct dynamical BH determinations have so far only been accessible to within about 100 Mpc \citep[see compilations by][]{Saglia16,vandenBosch16}. Beyond this distance, current facilities cannot resolve the SMBH's sphere of influence (SOI\footnote{The spherical region surrounding a BH where its gravitational force dominates, defined as the vicinity within which the enclosed mass equals twice the $M_{\rm BH}$: this can be approximately estimated as $R_{\rm SOI} \approx G M_{\rm BH}/\sigma^2_\star$, where $G$ represents the gravitational constant.}), the region where its gravity dominates. The angular size of the BH SOI ($R_{\rm SOI}$) in galaxies is typically much smaller than $0\farcs1$ \citep[eq.~2]{deZeeuw01}, falling below the diffraction limit of current world-leading telescopes equipped with adaptive optics (AO), such as the Gemini Telescope, the Very Large Telescope (VLT), and the Keck Telescope, which achieve a point spread function (PSF) full width at half maximum (FWHM) of approximately $0\farcs1$. 

On the other hand, the Atacama Large Millimeter/submillimeter Array (ALMA) is now capable of achieving angular resolutions of $\sim0\farcs01$ in Band~6 observations of \cotwo\ emission using the longest baseline antenna configuration (i.e. C-10), comparable to the spatial scales explored in the simulations presented in this work. However, such ultra–high-angular-resolution molecular gas observations often reveal resolved central cavities or deficits in the \cotwo\ circumnuclear disk emission, likely caused by active galactic nuclei \citep[AGN;][]{Imanishi20, Izumi20, Nguyen21}, which can degrade the accuracy of SMBH mass measurements, as demonstrated in the mm-Wave Interferometric Survey of Dark Object Masses \citep[WISDOM; e.g.][]{Smith19} and the Measuring Black Holes in Below Milky Way Stellar Mass (M$\star$) Galaxies \citep[MBHBM$\star$; e.g.][]{Nguyen20} projects. Another common limitation of using ALMA at ultra–high angular resolution to dynamically measure SMBH masses is that a significant fraction of the $^{12}$CO(2–1) emission from the circumnuclear gas disc can be resolved out. This loss of large-scale flux reduces the sensitivity and hampers the recovery of robust molecular gas kinematics. In principle, this effect can be mitigated by combining multiple measurement sets obtained with different array configurations, spanning a range of angular resolutions and maximum recoverable scales (e.g. combining the longest-baseline configuration, C-10, with more compact configurations such as C-8 and C-5). However, such multi-configuration observations are observationally expensive and typically require total on-source integration times in excess of $\sim$40 hours to achieve sufficient sensitivity, making this approach impractical for large samples.

Estimates of high-redshift AGN and quasars suggest that their \Mbh\ are much larger than predictions from the local galaxy stellar mass vs BH scaling relations \citep[e.g.,][]{Maiolino2024}. This implies that these correlations may have evolved over cosmic time. However, these BH estimates are not spatially resolved and must rely on strong assumptions and could suffer from significant measurement biases. Consequently, the evolution of the galaxy–BH scaling relation with redshift remains observationally unconstrained using direct dynamical methods, leaving open questions about how both BHs and their host galaxies grow in the context of galaxy–BH co-evolution.

\subsection{The Promise of the ELT for High-Redshift SMBH Demographics}

The advent of the Extremely Large Telescope (ELT) and its first-generation instruments—namely, the Multi-AO Imaging Camera for Deep Observations (MICADO) imager \citep{Davies10, Davies21} and the High Angular Resolution Monolithic Optical and Near-infrared Integral (HARMONI) field spectrograph \citep{Thatte16, Thatte20}—promises spatial resolutions around 10 milliarcseconds (mas). This capability potentially enables direct spatially-resolved mass measurements for $\gtrsim10^9$ \Msun\ SMBHs out to cosmological distances.

Dynamically weighing SMBHs requires resolving the line-of-sight (LOS) movements of stars or gas at scales corresponding to $R_{\rm SOI}$. The angular size subtended by a physical scale like $R_{\rm SOI}$ depends on the angular-diameter distance ($D_A$). Crucially, in standard cosmological models, $D_A(z)$ does not increase monotonically with redshift $z$. As shown in \autoref{fig:cosmological_dist}, it reaches a maximum value of $D_A\approx1750$ Mpc around $z\approx1.6$ and then decreases. This means that, counterintuitively, very distant objects of a given size can appear slightly larger in angular size than objects at intermediate redshifts. For example, a galaxy with $\sigma_\star\approx300$ \kms\ hosting a SMBH with \Mbh\ $\approx1.8\times10^9$ \Msun\ according to the \Mbh--$\sigma_\star$ relation \citep[eq.~3]{Kormendy13} has an $R_{\rm SOI}$ that would subtend $\approx10$ milliarcseconds (mas) at a distance $D_A\approx1800$ Mpc. But this distance is larger than the peak $D_A$. This implies that this $R_{\rm SOI}$ could in principle be resolved at any redshift. This angular scale is within the reach of the ELT's resolution.

However, even if the SOI is angularly resolved, observing high-redshift objects presents a significant challenge due to cosmological surface brightness dimming, which produces a steep decrease of the observed \emph{bolometric} surface brightness $\Sigma \propto (1 + z )^{-4}$. The expansion of the Universe causes the observed surface brightness of distant objects to decrease significantly, leading to lower signal-to-noise ratios (S/N) and increased measurement uncertainties. This sensitivity limit makes accurate dynamical measurements beyond the local Universe difficult with current facilities.

\begin{table*}
\begin{center}
\caption{Photometric properties of simulated galaxies.\label{tab:sampledata}}  
\begin{tabular}{cccccccccccc}
 \hline\hline 
Galaxy name&$\alpha$(J2000)&$\delta$(J2000)&$z$&$q$&$m$ (filter)&$R_\textrm{e}$&$n$&$\lg M_{\star}$&$\sigma_\textrm{e}$&$D_A$&Scale\\
&(h:m:s)&($^\circ:\arcmin:\arcsec$)& & ($b/a$)& (total mag)&(\arcsec)& &($M_{\odot}$)&(\kms)&(Mpc)&(kpc\,$\arcsec^{-1}$)\\
(1)&(2)&(3)&(4)&(5)&(6)&(7)&(8)&(9)&(10)&(11)&(12)\\	             
\hline 
LEGAC-86906($^{1}$) &10:02:13.375&   02:12:41.19&0.82&0.75&20.01 (F814W)&0.92&5.8&11.84&295.5&1562.3&7.57\\
LEGAC-227516($^{1}$)&10:00:52.644&   02:43:17.04&0.78&0.95&20.48 (F814W)&0.51&4.1&11.68&304.0&1534.7&7.44\\
S2F1-142($^{2}$)    &03:06:36.511&$-$00:03:01.00&1.39&0.74&18.65 (F160W)&0.36&3.5&11.52&347.0&1737.9&8.43\\ 
UDS-29410($^{3}$)   &02:17:51.220&$-$05:16:21.84&1.46&0.54&20.18 (F160W)&0.22&2.6&11.29&371.0&1743.0&8.45\\ 
CP-1243752($^{4}$)  &10:00:17.746&   02:17:52.71&2.09&0.79&20.82 (F160W)&0.34&4.5&11.66&350.0&1717.0&8.33\\  
\hline  
\end{tabular}
\end{center}
\parbox[t]{0.99\textwidth}{\textit{Notes:}  Column (1): Galaxy name with numerical superscripts indicate the primary reference for photometric parameters in Columns (1)--(8): ($^{1}$) \citet{vanderWel21}; ($^{2}$) \citet{Longhetti2014}; ($^{3}$) \citet{vandeSande13}; ($^{4}$) \citet{Stockmann20}. 
Column (2): Right Ascension (J2000).
Column (3): Declination (J2000).
Column (4): Redshift estimated for each galaxy with photometric redshift for LEGA-C galaxies and spectroscopic redshift for others.
Column (5): Observed axis ratio $q = b/a$.
Column (6): Total apparent AB magnitude $m$ in the specific observed band.
Column (7): Effective (half-light) radius $R_\mathrm{e}$ in arcseconds.
Column (8): Sérsic index $n$.
Columns (9)--(10): Logarithm of the stellar mass ($\lg(M_{\star}/M_{\odot})$, and stellar velocity dispersion ($\sigma_\textrm{e}$) in km~s$^{-1}$. Those values obtained from \citet{Cappellari23} for LEGAC-86906 and LEGAC-227516, from \citet{Longhetti2014} for S2F1-142, and from \citet{vandeSande13} for UDS 29410 and CP-1243752.
Column (11): Angular-diameter distance $D_A$ in Mpc.
Column (12): Physical scale in kpc arcsec$^{-1}$ at the galaxy's redshift.
Columns (11)--(12): $D_A$ and physical scale are calculated from $z_{\rm spec}$ (Column 4) assuming a flat concordance $\Lambda$CDM cosmology ($H_0=70$ km s$^{-1}$ Mpc$^{-1}$, $\Omega_m=0.3$, $\Omega_\Lambda=0.7$), using the calculator by \citet{Wright2006}.}  
\end{table*}

\begin{figure*} 
    \centering
    \includegraphics[width=0.95\textwidth]{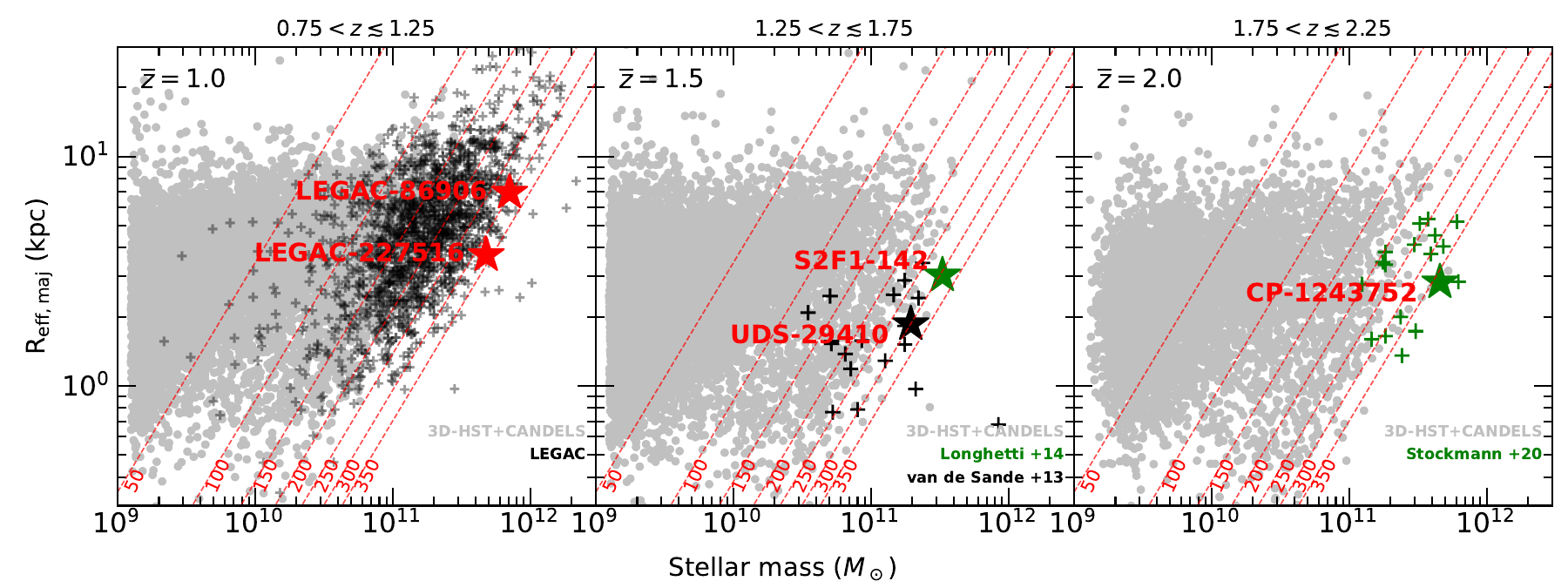}
    \caption{Our selected sample of five quiescent, massive, red-sequence galaxies (large colored stars with red name tags; \autoref{tab:sampledata}) is shown in the mass–size diagram, separated into three redshift panels: $z \approx 1$ (left), $z \approx 1.5$ (middle), and $z \approx 2$ (right). These galaxies are overlaid on the 3D-\hst\ CANDELS sample \citep[gray background points;][]{vanderWel14}, which is divided into corresponding redshift intervals ($0.75 < z \leq 1.25$, $1.25 < z \leq 1.75$, and $1.75 < z \leq 2.25$) and spans a stellar mass range of $2 \times 10^9 <$ \Mstar $ < 2 \times 10^{12}$ \Msun. Red dashed inclined lines represent contours of constant stellar velocity dispersion from 50 to 350 \kms, estimated using the virial relation \citep{Cappellari2006, Cappellari23, Krajnovic18a}.}
    \label{fig:highz_sample}
\end{figure*}

\begin{figure*}   
    \centering
    \includegraphics[width=0.99\textwidth]{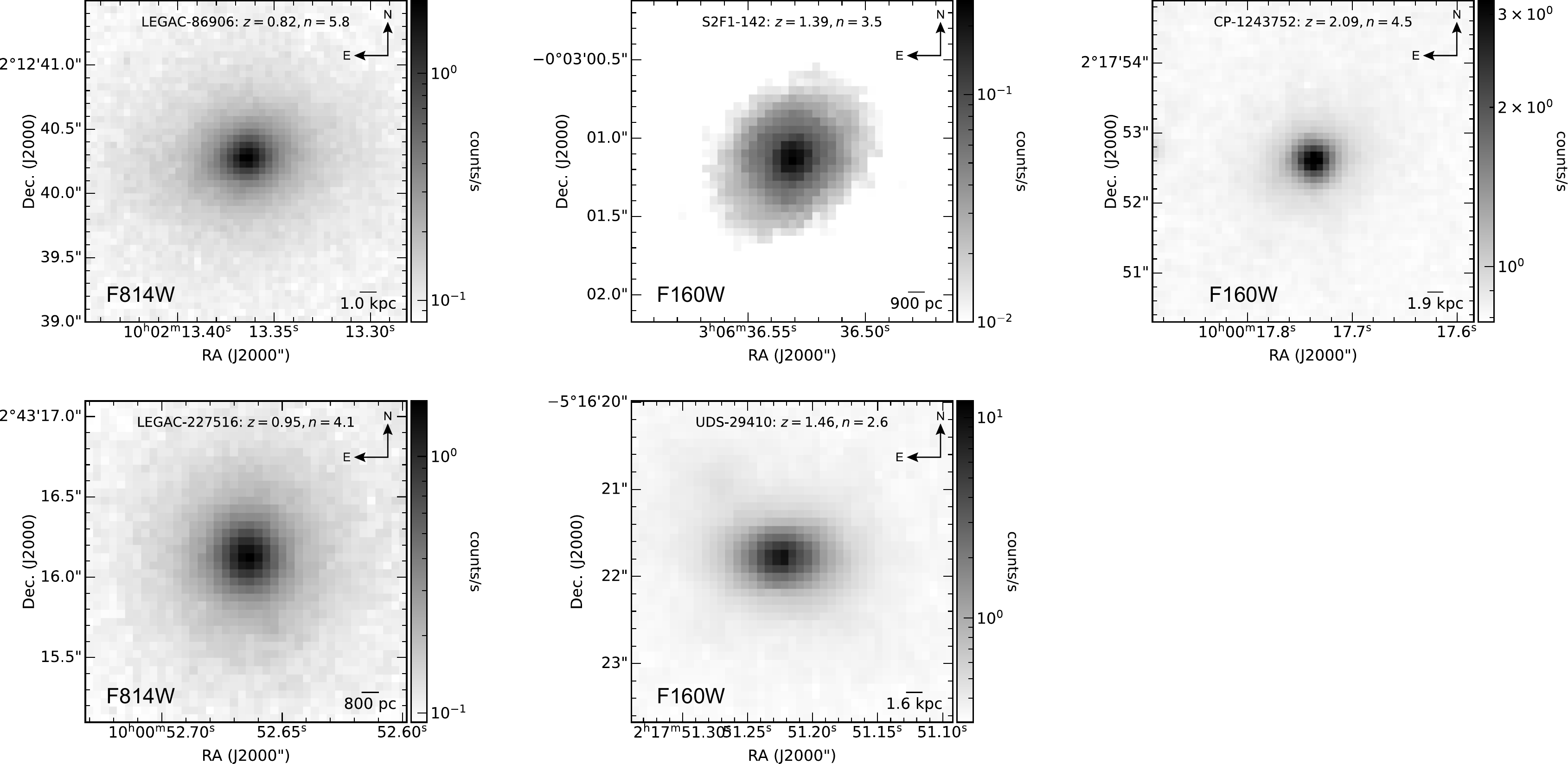}
    \caption{\hst/WFC3 F814W and F160W images for five galaxies in our simulated sample are displayed in grayscale.}
    \label{fig:HST_image}
\end{figure*}

In this paper, we evaluate the potential of the ELT's cutting-edge capabilities to overcome these limitations. We begin by describing our galaxy sample selection and the intrinsic Sérsic models that form the basis of our study in \autoref{sec:galaxy_sample_and_sersic_models}. We then detail the simulation methodology for generating mock ELT observations with both MICADO and HARMONI in \autoref{sec:simulating_mock_elt_observations}. The subsequent processing and analysis of these simulated data, including PSF determination from MICADO point source images, surface brightness parametrization of galaxy images, and stellar kinematic extraction from HARMONI data cubes, are presented in \autoref{sec:analysis_of_simulated_elt_data}. Following this, \autoref{sec:dynamical_modelling_and_smbh_mass_recovery} outlines the dynamical modelling techniques employed, using Jeans Anisotropic Models (JAM) and Markov chain Monte Carlo (MCMC) fitting, to recover SMBH masses from the processed data. Finally, in \autoref{sec:conclusions}, we summarize our key findings, discuss the implications for future SMBH research at high redshifts, and assess the overall feasibility of these challenging but crucial measurements for understanding galaxy–BH co-evolution.

Although a factor of more than 10 improvement in angular resolution adopted in this study with the ELT HARMONI and MICADO nominally compensates for the $\sim$10 increase in angular diameter distance at $z \sim 1$–2, our results demonstrate that spatial resolution alone is insufficient to guarantee successful black hole mass measurements at high redshift. By modeling physically motivated galaxy populations appropriate for cosmic noon, and by explicitly accounting for surface-brightness dimming, adaptive-optics performance and PSF structure, wavelength-dependent instrumental effects, and realistic signal-to-noise limitations, we show that only a subset of massive, favorable systems can be dynamically probed. Our simulations therefore provide quantitative constraints on which galaxies are viable targets and under what observational conditions, establishing a realistic framework for future ELT studies that goes well beyond simple angular-resolution scaling from the local Universe.

We adopt a standard flat concordance $\Lambda$CDM Universe model in all calculations with $H_0=70$ km s$^{-1}$ Mpc$^{-1}$, $\Omega_m=0.3$, and $\Omega_\Lambda=0.7$. To ensure consistency, we applied a foreground extinction correction using the method described by \citet{Schlafly11} and adopt the interstellar extinction law from \citet{Cardelli89} for all relevant quantities. Additionally, we used the AB photometric magnitude system \citep{Oke74} throughout our analysis.

\section{Galaxy Sample and Sérsic Models}\label{sec:galaxy_sample_and_sersic_models}

\begin{table*}
\caption{The Sérsic light-MGE models of five simulated galaxies\label{tab:mge_para}} 
\begin{tabular}{ccccccccc}
\hline\hline
$j$ & Galaxy & $\lg\Sigma_{\star,j}^{\rm lum}$ (\Lsun${\rm pc^{-2}})$ &$\lg\sigma_j$ ($\arcsec$) &$q'_j=b_j/a_i$&  Galaxy &$\lg\Sigma_{\star,j}^{\rm lum}$(\Lsun${\rm pc^{-2}})$ &$\lg\sigma_j$ ($\arcsec$) &$q'_j=b_j/a_i$\\
  (1)   &   (2)  & (3) &  (4) & (5)  &  (6) & (7) & (8) & (9)\\	         
\hline
1 &\multirow{15}{*}{LEGAC-8906}&4.463   &$-$3.000&0.75& \multirow{15}{*}{LEGAC-227516}&  4.053 &$-$3.000&0.95 \\
2 &                            &4.189   &$-$2.630&0.75&                               &  3.874 &$-$2.638&0.95 \\
3 &                            &3.951   &$-$2.318&0.75&                               &  3.712 &$-$2.336&0.95 \\
4 &                            &3.727   &$-$2.029&0.75&                               &  3.559 &$-$2.062&0.95 \\
5 &                            &3.488   &$-$1.751&0.75&                               &  3.392 &$-$1.802&0.95 \\
6 &                            &3.226   &$-$1.482&0.75&                               &  3.199 &$-$1.552&0.95 \\
7 &                            &2.936   &$-$1.220&0.75&                               &  2.977 &$-$1.312&0.95 \\
8 &                            &2.619   &$-$0.966&0.75&                               &  2.724 &$-$1.081&0.95 \\
9 &    ($z\approx0.8213$)      &2.274   &$-$0.720&0.75&     ($z\approx0.7794$)        &  2.438 &$-$0.858&0.95 \\
10&        ($I$ band)          &1.899   &$-$0.482&0.75&         ($I$ band)            &  2.118 &$-$0.644&0.95 \\
11&                            &1.498   &$-$0.252&0.75&                               &  1.763 &$-$0.438&0.95 \\
12&                            &1.068   &$-$0.028&0.75&                               &  1.373 &$-$0.238&0.95 \\
13&                            &0.615   & 0.193  &0.75&                               &  0.949 &$-$0.039&0.95 \\
14&                            &0.135   & 0.431  &0.75&                               &  0.477 & 0.178  &0.95 \\
15&                            &$-$0.502& 0.758  &0.75&                               &$-$0.209& 0.475  &0.95 \\
\hline
1 &\multirow{15}{*}{S2F1-142}&5.002&$-$3.000&0.74& \multirow{15}{*}{UDS-29410}&4.515&$-$3.000&0.54 \\
2 &                          &4.861&$-$2.642&0.74&                            &4.456&$-$2.650&0.54 \\
3 &                          &4.730&$-$2.347&0.74&                            &4.392&$-$2.364&0.54 \\
4 &                          &4.608&$-$2.083&0.74&                            &4.331&$-$2.113&0.54 \\
5 &                          &4.473&$-$1.834&0.74&                            &4.257&$-$1.879&0.54 \\
6 &                          &4.312&$-$1.596&0.74&                            &4.156&$-$1.658&0.54 \\
7 &                          &4.120&$-$1.368&0.74&                            &4.023&$-$1.447&0.54 \\
8 &                          &3.896&$-$1.149&0.74&                            &3.852&$-$1.247&0.54 \\
9 &    ($z\approx1.386$)     &3.636&$-$0.938&0.74&     ($z\approx1.456$)      &3.641&$-$1.056&0.54 \\
10&        ($H$ band)        &3.340&$-$0.737&0.74&         ($H$ band)         &3.386&$-$0.873&0.54 \\
11&                          &3.005&$-$0.543&0.74&                            &3.084&$-$0.700&0.54 \\
12&                          &2.632&$-$0.354&0.74&                            &2.734&$-$0.532&0.54 \\
13&                          &2.221&$-$0.166&0.74&                            &2.331&$-$0.363&0.54 \\
14&                          &1.745&  0.041 &0.74&                            &1.828&$-$0.176&0.54 \\
15&                          &1.027&  0.320 &0.74&                            &1.020&  0.068 &0.54 \\
\hline
1 &\multirow{15}{*}{CP-1243752}&6.101   &$-$3.000&0.79&   &  &  &  \\
2 &                            &5.861   &$-$2.640&0.79&   &  &  &  \\
3 &                            &5.643   &$-$2.349&0.79&   &  &  &  \\
4 &                            &5.448   &$-$2.092&0.79&   &  &  &  \\
5 &                            &5.252   &$-$1.852&0.79&   &  &  &  \\
6 &                            &5.039   &$-$1.622&0.79&   &  &  &  \\
7 &                            &4.802   &$-$1.399&0.79&   &  &  &  \\
8 &                            &4.538   &$-$1.184&0.79&   &  &  &  \\
9 &    ($z\approx2.090$)       &4.247   &$-$0.975&0.79&   &  &  &  \\
10&       ($H$ band)           &3.927   &$-$0.773&0.79&   &  &  &  \\
11&                            &3.578   &$-$0.577&0.79&   &  &  &  \\
12&                            &3.201   &$-$0.385&0.79&   &  &  &  \\
13&                            &2.797   &$-$0.192&0.79&   &  &  &  \\
14&                            &2.348   &  0.023 &0.79&   &  &  &  \\
15&                            &1.691   &  0.325 &0.79&   &  &  &  \\
\hline
\end{tabular}
\parbox[t]{0.99\textwidth}{\textit{Notes:} Column (1): Index $j$ of the Gaussian component. Columns (2) and (6): Galaxy name (and redshift). Columns (3) and (7): Logarithm of the stellar surface luminosity density either in $I$ or $H$ bands, $\lg\Sigma_{\star,j}^{\rm lum}$ (in \Lsun\,${\rm pc^{-2}}$). Columns (4) and (8): Logarithm of the Gaussian dispersion (width), $\lg\sigma_j$ (in arcseconds), along the major axis. Columns (5) and (9): Axial ratio, $q'_j = b_j/a_j$.}  

\end{table*}

\subsection{High-Redshift Galaxy Sample Selection}

To evaluate the capabilities of HARMONI and MICADO for measuring SMBH masses at high redshift, we selected a sample of galaxies within the range $1\lesssim z\lesssim2$. The primary selection criteria were the availability of essential observational data: spectroscopic redshift ($z$), high-resolution \hst\ imaging, photometric parameters (total apparent magnitude, Sérsic profile parameters $n$ and $R_\mathrm{e}$), stellar mass (\Mstar), and stellar velocity dispersion ($\sigma_\star$). Crucially, we targeted the brightest known quiescent, massive galaxies at these redshifts to maximize the S/N achievable in realistic mock observations, mitigating the challenges posed by cosmological surface brightness dimming.

Candidate galaxies were sourced from several surveys. For $z\approx1$, we utilized the Large Early Galaxy Astrophysics Census (LEGA-C) survey \citep{vanderWel16, vanderWel21}. For $z\gtrsim1.5$, we drew from various deep-field surveys targeting massive, quiescent galaxies, including the NEWFIRM Medium-Band Survey \citep[NMBS;][]{vanDokkum09,Whitaker10}, the UKIRT Infrared Deep Sky Survey \citep[UKIDSS;][]{Williams09, Lawrence07}, KMOS and X-shooter spectroscopic campaigns \citep{Stockmann20}, and VLT cluster surveys \citep{vandeSande13, Beifiori17}, ensuring the availability of necessary \hst/WFC3 imaging.

From these sources, we selected a representative sample of five galaxies, chosen as among the brightest known examples at their respective redshifts possessing the required data: two at $z\approx1$ (LEGAC-86906, LEGAC-227516), two at $z\approx1.5$ (S2F1-142, UDS 29410), and one at $z\approx2$ (CP-1243752). Their detailed properties are listed in \autoref{tab:sampledata}.    

Even with next-generation facilities such as HARMONI and MICADO on the ELT, it will not be feasible to sample the \Mbh–$\sigma$ relation across the full galaxy population in the same manner as in the local Universe. At high redshift, kinematically resolving the BH SOI will necessarily be restricted to the most massive, luminous, and structurally favorable systems, thereby probing only the “tip of the iceberg” of the underlying population. Accordingly, the aim of this work is not to define a target sample size or to statistically map the evolution of the scaling relation, but rather to demonstrate that direct, spatially resolved measurements of \Mbh\ will be achievable for a small yet critical subset of galaxies out to Cosmic Noon ($1\lesssim z\lesssim2$). Such measurements will provide essential anchor points for calibrating indirect \Mbh\ estimators and for assessing potential evolution in the \Mbh–galaxy scaling relations. Determining how far these observational limits can be extended in terms of sample size and galaxy properties will ultimately depend on early ELT results and is beyond the scope of the present study.

\autoref{fig:highz_sample} shows the position of our selected sample (large colored stars) on the mass–size diagram across three redshift bins, overlaid on the 3D-\hst\ CANDELS sample \citep{vanderWel14}. This confirms our targets are massive and relatively compact, characteristic of quiescent galaxies at these epochs. \autoref{fig:HST_image} displays the archival \hst/WFC3 F814W or F160W images for our sample. While large surveys contain numerous galaxies at these redshifts, most are fainter or lack the complete dataset needed for our detailed simulations, justifying our focus on these exceptionally bright systems.

\subsection{Adopted Intrinsic Galaxy Properties: Sérsic Models}\label{sec:adopted_intrinsic_galaxy_properties_sersic_models}

The light distribution of each galaxy is modeled using the S\'ersic profile \citep{Sersic68}:
\begin{equation} 
    I(R) = I_\mathrm{e}\exp \left\{-b_n\left[{\left(\frac{R}{R_\mathrm{e}}\right)^{1/n}}-1\right]\right\}	 
\end{equation} 
where $n$ is the S\'ersic index, $R_\mathrm{e}$ is the effective radius, and $I_e$ is the intensity at $R_\mathrm{e}$. The term $b_n=Q^{-1}(2n,1/2)$ is a coefficient, with $Q^{-1}$ being the inverse regularized incomplete gamma function\footnote{Implemented in \href{https://docs.scipy.org/doc/scipy/reference/generated/scipy.special.gammainccinv.html}{scipy.special.gammainccinv}} \citep[eq.~14]{Zhu2025}, ensuring that $R_\mathrm{e}$ encloses half the total light. The parameters for each galaxy were previously published in the literature and are reproduced in \autoref{tab:sampledata}. We assumed these Sérsic models represent the true intrinsic galaxy surface brightness as they were convolved with the \hst\ PSF, down to the smallest radii resolvable with the ELT. These intrinsic models form the basis for the simulations of mock ELT observations detailed in the subsequent sections, where they are used as input for generating both MICADO images and HARMONI data cubes. This approach allows us to test the capabilities of ELT instrumentation in recovering galaxy and BH properties starting from well-defined, albeit idealized, galaxy profiles.

\begin{figure*}   
    \centering
    \includegraphics[width=0.99\textwidth]{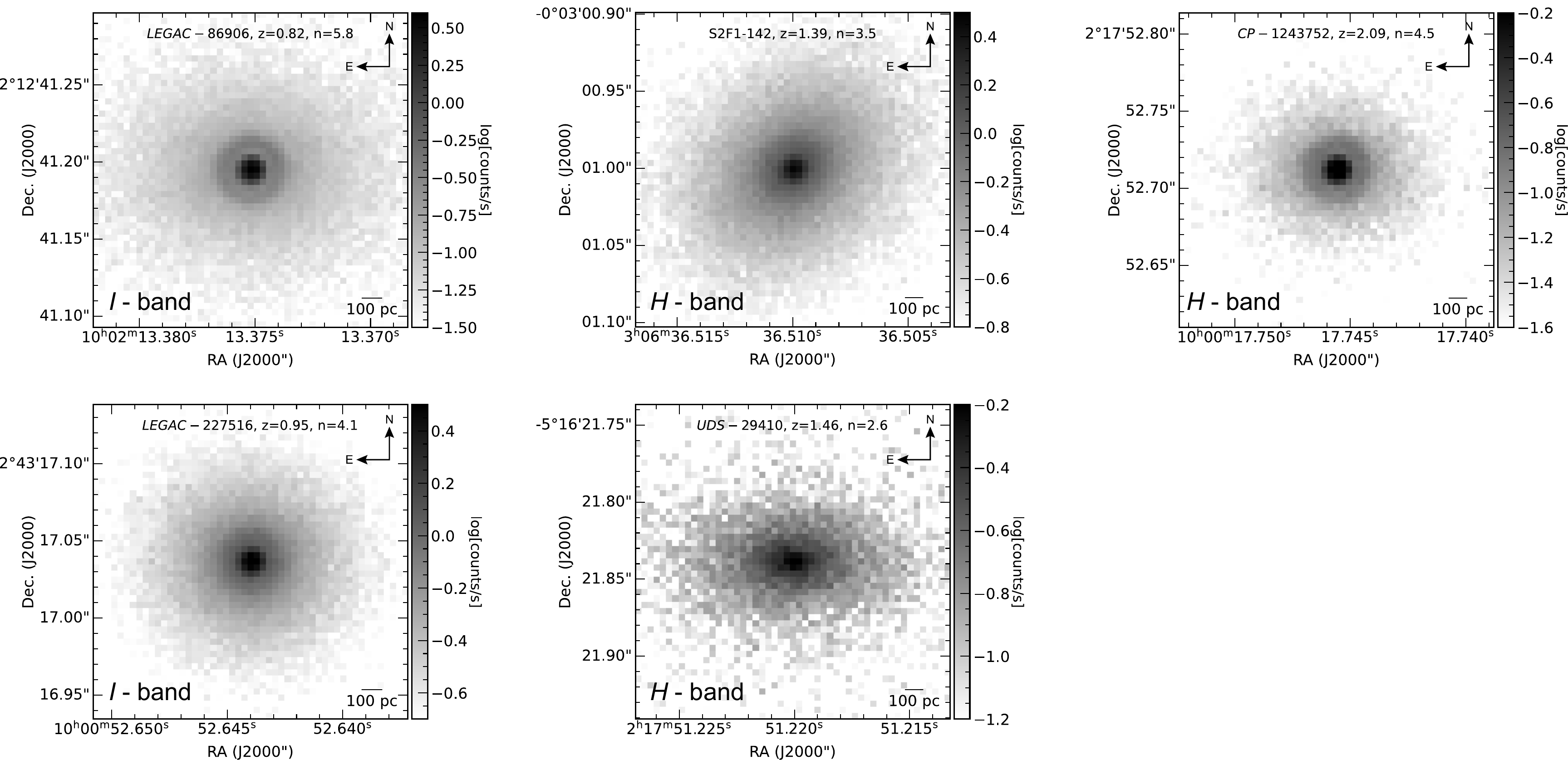}
        \caption{Grayscale mock MICADO images of the five simulated galaxies, produced using \textsc{SimCADO}.}
    \label{fig:simcado_MICADO}
\end{figure*}

\section{Simulating Mock ELT Observations}\label{sec:simulating_mock_elt_observations}

\subsection{Simulation Strategy Overview}\label{sec:simulation_strategy_overview}

The intrinsic galaxy properties, represented by the Sérsic models detailed in \autoref{sec:adopted_intrinsic_galaxy_properties_sersic_models}, form the foundation for our simulations of ELT observations. We use these models to generate realistic mock data for both the MICADO imager and the HARMONI integral-field spectrograph (IFS). This approach allows us to assess the ELT's capability to recover key galaxy parameters, including SMBH masses, from high-redshift galaxies. The following subsections describe the simulation process for each instrument.

\subsection{Simulating MICADO Imaging}\label{sec:simulating_micado_imaging}

MICADO (Multi-AO Imaging Camera for Deep Observations) is the ELT's first-light diffraction-limited imager, designed to deliver high sensitivity and a wide field of view (FoV, e.g., $\sim 50\arcsec \times 50\arcsec$ with a fine pixel scale of a few milliarcseconds) across near-infrared (NIR) wavelengths \citep{Davies2018, Davies21}. To simulate MICADO observations, we utilize the \textsc{SimCADO}\footnote{\url{https://pypi.org/project/SimCADO/}} software package \citep{Leschinski16}. \textsc{SimCADO} is a versatile tool that models the instrument's optical path, detector characteristics, and the effects of atmospheric turbulence corrected by AO.

The input to \textsc{SimCADO} for our target galaxies are the 1D Sérsic surface brightness profiles described in \autoref{sec:adopted_intrinsic_galaxy_properties_sersic_models}, with the corresponding observed axial ratio. \textsc{SimCADO} convolves these intrinsic light distributions with appropriate PSFs expected for MICADO, considering the chosen observing mode and the single-conjugate AO (SCAO\footnote{\url{https://simcado.readthedocs.io/en/latest/user_docs/9_PSFs.html}}) using a natural guide star (NGS) performance. 

In the SCAO mode, MICADO on the ELT is designed to deliver diffraction-limited NIR imaging (0.8–2.4 $\mu$m) by correcting atmospheric turbulence along the direction of a bright NGS. SCAO offers higher on-axis Strehl ratios than multi-conjugate AO (MCAO) over a smaller FoV, but performance depends on the guide star brightness and distance from the science target, with optimal results requiring a bright ($V\lesssim15-16$ mag) star within $\sim$15–20\arcsec\ of the field center. Under nominal conditions, SCAO is expected to achieve on-axis Strehl ratios of $\gtrsim$60\% and FWHM of $\sim$10 mas in $H$ band, with correspondingly sharp PSFs featuring a strong diffraction-limited core and extended wings due to residual wavefront errors. The PSF varies with field position and atmospheric conditions, and detailed PSF reconstruction algorithms aim to model these variations with $\sim$10--15\% precision in both Strehl and FWHM out to significant off-axis angles. This high-performance PSF is essential for accurately modeling the high–spatial-resolution surface brightness of galaxy nuclei, enabling the recovery of fine structural details and precise photometry required for robust SMBH mass measurements.

Realistic sky background levels and detector noise (read noise, dark current) are also incorporated to produce mock images that closely resemble actual MICADO observations. We produced images in either the $I$-band or $H$-band (\autoref{fig:simcado_MICADO}), depending on redshift, which is close to the wavelength of the mock HARMONI data cubes. This is the same band we would need to observe in real data to ensure we capture the distribution of the tracer population, which is required by the dynamical models.

To generate the mock MICADO images, we first constructed emission curves based on the spectral energy distribution (SED) of an elliptical galaxy, adopting either the $I$- or $H$-band filter. The spectral axis was adjusted to account for the redshifts of the simulated galaxies by shifting the SEDs to longer wavelengths. Each galaxy was treated as a single source and modelled according to Eq. (1).

We then used the optical train implemented in \textsc{SimCADO} to simulate the propagation of light from the galaxy through the atmosphere, the ELT primary mirror, the instrument optics, and the detector. Detector and sky noise were added using the HxRG Noise Generator (HxRG NG), originally described by \citet{Rauscher2015}, which models the dominant noise components in NIR detectors, including white read noise, correlated and uncorrelated $1/f$ noise, alternating column noise, residual bias drifts, and so-called “picture-frame” noise. The resulting noise frames, expressed in electrons, were combined with the simulated signal to produce realistic mock images consistent with expected detector performance.

All simulations adopted a diffraction-limited point-spread function in SCAO mode, with a characteristic FWHM$_{\rm PSF} \simeq 10$ mas \citep{Davies21}. To emulate realistic observing conditions, we fixed the exposure time per frame to {\tt obs\_dit = 15} minutes and varied the number of exposures ({\tt n\_dit}). We find that {\tt n\_dit = 4} provides optimal performance for all simulated galaxies, corresponding to a total on-source integration time of one hour, excluding target acquisition, overheads, and SCAO setup.

The resulting mock MICADO images are then used to derive detailed Multi-Gaussian Expansion (MGE) surface brightness models of the stellar tracer population for the JAM dynamical models.

\subsection{The HARMONI Instrument and HSIM Simulator}

HARMONI is planned as the first-light IFS for the ELT, designed to operate across optical and NIR wavelengths (0.458--2.469 $\mu$m). It utilizes an image slicer integral field unit (IFU).

It is important to note that HARMONI is currently undergoing a rescoping review. This review process may result in modifications to the instrument's final specifications, which are not yet settled at the time of writing. The simulations presented in this paper are based on the latest well-defined specifications available prior to the conclusion of the rescoping review.

Based on these pre-review specifications, HARMONI was designed to offer four spatial scales ($4\times4$, $10\times10$, $20\times20$, and $30\times60$ mas$^2$), corresponding to FoVs ranging from $0\farcs86\times0\farcs61$ to $9\farcs12\times6\farcs42$. It was planned to provide three spectral resolving powers ($\lambda/\Delta\lambda\approx3300$, 7100, and 17400) via 13 gratings \citep{Zieleniewski15}. HARMONI will be coupled with AO systems—either SCAO using a NGS or laser tomography AO (LTAO)—to correct for atmospheric turbulence and achieve near-diffraction-limited performance \citep{Thatte16, Thatte20}.

HARMONI aims to enable a wide range of science cases, including the study of galaxy kinematics and stellar populations at high redshift \citep{Kendrew16} and the measurement of black hole masses \citep{Nguyen23, Nguyen2025b, Ngo2025b, Ngo2025c}, which are the focus of this work.

We generated mock HARMONI observations using the \textsc{HSIM} software\footnote{v3.1 \url{https://github.com/HARMONI-ELT/HSIM}} \citep{Zieleniewski15}, a \textsc{Python} pipeline designed specifically for HARMONI. \textsc{HSIM} takes high-resolution, noise-free input data cubes, which represent the intrinsic properties of astronomical sources, and simulates the effects introduced by the atmosphere, telescope, and instrument based on their specified characteristics. This simulation process includes applying wavelength-dependent PSFs, detector effects, and various noise sources to produce realistic output data resembling actual HARMONI observations.

\subsection{Generating Input Noiseless Data Cubes}\label{sec:generating_input_noiseless_data_cubes}

To generate the input noiseless data cubes for \textsc{HSIM}, we developed a dedicated \textsc{Python} routine, \texttt{jam\_mock\_data\_cube}\footnote{Available from \url{https://github.com/micappe/jam_mock_data_cube}}. This routine takes a calibrated MGE model (in $L_\odot\text{arcsec}^{-2}$ for a given photometric band) and a stellar template spectrum as primary inputs. It then produces a mock noiseless 3-dimensional (3D) data cube, considering specified parameters for anisotropy, BH mass, and inclination.

\begin{figure*}
    \centering
    \includegraphics[width=0.99\textwidth]{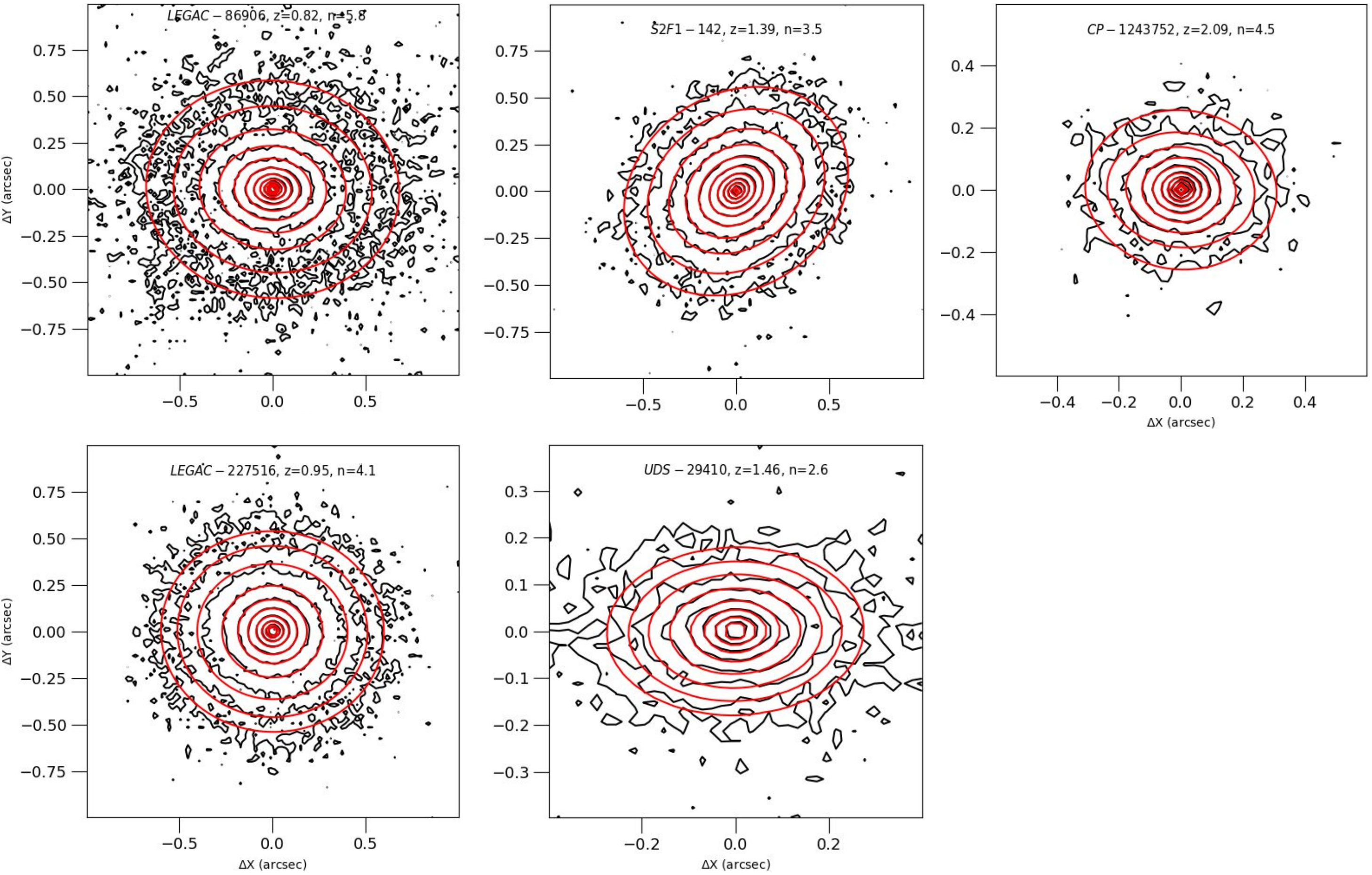}
    \caption{The comparison between the MICADO images produced by \textsc{SimCADO} and their best-fit MGE models for five simulated galaxies is presented in terms of 2D surface brightness density. Black contours represent the data, while red contours depict the model, illustrating the alignment between data and model at corresponding radii and contour levels which are spaced decreasing by 0.5 mag arcsec$^{-1}$ outward.}
    \label{fig:high_z_mge}
\end{figure*}

\begin{table*}
\caption{Mock \textsc{HSIM} IFS in the $H+K$ grating.\label{tab:hsimmock}}
\begin{tabular}{cccccccc}
 \hline\hline
Galaxy name &\hst\ image&Age&Metallicity&$\lg$\Mbh(2)&$\lg$\Mbh(3)&\textsc{HSIM}&\textsc{HSIM}$_{\rm sensitivity}$ \\
&archival &(Gyr)&[Fe/H]/MARCS&(\Msun)&(\Msun)&({\tt NDIT})&({\tt NDIT})\\
(1)&(2)&(3)&(4)&(5)&(6)&(7) &(8)\\	
\hline
LEGAC-86906($^{1}$) &($I$ band) F814W&4.5&0.0/{\tt z002}&9.10&9.64&50&30\\
LEGAC-227516($^{1}$)&($I$ band) F814W&4.0&0.0/{\tt z002}&9.18&9.54&80&50\\
S2F1-142($^{2}$)    &($H$ band) F160W&2.3&0.0/{\tt z002}&9.60&9.71&40&30\\
UDS 29410($^{3}$)   &($H$ band) F160W&0.7&0.0/{\tt z002}&9.48&9.93&40&30\\
CP-1243752($^{4}$)  &($H$ band) F160W&2.6&0.0/{\tt z002}&9.51&9.85&50&30\\
\hline
\end{tabular}
\parbox[t]{0.99\textwidth}{\textit{Notes:}
Column (1): Galaxy name. 
Column (2): Indicates if an archival \hst\ image is available. 
Column (3): Assumed median stellar population age in Gyr. Ages for LEGA-C galaxies are from ($^{1}$):\citet{Cappellari23}; ages for other galaxies are from ($^{2}$): \citet{Longhetti2014}; ($^{3}$): \citet{vandeSande13}; ($^{4}$): \citet{Stockmann20}.
Column (4): Metallicity in terms of [Fe/H] and the corresponding MARCS synthetic stellar spectra library identifier. Given that the MARCS synthetic stellar spectra library does not include stellar populations younger than 3 Gyr, we adopt its 3 Gyr stellar template as the SPS model for the three galaxies at $z \gtrsim 1.5$ in the mock HSIM IFS simulations those have ages less than this age.
Column (5): Estimated SMBH mass (in \Msun) using Eq. (2) from \citet{Krajnovic18b} (applicable for \Mstar\ $<M_{\rm crit}\approx2\times10^{11}$ \Msun). 
Column (6): Estimated SMBH mass (in \Msun) using Eq. (3) from \citet{Krajnovic18b} (applicable for \Mstar\ $>M_{\rm crit}$). 
Column (7): Total number of exposures ({\tt NDIT}) used in \textsc{HSIM} for the simulated HARMONI $H+K$ IFS observation. Each exposure has a duration of {\tt DIT = 15} minutes. 
Column (8): Sensitivity limit, expressed as the minimum number of exposures ({\tt NDIT}), representing the lowest S/N from the simulated IFS for which \textsc{pPXF} can still extract accurate kinematics. 
All listed exposure times are science time on target and do not include overheads for target acquisition, LTAO setup, or sky observations.} 
\end{table*}

\begin{table*}
\begin{center}
\caption{The mock MICADO mass-MGE models of five simulated galaxies\label{tab:mge}} 
\begin{tabular}{ccccccccc}
\hline\hline
$j$ & Galaxy &$\lg\Sigma_{\star,j}^{\rm mass}$(\Msun${\rm pc^{-2}})$ &$\lg\sigma_j$ ($\arcsec$) &$q'_j=b_j/a_i$& Galaxy &$\lg\Sigma_{\star,j}^{\rm mass}$(\Msun${\rm pc^{-2}})$ &$\lg\sigma_j$ ($\arcsec$) &$q'_j=b_j/a_i$\\
  (1)   &   (2)  & (3) &  (4) & (5)  &  (6) & (7)  &  (8) & (9) \\	         
\hline
1&\multirow{5}{*}{LEGAC-86906}&5.748&$-$2.097&0.95&\multirow{5}{*}{LEGAC-227516}&5.570&$-$2.353&0.95\\
2&                            &4.795&$-$1.656&0.95&                             &5.019&$-$1.84&0.95\\
3&                            &4.447&$-$1.391&0.95&                             &4.719&$-$1.464&0.94\\
4&    ($z\approx0.7794$)      &4.189&$-$1.022&0.95&    ($z\approx0.8213$)       &4.451&$-$1.057&0.93\\
5&         ($I$ band)         &3.827&$-$0.537&1.00&         ($I$ band)          &3.998&$-$0.488&0.90\\
\hline
1&\multirow{5}{*}{S2F1-142}&5.311&$-$2.097&0.80&\multirow{5}{*}{UDS-29410}&5.041&$-$2.210&0.75\\
2&                         &4.716&$-$1.641&0.80&                          &4.844&$-$1.648&0.75\\
3&                         &4.366&$-$1.371&0.80&                          &4.496&$-$1.260&0.75\\
4&    ($z\approx1.386$)    &4.267&$-$1.127&0.80&     ($z\approx1.456$)    &4.206&$-$0.840&0.75\\
5&         ($H$ band)      &3.961&$-$0.834&0.80&          ($H$ band)      &  -- &   --   & -- \\
6&                         &3.569&$-$0.510&0.80&                          &  -- &   --   & -- \\
7&                         &3.099&  0.222 &0.85&                          &  -- &   --   & -- \\
\hline
1&\multirow{7}{*}{CP-1243752}&5.819&$-$2.097&0.90&\multirow{9}{*}{---}& -- & -- & -- \\
2&                           &4.902&$-$1.620&0.90&                    & -- & -- & -- \\
3&                           &4.569&$-$1.296&0.90&                    & -- & -- & -- \\
4&                           &4.164&$-$0.952&0.90&                    & -- & -- & -- \\
5&    ($z\approx2.090$)      &3.552&$-$0.636&0.90&                    & -- & -- & -- \\
6&         ($H$ band)        &3.231&$-$0.381&0.90&                    & -- & -- & -- \\
7&                           &2.811&$-$0.222&0.95&                    & -- & -- & -- \\
\hline
\end{tabular}
\end{center}
\parbox[t]{0.99\textwidth}{\textit{Notes:} Column (1): Index $j$ of the Gaussian component. Columns (2) and (6): Galaxy name (and redshift). Columns (3) and (7): Logarithm of the stellar surface mass density, $\lg\Sigma_{\star,j}^{\rm mass}$ (in \Msun\,${\rm pc^{-2}}$). Columns (4) and (8): Logarithm of the Gaussian dispersion (width), $\lg\sigma_j$ (in arcseconds), along the major axis. Columns (5) and (9): Axial ratio, $q'_j = b_j/a_j$.}  
\end{table*}

In detail, we performed the following steps to create the noiseless data cubes:
\begin{enumerate}
    \item The initial 1-dimensional (1D) Sérsic profile, with parameters from \autoref{tab:sampledata}, is converted into an MGE using the \texttt{mge.fit\_1d} procedure from the \textsc{MgeFit} package\footnote{v5.0: \url{https://pypi.org/project/mgefit/}} \citep{Cappellari02}. The input logarithmically-sampled Sérsic profile is sampled from 0.01 to 100 times $R_\mathrm{e}$ and fitted with 15 Gaussians to ensure an accurate description of the galaxy's surface brightness profile \autoref{tab:mge_para}. 

    \item This MGE model serves as input for the \texttt{jam.axi.proj} procedure within the JAM package\footnote{v8.0: \url{https://pypi.org/project/jampy/}} \citep{Cappellari08, Cappellari20}. This step computes the 2-dimensional (2D) kinematic fields (mean line-of-sight velocity $V$ and velocity dispersion $\sigma$) for the given anisotropy, inclination, and BH mass. Specifically, JAM calculates the first ($V$) and second ($V_\mathrm{rms}$) velocity moments, from which the velocity dispersion is derived as $\sigma=\sqrt{V_{\rm rms}^2-V^2}$.

    \item The calculated kinematics are then used to construct a 3D data cube. This involves convolving the input stellar template spectrum with the line-of-sight velocity distribution (LOSVD) at each spatial position. The LOSVD is assumed to be a Gaussian function characterized by the mean velocity $V$ and dispersion $\sigma$ obtained in the previous step. This convolution is performed using the \texttt{varsmooth} function from the pPXF package\footnote{v9.4: \url{https://pypi.org/project/ppxf/}} \citep{Cappellari23}, which accurately handles potential undersampling of the LOSVD. The resolution of the input templates is from three times smaller to comparable to the HARMONI resolution. In all cases, we request an output sampling in \texttt{jam\_mock\_data\_cube} that is twice smaller than the HARMONI resolution, to ensure an accurate convolution with the line-spread function (LSF) within HSIM. 

    \item The routine then scales the flux of the input spectrum at each spatial pixel to match the observed surface brightness. This scaling utilizes the provided filter transmission curves along with the \texttt{ppxf\_util.mag\_sun} and \texttt{ppxf\_util.mag\_spectrum} functions from the pPXF package \citep{Cappellari23} to ensure correct photometric calibration in the specified band.

    \item Finally, the routine assembles the fully convolved and scaled spectra into a 3D data cube with the desired spatial and spectral sampling, ready to be used as input for the \textsc{HSIM} simulator.
\end{enumerate}

\begin{figure*}
    \centering
    \includegraphics[width=0.99\textwidth]{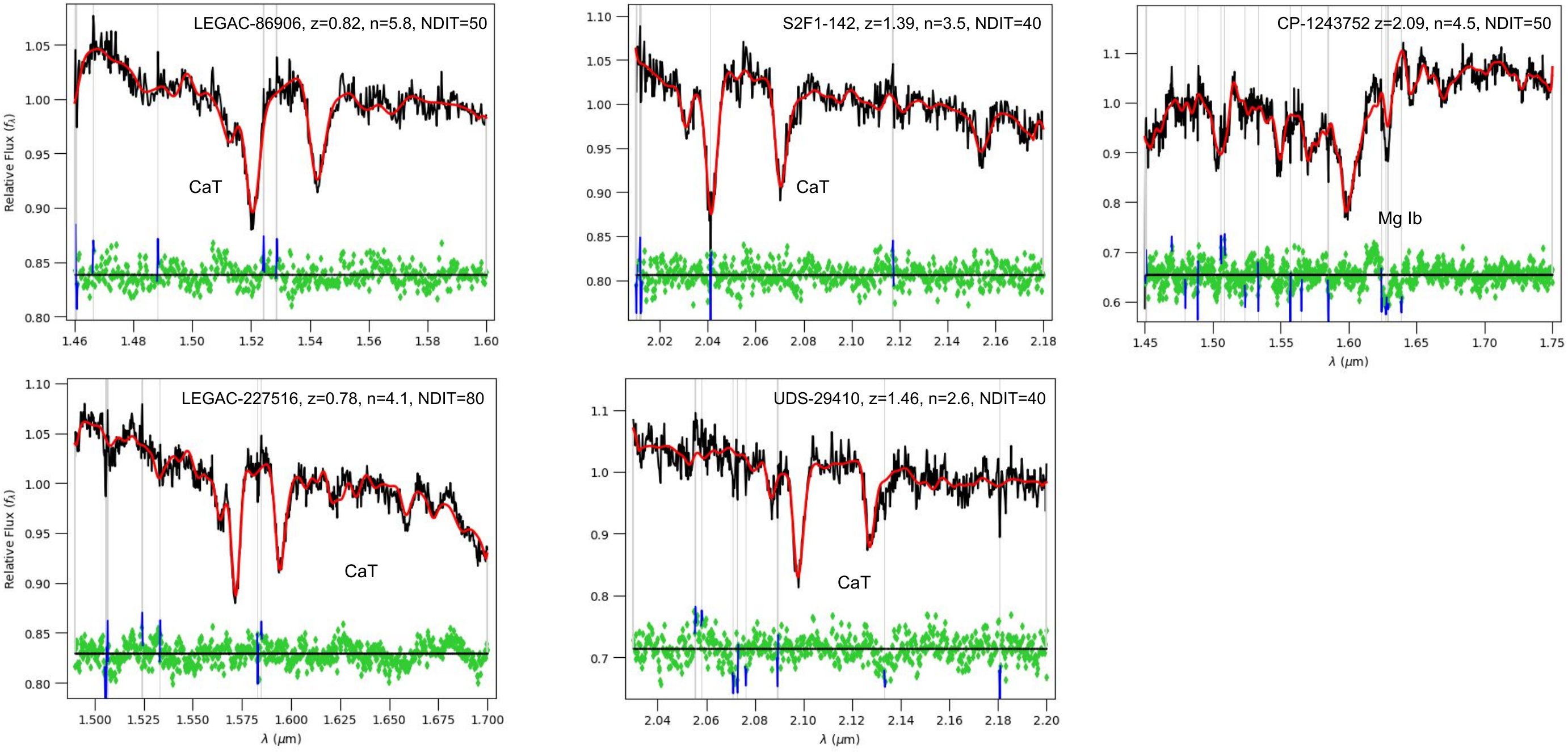}
    \caption{Examples of the $H+K$ HSIM mock spectra for five simulated galaxies are shown, including the targets at $z \approx 1$ (12.5-hour exposure for LEGAC-86906 and 20-hour exposure for LEGAC-27516), $z \approx 1.5$ (S2F1-142 and UDS-29410; 10-hour exposures for both), and $z \approx 2$ (CP-1243752; 12.5-hour exposure). The observed spectra (black) are fitted using \textsc{pPXF} (red), employing the stellar population models by \citet{Maraston11} based on the MARCS synthetic spectra \citep{Gustafsson08}. Prominent stellar absorption features, such as the CaT and \ion{Mg}{1}b lines, are marked in each panel. Fit residuals (i.e., {\tt data - model}) are shown in green below each spectrum, highlighting the fit quality. Vertical gray lines indicate the locations of sky emission lines, which are masked during the fit. These regions produce significant residuals, shown in blue, due to the exclusion from the fitting process.}
    \label{fig:spectrum_ppxf_hsim}
\end{figure*}

In our input to the \texttt{jam\_mock\_data\_cube} routine, we adopted the following specifications:
\begin{enumerate}

    \item \textbf{SMBH Mass Estimation:} For each galaxy, the central \Mbh\ for the HARMONI IFS simulation was estimated based on its \Mstar\ and $\sigma_e$. We used the relations from \citet{Krajnovic18b}: Eq.~(2) for $M_\star < M_{\rm crit}$ or Eq.~(3) for $M_\star > M_{\rm crit} = 2\times10^{11}$ \Msun.
    
    \item \textbf{Dynamical Anisotropy:} Consistent with observations that massive galaxies are generally characterized by low anisotropy \citep[fig.~10]{Cappellari2025}, we assumed an isotropic \textsc{JAM} model ($\beta_z=0$) for generating the kinematics.

    \item \textbf{Inclination:} Given the absence of published constraints on the inclinations of our five simulated targets, we adopted a uniform inclination of $i = 70^\circ$ for all systems in the \textsc{JAM} modelling when computing the 2D LOSVDs.
    
    \item \textbf{Simulation FoV and Spatial Sampling:} Considering the cosmall apparent size of the target galaxies and our objective to resolve kinematics near the SMBH's SOI, the mock kinematic fields were generated for a central $0\farcs4\times0\farcs4$ FoV. To ensure accurate subsequent convolution with the instrumental PSF by \textsc{HSIM}, these models were computed on a fine grid with $2\times2$ mas$^2$ pixels, which is five times smaller than the $10\times10$ mas$^2$ HARMONI spaxel size used in the simulations.
    
    \item \textbf{PSF Handling at Input Stage:} The LOSVDs generated by \textsc{JAM} were not convolved with any PSF at this stage, as PSF effects are incorporated later by the \textsc{HSIM} simulator.
    
    \item \textbf{Stellar Template Selection:} The input stellar template spectrum for \texttt{jam\_mock\_data\_cube} was taken from the  \citet{Maraston11} Stellar Population Synthesis (SPS) models, based on the MARCS synthetic library \citep{Gustafsson08}. For each galaxy, we selected a template with solar metallicity and an age that closely matched the known stellar population of the galaxy from \autoref{tab:hsimmock}.
    
\end{enumerate}

\subsection{Producing Mock HARMONI IFS Datacubes with \textsc{HSIM}}

The previously generated input-noiseless cubes were then processed with \textsc{HSIM} to simulate mock HARMONI IFS observations using the $H+K$ grating. \textsc{HSIM} performs several key steps: it convolves the input cubes with the HARMONI PSF, and then rebins and interpolates the spectral and spatial dimensions to match the HARMONI spectral resolution for the $H+K$ grating and the adopted pixel scale of $10\times10$ mas$^2$.

These simulations assumed median observational conditions for the Armazones site, including the LTAO mode with a NGS of 17.5 mag in the $H$-band within a 30$\arcsec$ radius. An optical zenith seeing at 0.5 $\mu$m with a FWHM of $0\farcs64$ and an airmass of 1.3 were assumed.

To emulate realistic observational strategies, multiple exposure frames and dithering patterns were applied. Each exposure utilized a Detector Integration Time ({\tt DIT}) of 15 minutes. The total exposure time for each simulation was calibrated to achieve a minimum S/N of 5 per \AA\ across the simulated FoV, and is determined by the number of exposures ({\tt NDIT}) as Total Exposure Time = {\tt DIT} $\times$ {\tt NDIT}. All specific \textsc{HSIM} simulation parameters and conditions are detailed in \autoref{tab:hsimmock}.

\section{Analysis of Simulated ELT Data}\label{sec:analysis_of_simulated_elt_data}

\begin{figure*}
    \centering
    \includegraphics[width=0.99\textwidth]{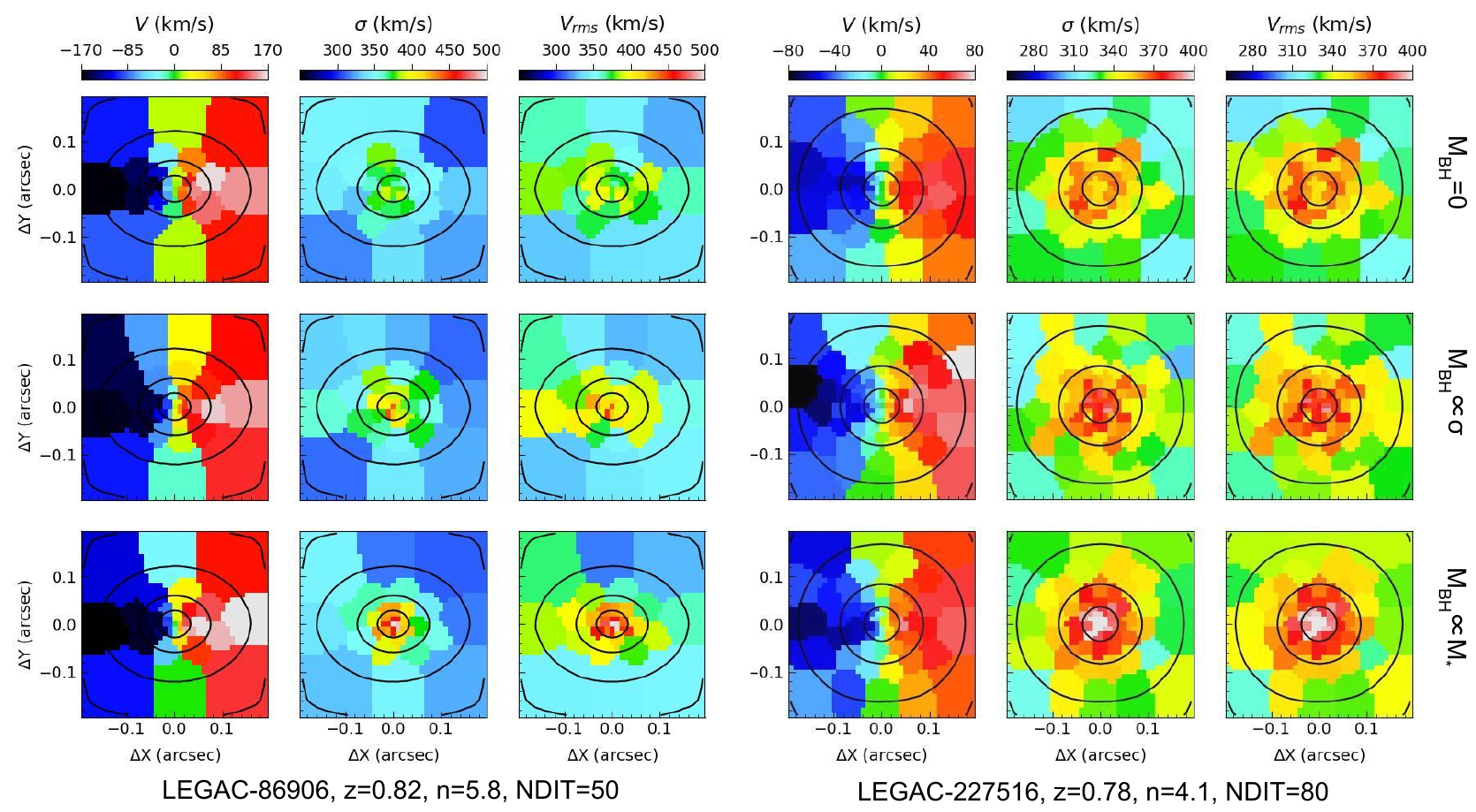}
    \caption{The kinematics of two simulated galaxies at redshift $z\approx1$, LEGAC-86906 and LEGAC-227516, are extracted from the mock $H+K$ IFS data using the CaT stellar absorption features. The LOSVDs are presented from left to right as rotation ($V$), velocity dispersion ($\sigma$), and root-mean-square velocity ($V_{\rm rms}$). From top to bottom, they are shown for different \Mbh\ cases: no black hole (top), \Mbh--$\sigma$ predicted black hole (middle), and \Mbh--\Mstar\ predicted black hole (bottom). The black contours in all maps correspond to the isophotes derived from the collapsed \textsc{HSIM} IFS cubes, spaced in intervals of 1 mag arcsec$^{-2}$.}
    \label{fig:high_z=1_mock_kin_H+K}
\end{figure*}

\begin{figure*}
    \centering
    \includegraphics[width=0.99\textwidth]{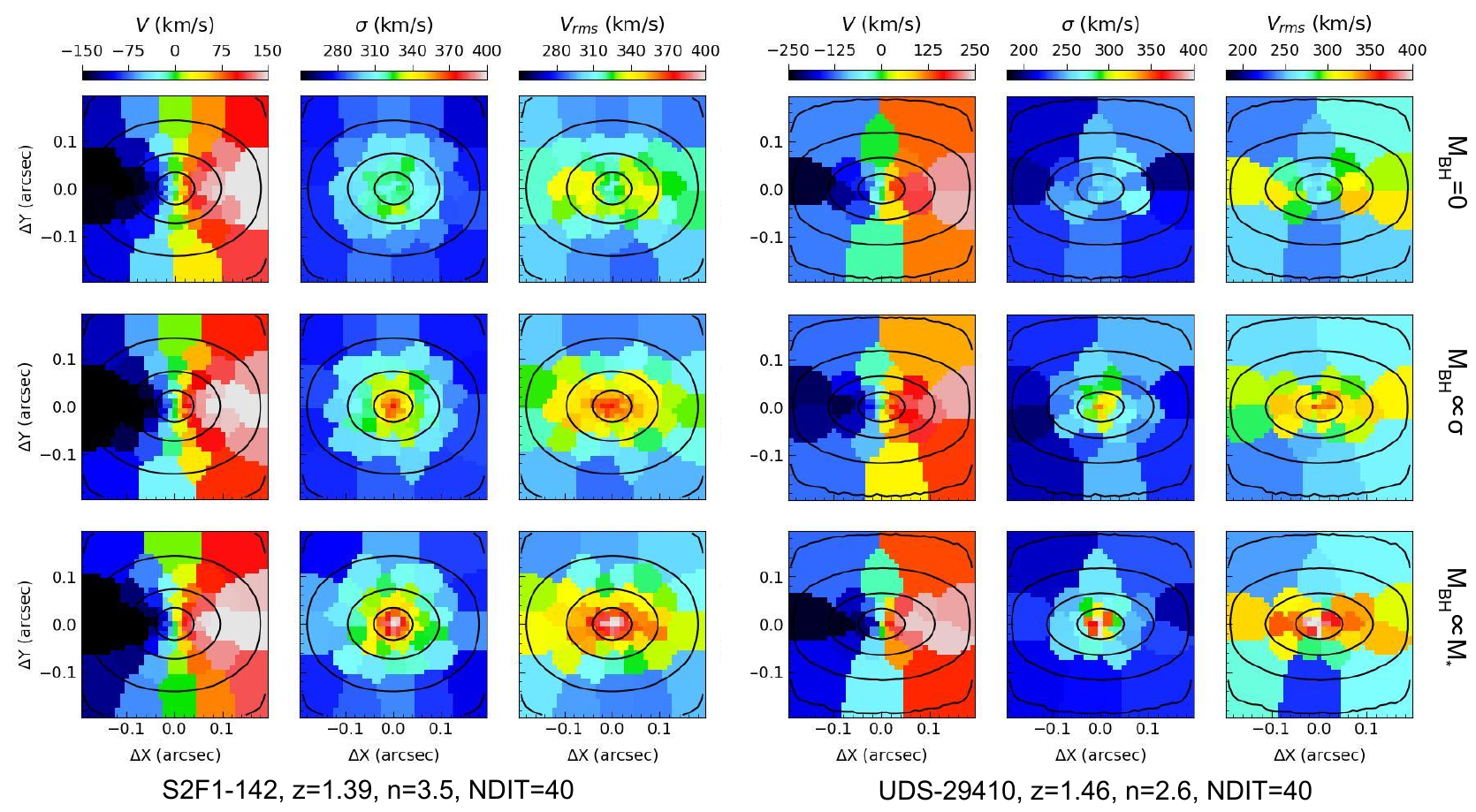}
    \caption{Same as \autoref{fig:high_z=1_mock_kin_H+K} but displayed for two simulated galaxies at redshift $z\approx1.5$: S2F1-142 and UDS 29410.}
    \label{fig:high_z=1.5_mock_kin_H+K} 
\end{figure*}

\begin{figure}
    \centering
    \includegraphics[width=0.45\textwidth]{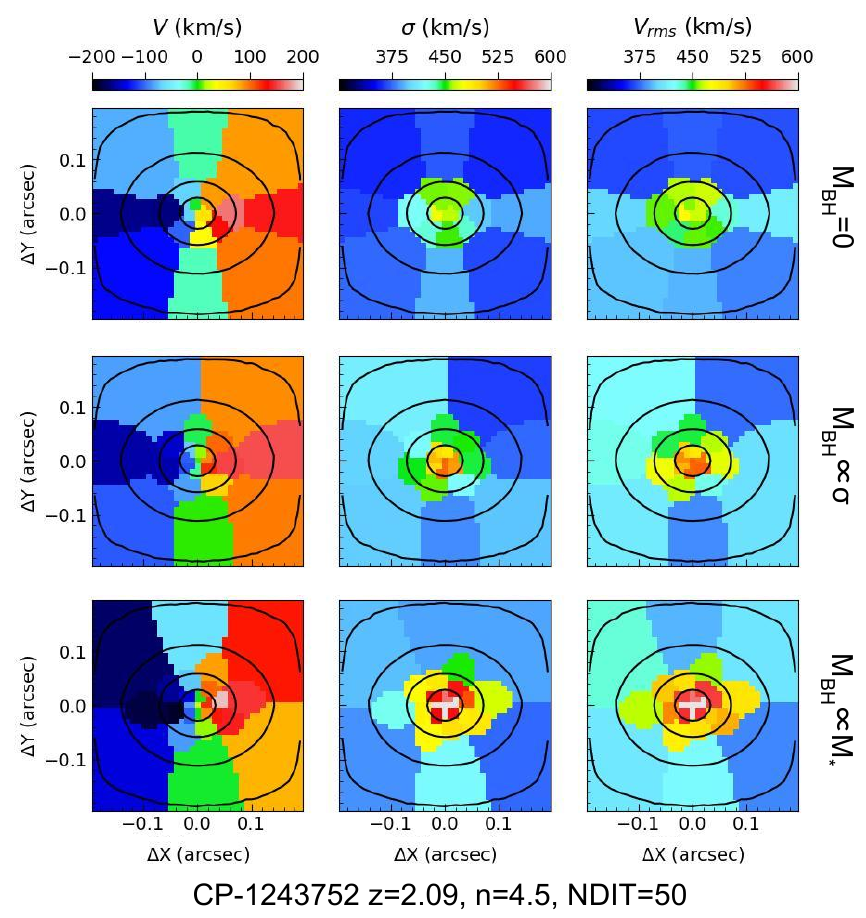}
    \caption{Same as \autoref{fig:high_z=1_mock_kin_H+K} but displayed for the simulated galaxy at redshift $z\approx2$: CP-1243752. Their LOSVDs are extracted from the mock $H+K$ IFS data using the \ion{Mg}{1}b stellar absorption features.}
    \label{fig:high_z=2_mock_kin_H+K}
\end{figure}

Once the mock ELT observations (MICADO images and HARMONI data cubes) are generated as described in \autoref{sec:simulating_mock_elt_observations}, the next crucial step is to process and analyze this simulated data, following the procedures we would apply to real data, to extract the physical parameters required for the dynamical modeling. This section outlines the methodologies used for deriving the PSF from simulated stellar images, parametrizing the galaxy surface brightness from simulated MICADO images, and extracting stellar kinematics from the simulated HARMONI data cubes.

\subsection{PSF Determination from Simulated MICADO Point Source Images}\label{sec:psf_determination_from_simulated_micado_point_source_images}

Accurate knowledge of the PSF is critical for deconvolving the observed galaxy light distribution and for reliable photometric and structural analysis. For MICADO, the PSF will be complex and dependent on the AO system performance and observing conditions. 

We simulate MICADO observations of point sources (stars) using \textsc{SimCADO} under the same conditions anticipated for the galaxy observations. These simulated stellar images are then analyzed to characterize the PSF. We use the MGE method, specifically the {\tt mge.fit\_sectors} routine from the \textsc{MgeFit} Python package \citep{Cappellari02}. The {\tt mge.fit\_sectors} algorithm models the 2D light distribution of the simulated star as a sum of co-axial Gaussian components. This MGE representation of the PSF is then used in the subsequent analysis of the simulated galaxy images to account for instrumental broadening.

\subsection{MGE Surface Brightness Parametrization from Simulated MICADO Galaxy Images}\label{sec:mge_surface_brightness_parametrization_from_simulated_micado_galaxy_images}

The simulated MICADO images of the galaxies, generated as described in \autoref{sec:simulating_micado_imaging}, provide high-resolution views of their stellar light distributions. To quantify their structure, we again employ the MGE method. The {\tt mge.fit\_sectors} routine is used to fit the 2D surface brightness distribution of each simulated galaxy. 

Crucially, during this fitting process, the MGE model of the PSF (derived in \autoref{sec:psf_determination_from_simulated_micado_point_source_images}) is convolved with the intrinsic MGE model of the galaxy before comparing with the simulated (PSF-convolved) MICADO galaxy image. This ensures that the derived MGE parameters represent the intrinsic galaxy structure, corrected for the effects of the PSF. The resulting MGE model (\autoref{fig:high_z_mge}) is given in \autoref{tab:mge} and provides a compact and accurate description of the galaxy's surface brightness distribution, which serves as the stellar photometric component in the dynamical models used for SMBH mass determination (see \autoref{sec:dynamical_modelling_and_smbh_mass_recovery}).

\subsection{Stellar Kinematic Extraction from Simulated HARMONI Data Cubes}\label{sec:stellar_kinematic_extraction_from_simulated_harmoni_data_cubes}

We derived the LOSVDs from the mock HARMONI IFS observations. For galaxies at redshifts $1\lesssim z\lesssim1.8$, the kinematics is mainly constrained by the \ion{Ca}{2} Triplet (CaT) stellar absorption features, while for those at $z \gtrsim 1.8$, the strongest constraints come from the \ion{Mg}{1}b features. These prominent absorption lines fall within the spectral range of the $H+K$ grating used in our \textsc{HSIM} simulations, as illustrated in \autoref{fig:spectrum_ppxf_hsim}.

Prior to kinematic extraction, we applied adaptive Voronoi spatial binning to the simulated IFS data cubes using the \texttt{vorbin} package\footnote{v3.2.1: \url{https://pypi.org/project/vorbin/}}  \citep{Cappellari03}. This procedure groups spaxels to achieve a target S/N per bin per \AA\ (detailed in Column 12 of \autoref{tab:jamresults}), minimizing uncertainties in the subsequent LOSVD measurements.

Following spatial binning, each binned spectrum was logarithmically rebinned along the spectral dimension. We then employed the Penalized PiXel-Fitting (\textsc{pPXF}) software \citep{Cappellari23} to fit these spectra using stellar population models by \citet{Maraston11} based on the MARCS synthetic spectra \citep{Gustafsson08}. To extract the mean velocity ($V$) and velocity dispersion ($\sigma$), we configured \textsc{pPXF} to fit only the first two velocity moments $(V,\sigma)$ ({\tt moments = 2}) while using additive polynomials ({\tt degree = 0}), and no multiplicative ones ({\tt mdegree = -1}). The instrumental broadening of the HARMONI IFS was accounted for by convolving the stellar templates with the differential instrumental dispersion. In the fit, we used 13 templates with ages spanning 3 to 15 Gyr (i.e., given that the youngest population of the MARCS synthetic spectra is 3 Gyr) and appropriate metallicities (see \autoref{tab:hsimmock} for details). \autoref{fig:spectrum_ppxf_hsim} displays the best-fit SPS template overlaid on segments of the mock $H+K$ IFS spectra from the central bin of each of the five simulated galaxies, with residuals ({\tt data - model}) shown as green dots. The resulting LOSVD maps ($V$, $\sigma$, and $V_{\rm rms}=\sqrt{V^2+\sigma^2}$) are presented in \autoref{fig:high_z=1_mock_kin_H+K}, \autoref{fig:high_z=1.5_mock_kin_H+K}, and \autoref{fig:high_z=2_mock_kin_H+K} for galaxy pairs at $z\approx1$, $1.5$, and $2$, respectively.

\begin{figure*}
    \centering
    \includegraphics[width=0.99\textwidth]{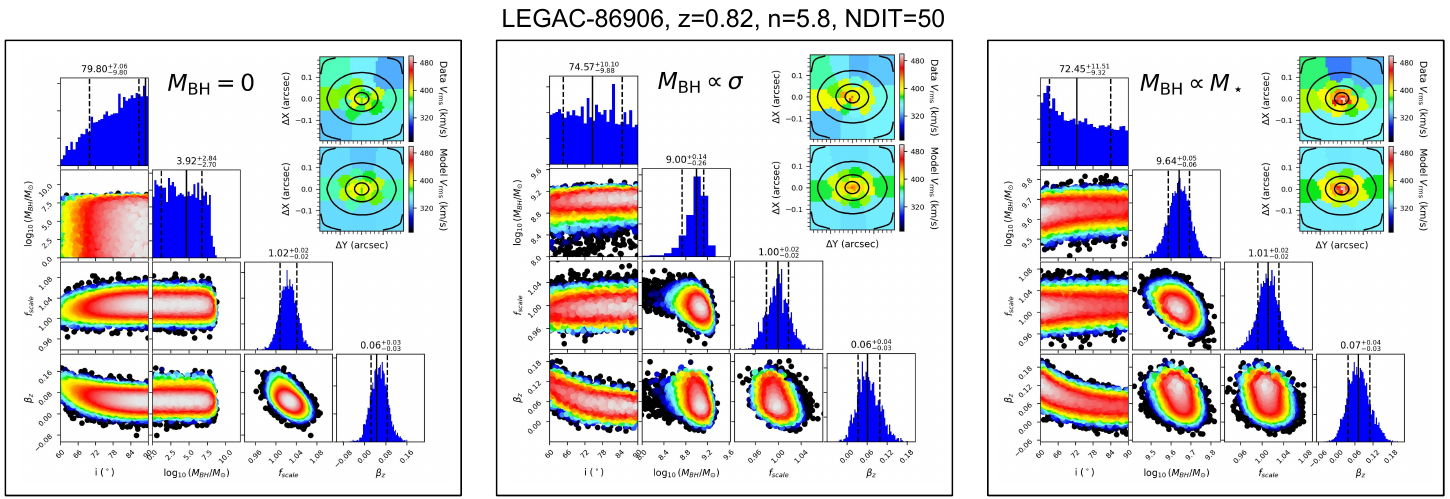}
    \includegraphics[width=0.99\textwidth]{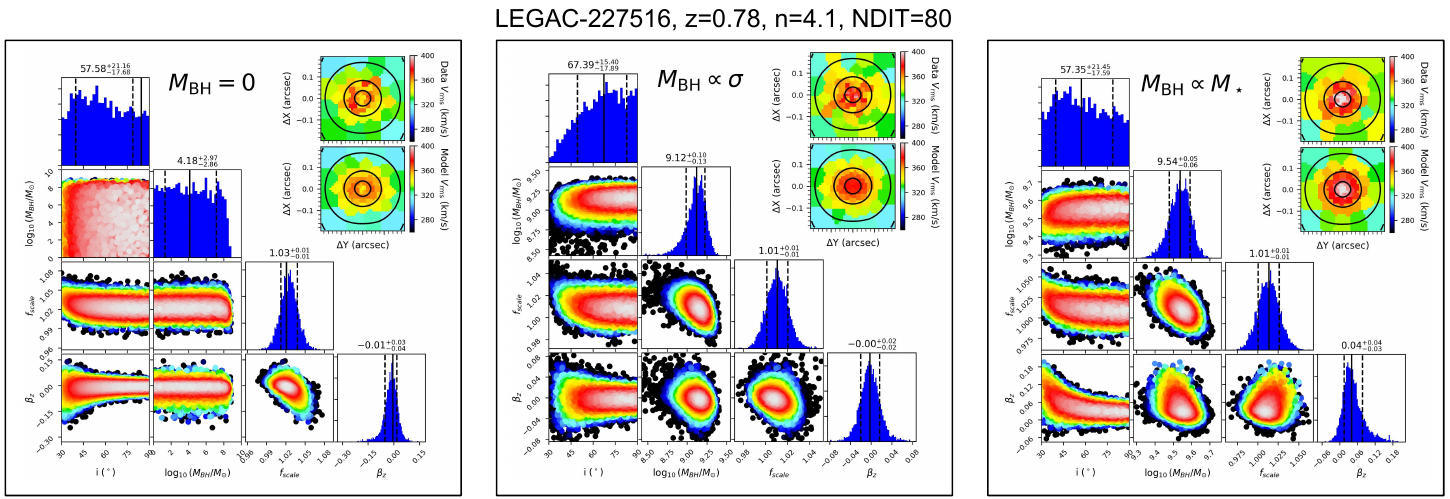}
    \caption{The PDFs obtained from the \textsc{adamet} MCMC optimization for the \textsc{JAM} models applied to the \textsc{HSIM} $H+K$ grating kinematics of two galaxies at $z\approx1$, LEGAC-86906 (top row) and LEGAC-227516; (bottom row), illustrate the projected 2D distributions and 1D histograms of four free parameters: $i$, \Mbh, $f_{\rm scale}$, and $\beta_z$.  Additionally, inset maps display the $V_{\rm rms}$ values. The top maps represent the kinematic maps extracted from the mock data cubes, while the bottom maps show the kinematic maps recovered from the best-fit \textsc{JAM} model. The strong agreement across the HARMONI-simulated FoV confirms the robustness of the model constraints. The best-fit JAM model is determined by the PDF with the highest likelihood, as presented in \autoref{tab:jamresults}.  Each row presents model constraints for different \Mbh\ cases: no black hole (left), \Mbh--$\sigma$ predicted black hole (middle), and \Mbh--\Mstar\ predicted black hole (right). The same LOSVD range is maintained in the color bars for each galaxy to visualize the impact of central black hole mass on stellar kinematics. The black contours in the $V_{\rm rms}$ maps correspond to the isophotes derived from the collapsed \textsc{HSIM} IFS cubes, as similar as \autoref{fig:high_z=1_mock_kin_H+K}. }
    \label{fig:z=1_kin_BHrecover_H+K}
\end{figure*}

For the targets at $z \approx 1$, we performed simulations with exposure times of 12.5 and 20~hours for LEGAC-86906 and LEGAC-227516, respectively. For the galaxies at $z \approx 1.5$ (S2F1-142 and UDS~29410), the simulations were carried out with exposure times of 10~hours, while for the galaxy at $z \approx 2$, an exposure time of 12.5~hours was adopted (\autoref{tab:hsimmock}, excluding overheads). The mock IFS data attained sufficient S/N across the fitting spectral ranges. The choice of the brightest known targets at each redshift maximized sensitivity and optimized exposure times, making these durations representative for observations at these redshifts.

To validate our kinematic measurements, we compared the LOSVDs obtained using the CaT or \ion{Mg}{1}b features with those derived by fitting the entire $H+K$ spectral range, showing differences of less than 5\%. We further tested the kinematics using the empirical X-shooter Stellar Library Data Release 3 \citep[XSL DR3;][]{Verro22}, which includes stellar spectra from 683 stars covering 3000--25,000 \AA\ at a resolution of $R \approx 10,000$ (a factor of two coarser than MARCS). The XSL encompasses a wide variety of stellar types. The LOSVD maps derived using XSL templates showed differences of less than 8\% compared to those from the MARCS SPS models.

\begin{figure*}
    \centering
    \includegraphics[width=0.99\textwidth]{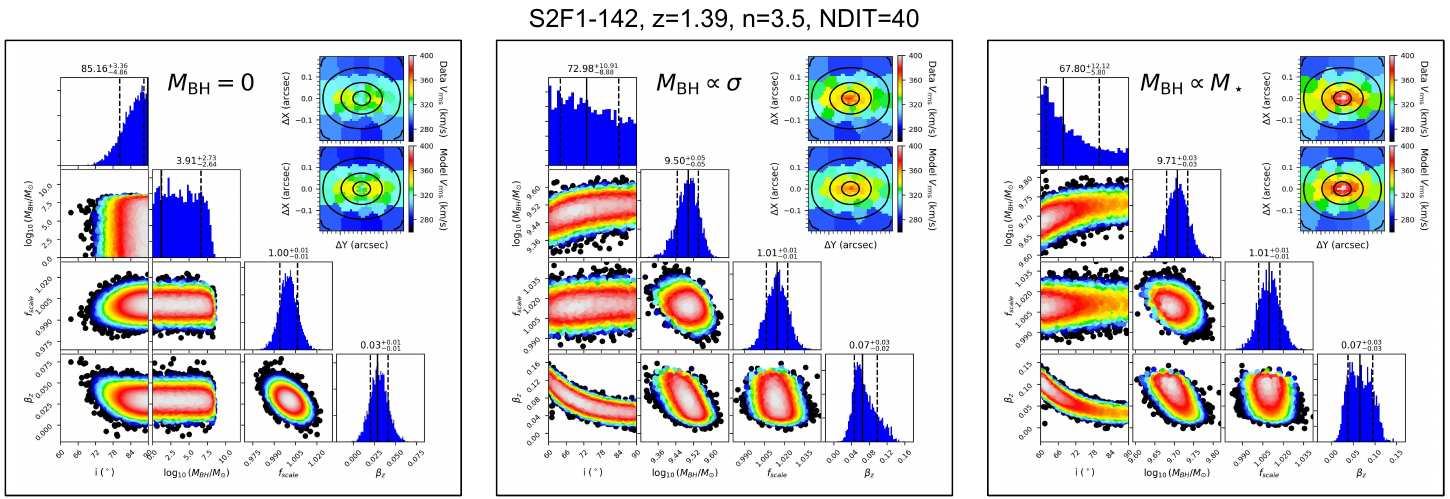}
    \includegraphics[width=0.99\textwidth]{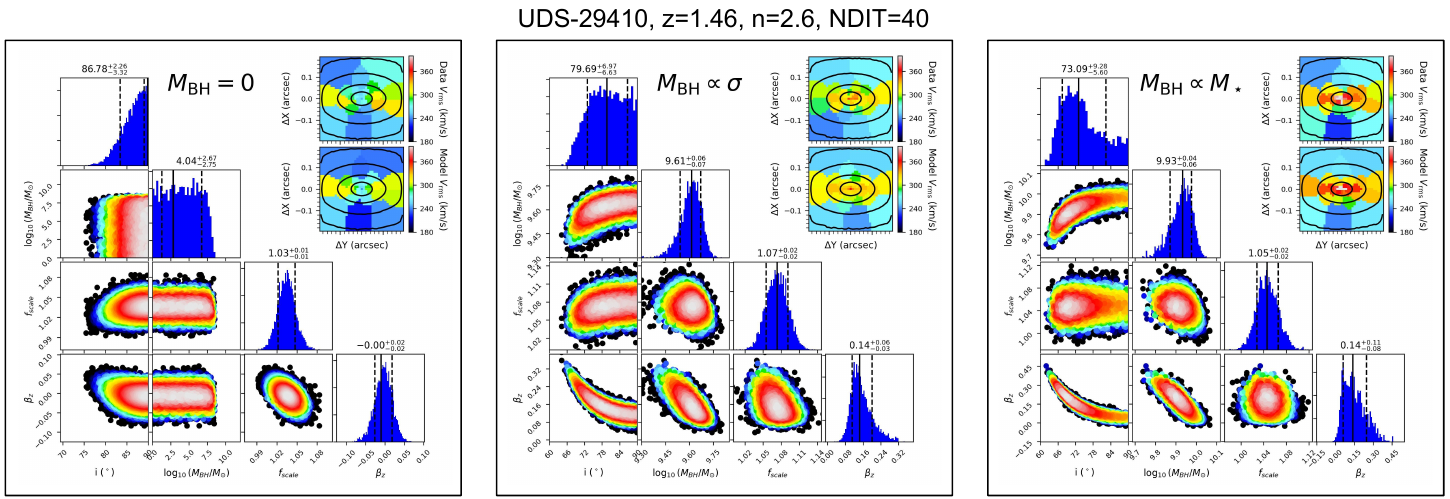}
    \caption{Same as \autoref{fig:z=1_kin_BHrecover_H+K} but for two simulated galaxies at redshift $z\approx1.5$: S2F1-142 (top row) and UDS 29410 (bottom row).}
    \label{fig:z=1.5_kin_BHrecover_H+K}
\end{figure*}

The signatures of central SMBHs are prominently seen in several central spaxels within the mock $V_{\rm rms}$ maps at the scales of the SMBH’s SOI (at least several times larger than the proposed observational scale of 10 mas in this work) where its gravity dominates, characterized by the central peaks/drops towards the center depending on whether its absence/presence or its large/small mass relative to the stellar mass distribution of the host galaxy. The contrast between the kinematics of these central spaxels and those at larger radii is visibly evident in all galaxies, irrespective of redshift. This highlights the exceptional spatial and spectral resolution and sensitivity capabilities of HARMONI in detecting the stellar kinematic signatures of SMBHs at redshifts of $1\lesssim z\lesssim2$, enabling their accurate dynamical mass measurements.

A detailed analysis of the $\sigma$ maps reveals central drops in both stellar velocity dispersion $\sigma$ and root-mean-squared velocity $V_{\rm rms}$ for targets at $z \lesssim 1$ (e.g., LEGAC-2346 and LEGAC-2408), consistent with those theorized in massive ellipticals by \citet{Tremaine94}, numerically predicted in local galaxies:  with $z \leq 0.3$ \citep{Nguyen23}, within $D\leq 10$ Mpc \citep{Nguyen2025b}, and within $10<D\leq 20$ Mpc \citep{Ngo2025b, Ngo2025c}, or observationally proven in NGC 4736 with the {\it James Webb Space Telescope (\jwst)} by  \citet{Nguyen2025a}. A key new finding is that for targets at $z \gtrsim 1.5$, no central $\sigma$ drop is observed. Instead, $V_{\rm rms}$ either exhibits a central drop along the major axis due to significant rotational support (e.g., S2F1-142 and UDS 29410 at $z \approx 1.5$) or continuously increases toward the center for galaxies with intermediate rotation (e.g., CP-1243752 at $z \approx 2$) similar to what found in the local universe very recently with \jwst\ \citep[e.g.,][]{Nguyen2025c, Tahmasebzadeh2025, Taylor2025}.

\section{Dynamical Modelling and SMBH Mass Recovery}\label{sec:dynamical_modelling_and_smbh_mass_recovery}

This section details the process of measuring SMBH masses from the simulated HARMONI IFS and MICADO imaging data and assesses the instrument's sensitivity for such observations at high redshift. We first outline the methodology for recovering SMBH masses, then present the results and their validation, and finally discuss the HARMONI sensitivity limits.

\subsection{Dynamical Modelling with JAM}

To determine the SMBH mass (\Mbh) and constrain other dynamical parameters, we compared the LOSVDs of the six simulated galaxies, derived in \autoref{sec:stellar_kinematic_extraction_from_simulated_harmoni_data_cubes}, with models generated by  \textsc{JAM} \citep{Cappellari08,Cappellari20}. Specifically, we focused on the maps of second velocity moments ($V_{\rm rms}$). 

\begin{figure*} 
    \centering
    \includegraphics[width=0.99\textwidth]{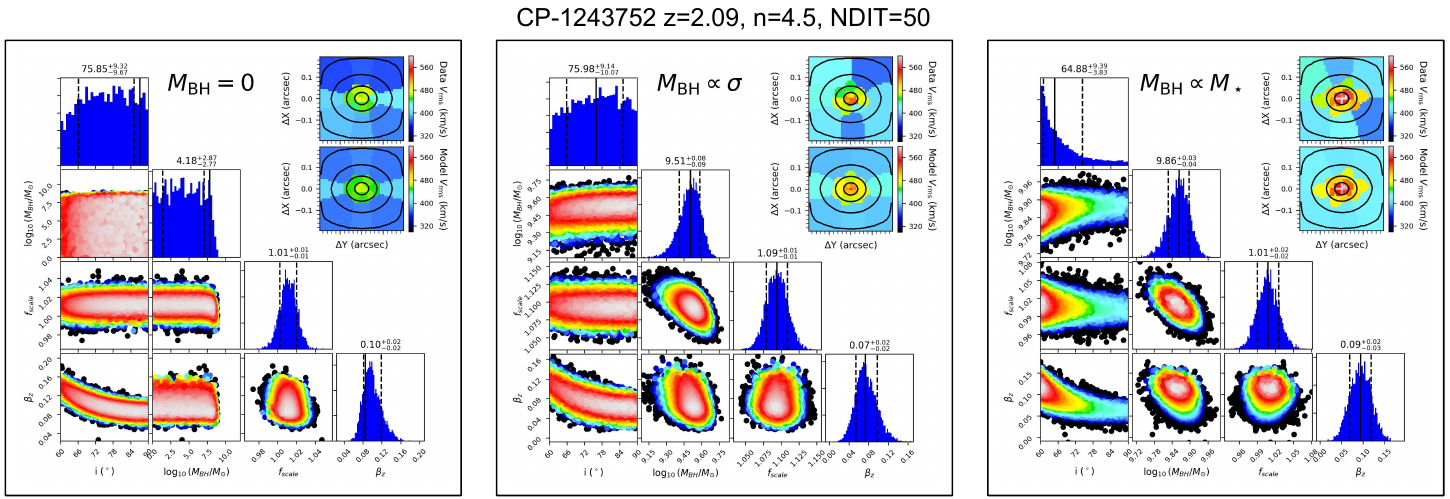}
    \caption{Same as \autoref{fig:z=1_kin_BHrecover_H+K} but for the simulated galaxy at redshift $z\approx2$: CP-1243752.}
    \label{fig:z=2_kin_BHrecover_H+K}
\end{figure*}

Our \textsc{JAM} models aimed to constrain four primary parameters: the \Mbh, the orbital anisotropy ($\beta_z$), a mass-scaling factor for the stellar component ($f_{\rm scale}$, expected to be close to 1), and the galaxy's inclination angle ($i$). To ensure efficient sampling of the parameter space, we adopted a logarithmic scale for \Mbh, while the other three parameters ($\beta_z$, $f_{\rm scale}$, $i$) were sampled on linear scales. The models accounted for the HARMONI LTAO PSF, characterized by a FWHM$_{\rm PSF}\approx12$ mas. The search ranges for these parameters, along with their initial guesses for the \textsc{JAM} optimization, are detailed in \autoref{tab:jamresults}.

We employed a MCMC approach, utilizing \textsc{JAM}, to explore the parameter space (\Mbh, $\beta_z$, $f_{\rm scale}$, $i$). This allowed us to identify the best-fit values and determine their associated statistical and measurement uncertainties, as constrained by the mock HARMONI kinematics and the MICADO-derived stellar mass models. The MCMC simulations were performed within a Bayesian framework using the adaptive Metropolis algorithm \citep{Haario01}, as implemented in the \textsc{AdaMet}\footnote{v2.0.9: \url{https://pypi.org/project/adamet/}} package \citep{Cappellari13a}. Each MCMC chain consisted of $3\times10^4$ iterations. The initial 20\% of these iterations were discarded as a burn-in phase, and the remaining 80\% were used to construct the posterior probability distribution function (PDF) for each parameter. The best-fit parameters correspond to the highest likelihood region of this PDF. Uncertainties for all four parameters were calculated at the 1$\sigma$ and 3$\sigma$ confidence levels (CL), representing the 16--84\% and 0.14--99.86\% ranges of the PDF, respectively.

\subsection{Results of SMBH Mass Recovery}\label{sec:results_smbh_recovery}

The best-fit \textsc{JAM} models and their associated statistical uncertainties are presented for two pairs and a single of simulated galaxies, grouped by redshift: $z\approx1$ (LEGAC-86906 and LEGAC-227516; see \autoref{fig:z=1_kin_BHrecover_H+K}), $z\approx1.5$ (S2F1-142 and UDS 29410; see \autoref{fig:z=1.5_kin_BHrecover_H+K}), and $z\approx2$ (CP-1243752; see \autoref{fig:z=2_kin_BHrecover_H+K}). These figures illustrate how well the best-fit \textsc{JAM} models describe the stellar kinematics derived from the mock \textsc{HSIM} $H+K$ IFS gratings for different input \Mbh\ values.

The figures feature 2D scatter plots for each pair of \textsc{JAM} parameters, where colored points indicate their likelihood (white for maximum likelihood and 1$\sigma$ CL, black for CLs larger than 3$\sigma$). Accompanying histograms show the 1D marginalized distributions for each parameter, from which the best-fit values and their 1$\sigma$ uncertainties (listed in \autoref{tab:jamresults}) were derived. Inset plots located at the top-right corner of each main panel in these figures provide a direct comparison between the input $V_{\rm rms}$ map (from mock IFS data) and the $V_{\rm rms}$ map generated by the best-fit \textsc{JAM} model. These comparisons use the same velocity scale for each galaxy and input \Mbh. The constrained models demonstrate good agreement with the mock kinematics across the simulated $0\farcs4\times0\farcs4$ HARMONI FoV, within the uncertainties of the mock kinematic measurements, irrespective of redshift. This finding is similar to previous studies by \citet{Nguyen23} for targets at $z\leq0.3$, \citet{Nguyen2025b} for targets within $D\leq10$ Mpc, and \citet{Ngo2025b, Ngo2025c} for galaxies in the distant range of $10 < D\leq20$ Mpc. The characteristic kinematic signatures in the mock LOSVDs of these high-redshift targets, influenced by their SMBHs, are successfully reproduced at the simulated 10 mas spaxel scale.

\begin{figure*}
    \centering
    \includegraphics[width=0.99\textwidth]{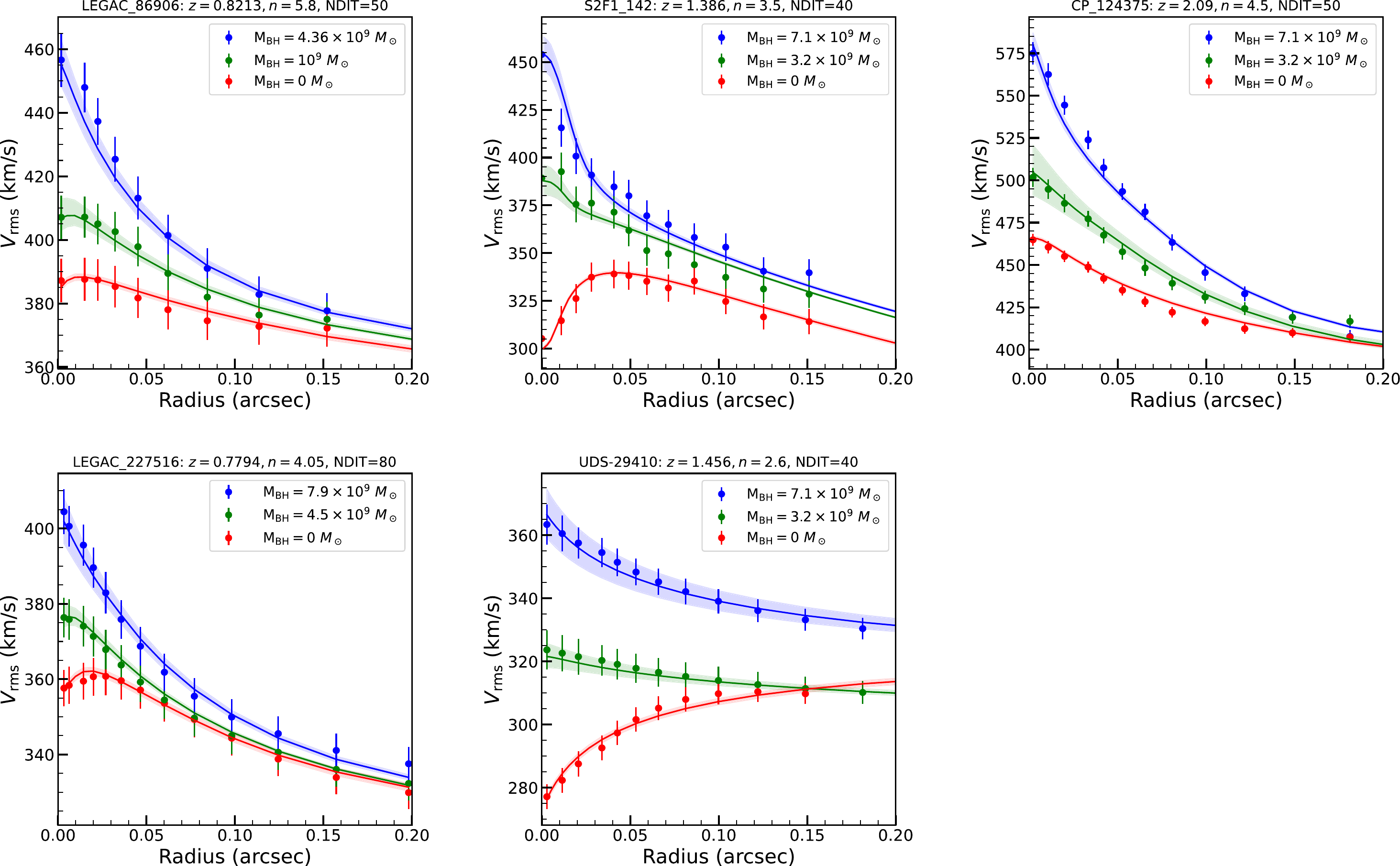}
    \caption{Comparisons of 1D kinematic profiles ($V_{\rm rms}$) extracted from the HSIM mock data cubes for all five selected galaxies. Profiles are shown for different input \Mbh: no BH (red), an \Mbh--$\sigma$ predicted BH (green), and an \Mbh--\Mstar\ predicted BH (blue). These are compared against their corresponding intrinsic kinematic profiles (same color solid lines) derived directly from the best-fitting JAM models. The error bars and the shaded regions represent the 1$\sigma$ uncertainties on the extracted kinematics of the HSIM mock data cubes and best-fitting JAM models, respectively. The profiles are extracted along the major axes of these galaxies.}
    \label{fig:1D_ppxf_kin}
\end{figure*}

Notably, in the absence of a central SMBH ($M_{\rm BH}=0$), the $V_{\rm rms}$ maps of galaxies at $z\lesssim1$ (e.g., LEGAC-86906 and LEGAC-227516) exhibit central drops across at least 5--7 spaxels. As the input \Mbh\ increases, these drops become shallower or transform into central peaks. Conversely, for the four galaxies at $z\gtrsim1.5$ (S2F1-142, UDS 29410, and CP-1243752), even without a central SMBH, both the mock kinematics and the best-fitting \textsc{JAM} models show slightly increasing velocity dispersion ($\sigma$) maps towards the galaxy center. In these cases, the presence of central drops or increases in the $V_{\rm rms}$ maps depends on the degree of rotational support.

The recovered \Mbh\ does not converge to zero when the input \Mbh\ is set to zero because the dynamical modeling is limited by uncertainties in the stellar kinematic measurements, as well as by other observational effects such as PSF convolution, throughput variations, AO performance, detector noise, and background noise. Measurement errors in the stellar LOSVDs (e.g. $V$ and $\sigma$)—propagate through these effects and introduce a nonzero central mass component in the dynamical fit, even in the absence of a true SMBH. In practice, these uncertainties impose a noise floor on the minimum \Mbh\ that can be reliably recovered. As a result, the modeling favors a small but nonzero \Mbh, which we interpret as an upper limit, to account for residual kinematic signatures within the central resolution element, thereby preventing convergence to exactly zero mass.

The recovered estimates for \Mbh\ and $f_{\rm scale}$ closely match their input values used in the \textsc{HSIM} simulations, with differences typically below 10\% for both parameters. The quoted uncertainties are derived from the MCMC analysis of the $V_{\rm rms}$ maps and include statistical effects. The slightly lower accuracy compared to previous, similar works \citep[e.g.,][]{Nguyen23, Nguyen2025b, Ngo2025b, Ngo2025c} can be attributed to the increased impact of background noise on observations of high-redshift sources, which leads to lower quality simulated kinematics. A well-known negative covariance between \Mbh\ and $f_{\rm scale}$, often described as a “banana shape” in the 2D PDF, is evident in the 3$\sigma$ CL regions. This degeneracy arises from the interplay between the gravitational potential of the central SMBH and that of the host galaxy's stellar component: a larger BH mass can be compensated by a smaller stellar mass-scaling factor, and vice versa. The ability to characterize this degeneracy reflects the high spatial resolution of our simulations (10 mas observational scale), which is sufficient to resolve the SMBHs' SOIs up to $z\lesssim2$. This success is also a testament to our selection of the brightest available targets from current surveys, ensuring adequate sensitivity and S/N to detect and measure the stellar kinematic signatures of SMBHs at these substantial distances. These results demonstrate that the ELT, with its advancements in spatial and spectral resolution, sensitivity, and AO capabilities, has the potential to extend dynamical SMBH measurements significantly beyond the local Universe, up to redshifts of $z\lesssim2$.

The recovered values for the orbital anisotropy, $\beta_z$, are well-constrained by the models, showing relatively small uncertainties. The results indicate a preference for slightly radial stellar orbits ($\beta_z \gtrsim 0$), with $3\sigma$ errors around $\pm 0.14$. This is consistent with the isotropic value ($\beta_z = 0$) assumed as input in the mock data generation (see \autoref{sec:generating_input_noiseless_data_cubes}). The inclination angle ($i$) is less tightly constrained by the simulated data, though there is a tendency towards disky rotation (higher $i$ values). These recovered inclinations are broadly consistent with the input values listed in \autoref{tab:sampledata}.

It is not necessary to explore trade-offs in spectral (velocity) resolution in this work, as they are not expected to have a significant impact on the \Mbh\ constraints for the systems considered here. In our modeling, all spectral pixels already contribute to the total $\chi^2$ of the kinematic fits, such that additional spectral binning would primarily redistribute existing information rather than improve the effective S/N relevant for constraining the LOSVDs. Furthermore, for the massive galaxies targeted in this study, the intrinsic stellar velocity dispersions are large (\autoref{tab:sampledata}, Column~10), and the adopted spectral resolution ($H+K$ grating with $R\sim 3300$) is already more than sufficient to resolve the relevant kinematic features. Consequently, the velocity resolution does not constitute a limiting factor in our analysis, and varying it within a reasonable range would not qualitatively alter our inferred \Mbh\ or the conclusions drawn from these simulations.

\subsection{Comparison of 1D Kinematic Profiles}

\begin{figure} 
    \centering
    \includegraphics[width=0.45\textwidth]{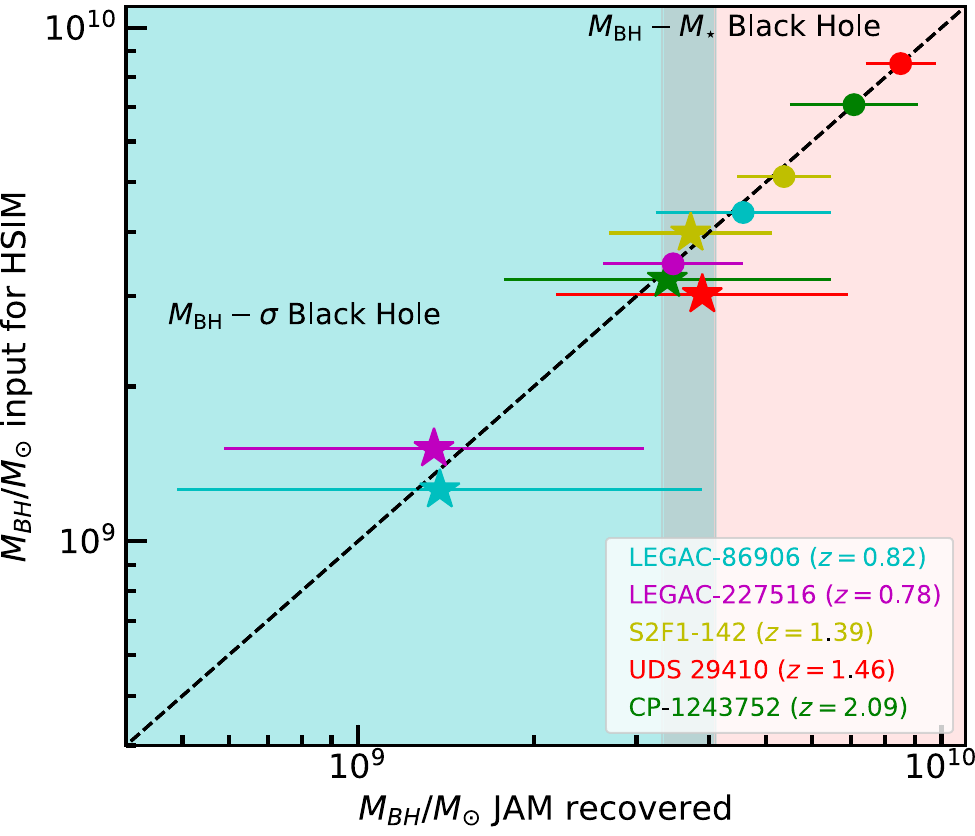}
    \caption{A summarized comparison of the input \Mbh\ for \textsc{HSIM} and our recovered values at $3\sigma$ uncertainties. The black dashed line indicates the line of equality between the input \Mbh\ for \textsc{HSIM} and our recoveries. Stars are \Mbh\ that follows the \Mbh--$\sigma$ relation, while dots are \Mbh\ that follows the \Mbh--\Mstar\ relation \citep[or Eq. (2) or (3) of ][respectively]{Krajnovic18b}.}
    \label{fig:summaryBH}
\end{figure}

To illustrate the agreement between our simulations and dynamical modeling, we compared the 1D radial profiles of $V_{\rm rms}$, as shown in \autoref{fig:1D_ppxf_kin}. In this figure, data points represent the mean $V_{\rm rms}$ values extracted from the mock kinematic maps in concentric annuli (with 10 mas steps from the galaxy center). The solid lines correspond to the $V_{\rm rms}$ values derived from the best-fitting \textsc{JAM} kinematic maps using the same annular extraction method. Kinematics for different input \Mbh\ values are shown in distinct colors, applied consistently to both the mock data and the models. A key observation from these profiles is that the kinematic signatures of SMBHs are typically distinguishable within a radius of $\approx$50 mas.

\autoref{fig:summaryBH} provides a consolidated comparison between the input \Mbh\ values (see \autoref{tab:hsimmock}: column~5 for the predicted \Mbh\ from the \citet{Krajnovic18a} \Mbh–$\sigma$ relation and column~6 for the predicted \Mbh\ from the \citet{Krajnovic18a} \Mbh–$M_\star$ relation) used in the \textsc{HSIM} simulations and the corresponding values recovered by the \textsc{AdaMet} MCMC algorithm coupled with \textsc{JAM} (see \autoref{tab:jamresults}). The error bars represent the $3\sigma$ uncertainties, including both statistical and kinematic measurement uncertainties.

\subsection{HARMONI Sensitivity Limit}\label{sec:sensitivity}

We tested the sensitivity of HARMONI IFS for our selected galaxies at their respective redshifts by varying the on-source exposure time. Specifically, we repeated simulations, systematically reducing the {\tt NDIT}, while fixed each {\tt DIT = 15} minutes (thereby decreasing both the total exposure time, {\tt NDIT} $\times$ {\tt DIT}, and spectral S/N) until the resulting HARMONI IFS cubes produced reliable kinematic maps as shown in \autoref{fig:high_z=1_mock_kin_H+K}, \autoref{fig:high_z=1.5_mock_kin_H+K}, and \autoref{fig:high_z=2_mock_kin_H+K} using \textsc{pPXF}. To enhance sensitivity when measuring the stellar kinematics using these newly testing cubes, we applied adaptive \textsc{VORONOI} binning, ensuring a targeted S/N $\approx30$/\AA/measurement bin. This analysis allowed us to determine the required exposure times for each galaxy (Column 8 of \autoref{tab:hsimmock}). We note that, in these tests, the total apparent magnitudes of the five simulated galaxies were fixed to the observed values in the relevant photometric bands, as listed in Column~6 of \autoref{tab:sampledata}, when constructing the input noiseless data cubes (see \autoref{sec:generating_input_noiseless_data_cubes}). When generating the mock observations with \textsc{HSIM}, the only parameter allowed to vary was {\tt NDIT}.

The galaxy surface brightness profiles were derived from \hst\ imaging (both parametric fits and direct measurements) and interpolated to match the proposed observational scales of 10 mas for HARMONI IFS and 4 mas for MICADO imaging using a Sérsic profile. Our analysis shows that the required on-source exposure time for reliable kinematic measurements scales with apparent brightness in the relevant filter.   To establish the correspondence between on-source exposure time and total apparent magnitude, we artificially dimmed the total apparent magnitudes of our five simulated targets in either the $I$- or $H$-band images. Using the same procedure described in \autoref{sec:generating_input_noiseless_data_cubes}, we generated the corresponding input cubes for \textsc{HSIM}, thereby reducing the flux levels in the spectra of the noiseless data cubes. The resulting mock \textsc{HSIM} integral-field data exhibit lower spectral S/N, which in turn require larger values of {\tt NDIT} (i.e. longer total on-source integration times) to recover the minimum spectral S/N necessary for reliable LOSVD measurements.   We therefore varied {\tt NDIT} in \textsc{HSIM} until the simulated HARMONI data cubes yielded robust stellar kinematic maps when analysed with \textsc{pPXF}. From these tests, we find that for galaxies at $z \approx 1$ observed in the F814W band, total apparent magnitudes of 20 and 20.5 correspond to on-source integration times of approximately 5 and 7.5 hours, respectively.  In contrast, for galaxies observed in F160W at $1<z\lesssim2$, 5~hours of on-source time is adequate to detect systems as faint as 20.8~mag. These results confirm the suitability of these targets for observations with the ELT. Given the careful simulation of \textsc{MICADO} imaging, our HARMONI IFS sensitivity estimates for the required exposure times are robust, supporting our SMBH survey at redshifts $z\approx2$.

\subsection{Importance of Dynamically Weighing SMBHs in Distant Universe}\label{sec:smbh_distant}

Spatially resolved dynamical measurements of black hole masses, using either stellar or cold gas kinematics, play a uniquely important role in anchoring our understanding of black hole growth across cosmic time. Unlike virial mass estimates for AGN and quasars, which depend on local empirical calibrations (e.g., reverberation mapping) and assumptions about geometry and dynamics, dynamical measurements rely directly on resolved gravitational motions and are therefore fundamentally calibration-independent. As such, they provide essential benchmarks for testing and refining indirect black hole mass estimators at high redshift. Establishing robust, kinematically measured black hole masses in increasingly distant galaxies—enabled by future facilities such as the 39m ELT and its advanced instruments, including the HARMONI spectrograph and the MICADO imager—is therefore critical for shaping black hole physics studies in the next decades by (i) assessing the true evolution of galaxy–black hole scaling relations and (ii) disentangling physical growth from observational bias in studies of black hole–galaxy co-evolution.

\section{Conclusions}\label{sec:conclusions}

This paper investigated the feasibility of extending direct SMBH mass measurements to high-redshift ($1\lesssim z\lesssim2$) galaxies using the capabilities of the upcoming ELT with its first-light instruments, MICADO and HARMONI. We developed a comprehensive simulation pipeline, generating mock MICADO images and HARMONI IFS data for a sample of five bright, massive, quiescent galaxies. These simulations incorporated realistic instrumental effects, observational conditions, and \textsc{JAM} for stellar dynamics.

Our principal findings are:

\begin{table*}
\caption{Best-fit \textsc{JAM} parameters and their associated uncertainties for six simulated galaxies.\label{tab:jamresults}}
\begin{center}
\begin{tabular}{c|ccc|ccc|ccc|cc}
\hline\hline
Model &\multicolumn{3}{c|}{input $M_{\rm BH}=0$~M$_\odot$}&\multicolumn{3}{c|}{input $\log_{10}M_{\rm BH}$(2)}&\multicolumn{3}{c|}{input $\log_{10}M_{\rm  BH}$(3)}&\textsc{VORONOI}&Number of \\ 
Parameters & Best fit & 1$\sigma$ & 3$\sigma$ & Best fit & 1$\sigma$ & 3$\sigma$ &Best fit&1$\sigma$&3$\sigma$&S/N&bins\\  
        (1) & (2) & (3) & (4) & (5) & (6) & (7) & (8) & (9) & (10) & (11) & (12) \\ 
        \hline
        \multicolumn{12}{c}{\underline{LEGAC-86906}}  \\[1mm]
$\log_{10}(M_{\rm BH}/$\Msun)& 3.92 & $\pm{2.84}$ & $\pm{4.22}$ & 9.00 & $\pm{0.26}$ & $\pm{0.45}$ & 9.64 & $\pm{0.06}$ & $\pm{0.15}$ &\multirow{4}{*}{76}&\multirow{4}{*}{32} \\
$f_{\rm scale}$              & 1.02 & $\pm{0.02}$ & $\pm{0.06}$ & 1.00 & $\pm{0.02}$ & $\pm{0.05}$ & 1.01 & $\pm{0.02}$ & $\pm{0.06}$ & &\\
$i$($^\circ$)                & 79.80& $\pm{9.80}$ & $\pm{14.80}$& 72.57 & $\pm{10.10}$& $\pm{14.97}$& 72.45& $\pm{11.51}$& $\pm{14.95}$& &\\
 $\beta_z$                   & 0.06 & $\pm{0.03}$ & $\pm{0.10}$ & 0.06 & $\pm{0.04}$ & $\pm{0.09}$ & 0.07 & $\pm{0.04}$ & $\pm{0.10}$ &\\ 
        \hline
        \multicolumn{12}{c}{\underline{LEGAC-27516}} \\[1mm]
$\log_{10}(M_{\rm BH}/$\Msun)& 4.18 & $\pm{2.97}$ & $\pm{4.17}$ &  9.12 & $\pm{0.13}$ & $\pm{0.36}$ & 9.54 & $\pm{0.06}$ & $\pm{0.12}$ &\multirow{4}{*}{63}&\multirow{4}{*}{43} \\
$f_{\rm scale}$              & 1.03 & $\pm{0.01}$ & $\pm{0.03}$ &  1.01 & $\pm{0.01}$ & $\pm{0.03}$ & 1.01 & $\pm{0.01}$ & $\pm{0.03}$ & &\\
$i$($^\circ$)                & 57.58& $\pm{21.16}$& $\pm{29.87}$& 67.37 & $\pm{17.89}$& $\pm{29.70}$& 57.35& $\pm{21.45}$& $\pm{29.75}$& &\\
$\beta_z$                    &  -0.01& $\pm{0.04}$ & $\pm{0.16}$ &$-$0.00& $\pm{0.02}$ & $\pm{0.06}$ & 0.04 & $\pm{0.04}$ & $\pm{0.10}$ & & \\
        \hline
        \multicolumn{12}{c}{\underline{S2F1-142}}\\[1mm]
$\log_{10}(M_{\rm BH}/$\Msun)& 3.91 & $\pm{2.73}$ & $\pm{4.18}$ & 9.50 & $\pm{0.05}$ & $\pm{0.14}$ & 9.71 & $\pm{0.03}$ & $\pm{0.08}$ &\multirow{4}{*}{55}&\multirow{4}{*}{64}\\
$f_{\rm scale}$         & 1.00 & $\pm{0.01}$ & $\pm{0.03}$ & 1.01 & $\pm{0.01}$ & $\pm{0.04}$ & 1.01 & $\pm{0.01}$ & $\pm{0.04}$ & & \\
$i$($^\circ$)           & 85.16& $\pm{4.86}$ & $\pm{14.37}$& 72.98& $\pm{10.91}$& $\pm{14.91}$& 67.80 & $\pm{12.12}$ & $\pm{14.89}$& & \\
$\beta_z$               & 0.03 & $\pm{0.01}$ & $\pm{0.06}$ &  0.07 & $\pm{0.03}$ & $\pm{0.08}$ & 0.07 & $\pm{0.03}$ & $\pm{0.07}$ & & \\
        \hline
        \multicolumn{12}{c}{\underline{UDS 29410}} \\[1mm]
$\log_{10}(M_{\rm BH}/$\Msun)& 4.04 & $\pm{2.75}$ & $\pm{4.20}$ & 9.61 & $\pm{0.07}$ & $\pm{0.25}$ & 9.93 & $\pm{0.06}$ & $\pm{0.15}$ &\multirow{4}{*}{26}&\multirow{4}{*}{34}\\
$f_{\rm scale}$              & 1.03 & $\pm{0.01}$ & $\pm{0.04}$ & 1.07 & $\pm{0.02}$ & $\pm{0.05}$ & 1.05 & $\pm{0.02}$ & $\pm{0.06}$ & & \\
$i$~($^\circ$)               & 86.78& $\pm{3.32}$ & $\pm{6.68}$ & 79.69& $\pm{6.97}$ & $\pm{12.08}$& 73.09& $\pm{9.28}$ & $\pm{13.97}$& & \\
$\beta_z$                    & 0.00 & $\pm{0.02}$ & $\pm{0.06}$ & 0.14 & $\pm{0.06}$ & $\pm{0.013}$& 0.14 & $\pm{0.11}$ & $\pm{0.23}$ & & \\
        \hline
        \multicolumn{12}{c}{\underline{CP-1243752}} \\[1mm]
$\log_{10}(M_{\rm BH}/$\Msun)& 4.18 & $\pm{2.87}$ & $\pm{4.41}$ & 9.51 & $\pm{0.09}$ & $\pm{0.28}$ & 9.86 & $\pm{0.04}$ & $\pm{0.11}$ &\multirow{4}{*}{46}&\multirow{4}{*}{26}\\
$f_{\rm scale}$              & 1.01 & $\pm{0.01}$ & $\pm{0.03}$ & 1.09 & $\pm{0.01}$ & $\pm{0.05}$ & 1.01 & $\pm{0.02}$ & $\pm{0.05}$ & & \\
$i$~($^\circ$)               & 75.85& $\pm{9.67}$ & $\pm{14.95}$& 75.98& $\pm{10.07}$& $\pm{14.95}$& 64.88& $\pm{9.39}$ & $\pm{14.92}$& &\\
$\beta_z$                    & 0.10 & $\pm{0.02}$ & $\pm{0.06}$ & 0.07 & $\pm{0.02}$ & $\pm{0.07}$ & 0.09 & $\pm{0.03}$ & $\pm{0.09}$ & & \\
        \hline
    \end{tabular}
\end{center}
\parbox[t]{0.99\textwidth}{\textit{Notes:} We established consistent search ranges for the JAM parameters as follows: $\lg(M_{\rm BH}/$M$_\odot)$: $0 \rightarrow 13$, $f_{\rm scale}$: $0.7 \rightarrow 1.3$, $\beta_z$: $-0.99 \rightarrow 1$, and $i$: $60^{\circ} \rightarrow 90^{\circ}$. Initial guesses: $M_{\rm BH}=0$~\Msun\ and $\lg(M_{\rm BH}/$M$_\odot)=9.5$ for cases where the input mock kinematics without and with a SMBH, respectively, $f_{\rm scale}=1.0$, $i (^{\circ})=70$, and $\beta_z=0$. Column (1): Galaxy name. Columns (2)--(4) present the best-fit parameters along with the 1$\sigma$ (16–84\%) and 3$\sigma$ (0.14–99.86\%) statistical and systematic uncertainties obtained from JAM when constrained using the HARMONI kinematics and MICADO stellar mass models. Here, the derived uncertainties are asymmetric as seen in \autoref{fig:z=1_kin_BHrecover_H+K}, \autoref{fig:z=1.5_kin_BHrecover_H+K}, and \autoref{fig:z=2_kin_BHrecover_H+K}; for consistency and transparency, we report symmetric errors corresponding to the larger asymmetric deviation in this table. These model constraints correspond to the case of no input black hole ($M_{\rm BH} = 0$~M$_\odot$). Columns (5)--(7) and (8)--(10) show the same information as Columns (2)--(4) but for cases where the input \Mbh\ is computed using Eqs. (2) and (3) from \citet{Krajnovic18b}, corresponding to \Mstar\ $< M_{\rm crit} \approx 2 \times 10^{11}$~\Msun\ and \Mstar\ $> M_{\rm crit}$, respectively. Columns (11)--(12) provide the target S/N per \AA\ required for each bin (obtained in the spectral range around either the CaT ($z\lesssim1.8$) or \ion{Mg}{1}b ($z>1.8$) stellar absorption features) and the number of \textsc{VORONOI} kinematic bins that exceed the target S/N, respectively.} 
\end{table*}

\begin{enumerate}

    \item \textbf{High-Fidelity Mass Models with MICADO:} Simulated MICADO observations, with an angular resolution of $\approx 10$ mas (FWHM$_{\rm PSF}$) and a 4 mas pixel scale, demonstrate that detailed stellar mass models, essential for dynamical studies, can be derived from approximately one hour of on-source exposure time per galaxy in the $I$ or $H$ band. The resulting surface brightness profiles are consistent with existing \hst\ data and provide the necessary resolution to probe the central regions of high-redshift galaxies.

    \item \textbf{Resolving Kinematics with HARMONI:} Mock HARMONI IFS observations, simulated using the $H+K$ grating at a 10 mas spaxel scale, successfully yielded stellar kinematic maps ($V$, $\sigma$, $V_{\rm rms}$) of sufficient quality. These kinematics, primarily extracted from CaT features (for $z \lesssim 1.8$) or \ion{Mg}{1}b features (for $z \approx 2$), clearly revealed the gravitational influence of central SMBHs. We observed distinct kinematic signatures, such as central drops or peaks in $V_{\rm rms}$ maps, depending on the SMBH mass and host galaxy properties, even at these substantial distances.

    \item \textbf{Accurate SMBH Mass Recovery:} By applying \textsc{JAM}-based dynamical models within an MCMC framework to the simulated HARMONI $V_{\rm rms}$ maps, we demonstrated that SMBH masses can be recovered with good accuracy. For our sample of bright, massive galaxies, the recovered \Mbh\ values typically agreed with the input simulation values to within $\sim$10\% up to $z\approx2$. The models also constrained other dynamical parameters, such as the stellar mass-scaling factor and orbital anisotropy, highlighting the robustness of the method.

    \item \textbf{Observational Requirements and Sensitivity:} We quantified the sensitivity of HARMONI, determining the minimum on-source exposure times required to obtain reliable stellar kinematics for SMBH mass measurements. These exposure times scale with the apparent brightness in the observed filter. For galaxies at $z\approx1$ observed in F814W, total magnitudes of 20–20.5 correspond to integration times of approximately 5–7.5~hours, while for F160W at $1<z\lesssim2$, 5~hours is sufficient to reach galaxies as faint as 20.8~mag. These limits establish practical exposure thresholds for future kinematic studies of distant galaxies and ensure sufficient S/N in the extracted spectra to mitigate the effects of cosmological surface brightness dimming and sky background.

    \item \textbf{A Framework for Future ELT Studies:} The simulation tools and methodologies presented, including the \texttt{jam\_mock\_data\_cube} routine, provide a robust framework for planning and interpreting future ELT observations. This work supports the potential of the ELT to significantly advance our understanding of SMBH demographics and their co-evolution with galaxies by extending direct dynamical measurements well beyond the local Universe.

\end{enumerate}

In summary, our simulations affirm that the ELT, with MICADO and HARMONI, possesses the transformative capability to directly measure SMBH masses in massive quiescent galaxies at $1\lesssim z\lesssim2$. These pioneering observations will be crucial for empirically testing the evolution of SMBH-galaxy scaling relations and providing new insights into the processes that govern galaxy formation and evolution across cosmic time.

\section*{Acknowledgements}
The authors would like to thank the anonymous referee for their careful reading and useful comments, that helped to improve the paper greatly.  We are grateful to Prof. van de Sande for providing the 3D-HST+CANDELS data used in \autoref{fig:highz_sample}. N.T. would like to acknowledge partial support from UKRI grant ST/X002322/1 for UK ELT Instrument Development at Oxford. M.P.S. acknowledges support under grants RYC2021-033094-I, CNS2023-145506 and PID2023-146667NB-I00 funded by MCIN/AEI/10.13039/501100011033 and the European Union Next Generation EU/PRTR.

{\it Facilities:} HST/WFC3.

{\it Software:} 
{\tt Python~3.12} \citep{VanRossum2009}, 
{\tt Matplotlib~3.6} \citep{Hunter2007}, 
{\tt NumPy~1.22} \citep{Harris2020}, 
{\tt SciPy~1.3} \citep{Virtanen2020},  
{\tt photutils~0.7} \citep{bradley2024}, 
{\tt AstroPy~5.1} \citep{AstropyCollaboration22}, 
{\tt AdaMet 2.0} \citep{Cappellari13a}, 
{\tt JamPy~7.2} \citep{Cappellari20}, 
{\tt pPXF~8.2} \citep{Cappellari23}, 
{\tt vorbin~3.1} \citep{Cappellari03}, 
{\tt MgeFit~5.0} \citep{Cappellari02}, 
{\tt HSIM~3.11} \citep{Zieleniewski15}, and 
{\tt SimCADO} \citep{Leschinski16}.
 
\section*{Data Avaibility}
All data and software used in this paper are public. We provided their links in the text when discussed. 

\bibliographystyle{mnras}
\bibliography{highz_SMBHs} 

@ARTICLE{Taylor2025,
       author = {{Taylor}, Matthew A. and {Tahmasebzadeh}, Behzad and {Thompson}, Solveig and {Vasiliev}, Eugene and {Valluri}, Monica and {Drinkwater}, Michael J. and {C{\^o}t{\'e}}, Patrick and {Ferrarese}, Laura and {Roediger}, Joel and {Baumgardt}, Holger and {Bentz}, Misty C. and {Dage}, Kristen and {Peng}, Eric W. and {Lapeer}, Drew and {Liu}, Chengze and {Sumners}, Zach and {Wang}, Kaixiang and {Baldassare}, Vivienne and {Blakeslee}, John P. and {Ko}, Youkyung and {Woods}, Tyrone E.},
        title = "{A Supermassive Black Hole in a Diminutive Ultracompact Dwarf Galaxy Discovered with JWST/NIRSpec+IFU}",
      journal = {\apjl},
         year = 2025,
        month = sep,
       volume = {991},
       number = {1},
          eid = {L24},
        pages = {L24},
          doi = {10.3847/2041-8213/ae028e},
archivePrefix = {arXiv},
       eprint = {2503.00113},
 primaryClass = {astro-ph.GA},
       adsurl = {https://ui.adsabs.harvard.edu/abs/2025ApJ...991L..24T}
}

@INPROCEEDINGS{Cappellari2025,
       author = {{Cappellari}, Michele},
        title = "{Early-type galaxies: Elliptical and S0 galaxies, or fast and slow rotators}",
    booktitle = {Encyclopedia of Astrophysics, Volume 4},
         year = 2026,
       volume = {4},
        month = jan,
        pages = {122-152},
          doi = {10.1016/B978-0-443-21439-4.00109-7},
archivePrefix = {arXiv},
       eprint = {2503.02746},
 primaryClass = {astro-ph.GA},
       adsurl = {https://ui.adsabs.harvard.edu/abs/2026enap....4..122C}
}

@ARTICLE{Tahmasebzadeh2025,
       author = {{Tahmasebzadeh}, Behzad and {Taylor}, Matthew A. and {Valluri}, Monica and {Yoshino}, Haruka and {Vasiliev}, Eugene and {Drinkwater}, Michael J. and {Thompson}, Solveig and {Dage}, Kristen and {C{\^o}t{\'e}}, Patrick and {Ferrarese}, Laura and {Akiba}, Tatsuya and {Baldassare}, Vivienne and {Bentz}, Misty C. and {Blakeslee}, John P. and {Baumgardt}, Holger and {Ko}, Youkyung and {Liu}, Chengze and {Madigan}, Ann-Marie and {Peng}, Eric W. and {Roediger}, Joel and {Wang}, Kaixiang and {Woods}, Tyrone E.},
        title = "{A JWST View of the Overmassive Black Hole in NGC 4486B}",
      journal = {\apjl},
         year = 2025,
        month = aug,
       volume = {989},
       number = {2},
          eid = {L42},
        pages = {L42},
          doi = {10.3847/2041-8213/adf728},
archivePrefix = {arXiv},
       eprint = {2505.14676},
 primaryClass = {astro-ph.GA},
       adsurl = {https://ui.adsabs.harvard.edu/abs/2025ApJ...989L..42T}
}

@ARTICLE{Nguyen2025c,
       author = {{Nguyen}, Dieu D. and {Ngo}, Hai N. and {Cappellari}, Michele and {Le}, Tinh Q.~T. and {Ho}, Tien H.~T. and {Le}, Tuan N. and {Gallo}, Elena and {Thatte}, Niranjan and {Zou}, Fan and {Perna}, Michele and {Pereira-Santaella}, Miguel},
        title = "{Measuring the Central Dark Mass in NGC 4258 with JWST/NIRSpec Stellar Kinematics}",
      journal = {arXiv e-prints},
         year = 2025,
        month = sep,
          eid = {arXiv:2509.20519},
        pages = {arXiv:2509.20519},
archivePrefix = {arXiv},
       eprint = {2509.20519},
 primaryClass = {astro-ph.GA},
       adsurl = {https://ui.adsabs.harvard.edu/abs/2025arXiv250920519N}
}

@ARTICLE{Ngo2025a,
       author = {{Ngo}, Hai N. and {Nguyen}, Dieu D. and {Le}, Tinh Q.~T. and {Ho}, Khue N.~H. and {Ho}, Tien H.~T. and {Gallo}, Elena and {Nyland}, Kristina and {Imanishi}, Masatoshi and {Nakanishi}, Kouichiro and {Le}, Que T. and {Pacucci}, Fabio and {Girma}, Eden},
        title = "{Revisiting the Supermassive Black Hole Mass of NGC 7052 Using High Spatial Resolution Molecular Gas Observed with ALMA}",
      journal = {\apj},
         year = 2025,
        month = oct,
       volume = {992},
       number = {2},
          eid = {211},
        pages = {211},
          doi = {10.3847/1538-4357/ae0455},
archivePrefix = {arXiv},
       eprint = {2509.02956},
 primaryClass = {astro-ph.GA},
       adsurl = {https://ui.adsabs.harvard.edu/abs/2025ApJ...992..211N}
}

@Article{Ngo2025c,
AUTHOR = {Ngo, Hai N. and Nguyen, Dieu D. and Le, Tinh T. Q. and Ho, Tien H. T. and Nguyen, Truong N. and Dang, Trung H.},
TITLE = {Detecting Intermediate-Mass Black Holes out to 20 Mpc with ELT/HARMONI: The Case of FCC 119},
JOURNAL = {Universe},
VOLUME = {11},
YEAR = {2025},
NUMBER = {11},
ARTICLE-NUMBER = {360},
URL = {https://www.mdpi.com/2218-1997/11/11/360},
ISSN = {2218-1997},
DOI = {10.3390/universe11110360}
}

@ARTICLE{Ngo2025b,
       author = {{Ngo}, Hai N and {Nguyen}, Dieu D. and {Nguyen}, Truong N. and {Dang}, Trung H. and {Ho}, Tien H.~T.},
        title = "{Extending the simulations of intermediate-mass black hole mass measurements to Virgo Cluster using ELT/HARMONI high resolution integral-field stellar kinematics}",
      journal = {arXiv e-prints},
         year = 2025,
        month = sep,
          eid = {arXiv:2509.03364},
        pages = {arXiv:2509.03364},
          doi = {10.48550/arXiv.2509.03364},
archivePrefix = {arXiv},
       eprint = {2509.03364},
 primaryClass = {astro-ph.GA},
       adsurl = {https://ui.adsabs.harvard.edu/abs/2025arXiv250903364N}
}

@ARTICLE{Longhetti2014,
       author = {{Longhetti}, M. and {Saracco}, P. and {Gargiulo}, A. and {Tamburri}, S. and {Lonoce}, I.},
        title = "{Large Binocular Telescope/LUCIFER spectroscopy: kinematics of a compact early-type galaxy at z ≃ 1.4}",
      journal = {\mnras},
         year = 2014,
        month = apr,
       volume = {439},
       number = {4},
        pages = {3962-3968},
          doi = {10.1093/mnras/stu252},
       adsurl = {https://ui.adsabs.harvard.edu/abs/2014MNRAS.439.3962L}
}

@ARTICLE{vanderWel14,
       author = {{van der Wel}, A. and {Franx}, M. and {van Dokkum}, P.~G. and {Skelton}, R.~E. and {Momcheva}, I.~G. and {Whitaker}, K.~E. and {Brammer}, G.~B. and {Bell}, E.~F. and {Rix}, H. -W. and {Wuyts}, S. and {Ferguson}, H.~C. and {Holden}, B.~P. and {Barro}, G. and {Koekemoer}, A.~M. and {Chang}, Yu-Yen and {McGrath}, E.~J. and {H{\"a}ussler}, B. and {Dekel}, A. and {Behroozi}, P. and {Fumagalli}, M. and {Leja}, J. and {Lundgren}, B.~F. and {Maseda}, M.~V. and {Nelson}, E.~J. and {Wake}, D.~A. and {Patel}, S.~G. and {Labb{\'e}}, I. and {Faber}, S.~M. and {Grogin}, N.~A. and {Kocevski}, D.~D.},
        title = "{3D-HST+CANDELS: The Evolution of the Galaxy Size-Mass Distribution since z = 3}",
      journal = {\apj},
         year = 2014,
        month = jun,
       volume = {788},
       number = {1},
          eid = {28},
        pages = {28},
          doi = {10.1088/0004-637X/788/1/28},
archivePrefix = {arXiv},
       eprint = {1404.2844},
 primaryClass = {astro-ph.GA},
       adsurl = {https://ui.adsabs.harvard.edu/abs/2014ApJ...788...28V}
}

@ARTICLE{Beifiori17,
       author = {{Beifiori}, Alessandra and {Mendel}, J. Trevor and {Chan}, Jeffrey C.~C. and {Saglia}, Roberto P. and {Bender}, Ralf and {Cappellari}, Michele and {Davies}, Roger L. and {Galametz}, Audrey and {Houghton}, Ryan C.~W. and {Prichard}, Laura J. and {Smith}, Russell and {Stott}, John P. and {Wilman}, David J. and {Lewis}, Ian J. and {Sharples}, Ray and {Wegner}, Michael},
        title = "{The KMOS Cluster Survey (KCS). I. The Fundamental Plane and the Formation Ages of Cluster Galaxies at Redshift 1.4 < z < 1.6}",
      journal = {\apj},
         year = 2017,
        month = sep,
       volume = {846},
       number = {2},
          eid = {120},
        pages = {120},
          doi = {10.3847/1538-4357/aa8368},
archivePrefix = {arXiv},
       eprint = {1708.00454},
 primaryClass = {astro-ph.GA},
       adsurl = {https://ui.adsabs.harvard.edu/abs/2017ApJ...846..120B}
}

@ARTICLE{Stockmann20,
       author = {{Stockmann}, Mikkel and {Toft}, Sune and {Gallazzi}, Anna and {Zibetti}, Stefano and {Conselice}, Christopher J. and {Margalef-Bentabol}, Berta and {Zabl}, Johannes and {J{\o}rgensen}, Inger and {Magdis}, Georgios E. and {G{\'o}mez-Guijarro}, Carlos and {Valentino}, Francesco M. and {Brammer}, Gabriel B. and {Ceverino}, Daniel and {Cortzen}, Isabella and {Davidzon}, Iary and {Demarco}, Richardo and {Faisst}, Andreas and {Hirschmann}, Michaela and {Krogager}, Jens-Kristian and {Lagos}, Claudia D. and {Man}, Allison W.~S. and {Mundy}, Carl J. and {Peng}, Yingjie and {Selsing}, Jonatan and {Steinhardt}, Charles L. and {Whitaker}, Kathrine E.},
        title = "{X-shooter Spectroscopy and HST Imaging of 15 Massive Quiescent Galaxies at z {\ensuremath{\gtrsim}} 2}",
      journal = {\apj},
         year = 2020,
        month = jan,
       volume = {888},
       number = {1},
          eid = {4},
        pages = {4},
          doi = {10.3847/1538-4357/ab5af4},
archivePrefix = {arXiv},
       eprint = {1912.01619},
 primaryClass = {astro-ph.GA},
       adsurl = {https://ui.adsabs.harvard.edu/abs/2020ApJ...888....4S}
}

@ARTICLE{Thater22,
       author = {{Thater}, Sabine and {Krajnovi{\'c}}, Davor and {Weilbacher}, Peter M. and {Nguyen}, Dieu D. and {Bureau}, Martin and {Cappellari}, Michele and {Davis}, Timothy A. and {Iguchi}, Satoru and {McDermid}, Richard and {Onishi}, Kyoko and {Sarzi}, Marc and {van de Ven}, Glenn},
        title = "{Cross-checking SMBH mass estimates in NGC 6958 - I. Stellar dynamics from adaptive optics-assisted MUSE observations}",
      journal = {\mnras},
         year = 2022,
        month = feb,
       volume = {509},
       number = {4},
        pages = {5416-5436},
          doi = {10.1093/mnras/stab3210},
archivePrefix = {arXiv},
       eprint = {2111.01620},
 primaryClass = {astro-ph.GA},
       adsurl = {https://ui.adsabs.harvard.edu/abs/2022MNRAS.509.5416T}
}

@ARTICLE{vandenBosch16,
       author = {{van den Bosch}, Remco C.~E.},
        title = "{Unification of the fundamental plane and Super Massive Black Hole Masses}",
      journal = {\apj},
         year = 2016,
        month = nov,
       volume = {831},
       number = {2},
          eid = {134},
        pages = {134},
          doi = {10.3847/0004-637X/831/2/134},
archivePrefix = {arXiv},
       eprint = {1606.01246},
 primaryClass = {astro-ph.GA},
       adsurl = {https://ui.adsabs.harvard.edu/abs/2016ApJ...831..134V}
}

@article{Haario01,
author = {Heikki Haario and Eero Saksman and Johanna Tamminen},
title = {{An adaptive Metropolis algorithm}},
year = 2001,
month = Apr,
volume = {7},
journal = {Bernoulli},
number = {2},
publisher = {Bernoulli Society for Mathematical Statistics and Probability},
pages = {223 - 242},
adsurl = {https://projecteuclid.org/journals/bernoulli/volume-7/issue-2/An-adaptive-Metropolis-algorithm/bj/1080222083.full},
}

@Article{Cappellari08,
  author        = {Cappellari, M.},
  title         = {Measuring the inclination and mass-to-light ratio of axisymmetric galaxies via anisotropic Jeans models of stellar kinematics},
  doi           = {10.1111/j.1365-2966.2008.13754.x},
  eprint        = {0806.0042},
  pages         = {71--86},
  volume        = {390},
  adsurl        = {https://ui.adsabs.harvard.edu/abs/2008MNRAS.390...71C},
  archiveprefix = {arXiv},
  journal       = {\mnras},
  month         = oct,
  year          = {2008},
}

@Book{Sersic68,
  author    = {Sersic, J. L.},
  title     = {Atlas de galaxias australes},
  publisher = {Obs. Astron. Univ. Nacional de C\'ordoba},
  address   = {C\'ordoba},
  adsurl    = {https://ui.adsabs.harvard.edu/abs/1968adga.book.....S},
  year      = {1968},
}

@Article{Tremaine94,
  author  = {Tremaine, S. and Richstone, D. O. and Byun, Y.-I. and Dressler, A. and Faber, S. M. and Grillmair, C. and Kormendy, J. and Lauer, T. R.},
  title   = {A family of models for spherical stellar systems},
  doi     = {10.1086/116883},
  eprint  = {arXiv:astro-ph/9309044},
  pages   = {634--644},
  volume  = {107},
  adsurl  = {https://ui.adsabs.harvard.edu/abs/1994AJ....107..634T},
  journal = {\aj},
  month   = feb,
  year    = {1994},
}

@Article{Kormendy13,
  author        = {Kormendy, J. and Ho, L. C.},
  title         = {{Coevolution (Or Not) of Supermassive Black Holes and Host Galaxies}},
  doi           = {10.1146/annurev-astro-082708-101811},
  eprint        = {1304.7762},
  pages         = {511--653},
  volume        = {51},
  adsurl        = {https://ui.adsabs.harvard.edu/abs/2013ARA%26A..51..511K},
  archiveprefix = {arXiv},
  journal       = {\araa},
  month         = aug,
  primaryclass  = {astro-ph.CO},
  year          = {2013},
}

@Article{Cappellari20,
  author        = {Cappellari, Michele},
  title         = {{Efficient solution of the anisotropic spherically aligned axisymmetric Jeans equations of stellar hydrodynamics for galactic dynamics}},
  doi           = {10.1093/mnras/staa959},
  eprint        = {1907.09894},
  number        = {4},
  pages         = {4819--4837},
  volume        = {494},
  adsurl        = {https://ui.adsabs.harvard.edu/abs/2020MNRAS.494.4819C},
  archiveprefix = {arXiv},
  journal       = {\mnras},
  month         = apr,
  primaryclass  = {astro-ph.GA},
  year          = {2020},
}

@Article{Cappellari02,
  author  = {Cappellari, M.},
  title   = {Efficient multi-Gaussian expansion of galaxies},
  doi     = {10.1046/j.1365-8711.2002.05412.x},
  eprint  = {arXiv:astro-ph/0201430},
  pages   = {400--410},
  volume  = {333},
  adsurl  = {https://ui.adsabs.harvard.edu/abs/2002MNRAS.333..400C},
  journal = {\mnras},
  month   = jun,
  year    = {2002},
}

@ARTICLE{Onishi17,
       author = {{Onishi}, Kyoko and {Iguchi}, Satoru and {Davis}, Timothy A. and
         {Bureau}, Martin and {Cappellari}, Michele and {Sarzi}, Marc and
         {Blitz}, Leo},
        title = "{WISDOM project - I. Black hole mass measurement using molecular gas kinematics in NGC 3665}",
      journal = {\mnras},
         year = "2017",
        month = "Jul",
       volume = {468},
       number = {4},
        pages = {4663-4674},
          doi = {10.1093/mnras/stx631},
archivePrefix = {arXiv},
       eprint = {1703.05247},
 primaryClass = {astro-ph.GA},
       adsurl = {https://ui.adsabs.harvard.edu/abs/2017MNRAS.468.4663O}
}

@ARTICLE{Rauscher2015,
       author = {{Rauscher}, Bernard J.},
        title = "{Teledyne H1RG, H2RG, and H4RG Noise Generator}",
      journal = {\pasp},
         year = 2015,
        month = nov,
       volume = {127},
       number = {957},
        pages = {1144},
          doi = {10.1086/684082},
archivePrefix = {arXiv},
       eprint = {1509.06264},
 primaryClass = {astro-ph.IM},
       adsurl = {https://ui.adsabs.harvard.edu/abs/2015PASP..127.1144R}
}

@ARTICLE{Imanishi20,
       author = {{Imanishi}, Masatoshi and {Nguyen}, Dieu D. and {Wada}, Keiichi and {Hagiwara}, Yoshiaki and {Iguchi}, Satoru and {Izumi}, Takuma and {Kawakatu}, Nozomu and {Nakanishi}, Kouichiro and {Onishi}, Kyoko},
        title = "{ALMA 0''02 Resolution Observations Reveal HCN-abundance-enhanced Counter-rotating and Outflowing Dense Molecular Gas at the NGC 1068 Nucleus}",
      journal = {\apj},
         year = 2020,
        month = oct,
       volume = {902},
       number = {2},
          eid = {99},
        pages = {99},
          doi = {10.3847/1538-4357/abaf50},
archivePrefix = {arXiv},
       eprint = {2008.08101},
 primaryClass = {astro-ph.GA},
       adsurl = {https://ui.adsabs.harvard.edu/abs/2020ApJ...902...99I}
}

@ARTICLE{Izumi20,
       author = {{Izumi}, Takuma and {Nguyen}, Dieu D. and {Imanishi}, Masatoshi and {Kawamuro}, Taiki and {Baba}, Shunsuke and {Nakano}, Suzuka and {Kohno}, Kotaro and {Matsushita}, Satoki and {Meier}, David S. and {Turner}, Jean L. and {Michiyama}, Tomonari and {Harada}, Nanase and {Mart{\'\i}n}, Sergio and {Nakanishi}, Kouichiro and {Takano}, Shuro and {Wiklind}, Tommy and {Nakai}, Naomasa and {Hsieh}, Pei-Ying},
        title = "{ALMA Observations of Multiple CO and C Lines toward the Active Galactic Nucleus of NGC 7469: An X-Ray-dominated Region Caught in the Act}",
      journal = {\apj},
         year = 2020,
        month = jul,
       volume = {898},
       number = {1},
          eid = {75},
        pages = {75},
          doi = {10.3847/1538-4357/ab9cb1},
archivePrefix = {arXiv},
       eprint = {2006.09406},
 primaryClass = {astro-ph.GA},
       adsurl = {https://ui.adsabs.harvard.edu/abs/2020ApJ...898...75I}
}

@ARTICLE{Nguyen20,
       author = {{Nguyen}, Dieu D. and {den Brok}, Mark and {Seth}, Anil C. and
         {Davis}, Timothy A. and {Greene}, Jenny E. and {Cappellari}, Michelle and
         {Jensen}, Joseph B. and {Thater}, Sabine and {Iguchi}, Satoru and
         {Imanishi}, Masatoshi and {Izumi}, Takuma and {Nyland}, Kristina and
         {Neumayer}, Nadine and {Nakanishi}, Kouichiro and {Nguyen}, Phuong M. and
         {Tsukui}, Takafumi and {Bureau}, Martin and {Onishi}, Kyoko and
         {Nguyen}, Quang L. and {Le}, Ngan M.},
        title = "{The MBHBM$_{{\ensuremath{\star}}}$ Project. I. Measurement of the Central Black Hole Mass in Spiral Galaxy NGC 3504 Using Molecular Gas Kinematics}",
      journal = {\apj},
         year = 2020,
        month = mar,
       volume = {892},
       number = {1},
          eid = {68},
        pages = {68},
          doi = {10.3847/1538-4357/ab77aa},
archivePrefix = {arXiv},
       eprint = {1902.03813},
 primaryClass = {astro-ph.GA},
       adsurl = {https://ui.adsabs.harvard.edu/abs/2020ApJ...892...68N}
}

@ARTICLE{Nguyen21,
       author = {{Nguyen}, Dieu D. and {Izumi}, Takuma and {Thater}, Sabine and {Imanishi}, Masatoshi and {Kawamuro}, Taiki and {Baba}, Shunsuke and {Nakano}, Suzuka and {Turner}, Jean L. and {Kohno}, Kotaro and {Matsushita}, Satoki and {Mart{\'\i}n}, Sergio and {Meier}, David S. and {Nguyen}, Phuong M. and {Nguyen}, Lam T.},
        title = "{Black hole mass measurement using ALMA observations of [CI] and CO emissions in the Seyfert 1 galaxy NGC 7469}",
      journal = {\mnras},
         year = 2021,
        month = jul,
       volume = {504},
       number = {3},
        pages = {4123-4142},
          doi = {10.1093/mnras/stab1002},
archivePrefix = {arXiv},
       eprint = {2104.03539},
 primaryClass = {astro-ph.GA},
       adsurl = {https://ui.adsabs.harvard.edu/abs/2021MNRAS.504.4123N}
}

@ARTICLE{Nguyen22,
       author = {{Nguyen}, Dieu D. and {Bureau}, Martin and {Thater}, Sabine and {Nyland}, Kristina and {den Brok}, Mark and {Cappellari}, Michele and {Davis}, Timothy A. and {Greene}, Jenny E. and {Neumayer}, Nadine and {Imanishi}, Masatoshi and {Izumi}, Takuma and {Kawamuro}, Taiki and {Baba}, Shunsuke and {Nguyen}, Phuong M. and {Iguchi}, Satoru and {Tsukui}, Takafumi and {Lam}, T.~N. and {Ho}, Than},
        title = "{The MBHBM$^{{\ensuremath{\star}}}$ Project - II. Molecular gas kinematics in the lenticular galaxy NGC 3593 reveal a supermassive black hole}",
      journal = {\mnras},
         year = 2022,
        month = jan,
       volume = {509},
       number = {2},
        pages = {2920-2939},
          doi = {10.1093/mnras/stab3016},
archivePrefix = {arXiv},
       eprint = {2110.08476},
 primaryClass = {astro-ph.GA},
       adsurl = {https://ui.adsabs.harvard.edu/abs/2022MNRAS.509.2920N}
}

@INPROCEEDINGS{Thatte20,
       author = {{Thatte}, Niranjan A. and {Bryson}, Ian and {Clarke}, Fraser and {Ferraro-Wood}, Vanessa and {Fusco}, Thierry and {Le Mignant}, David and {Melotte}, Dave and {Neichel}, Benoit and {Schnetler}, Hermine and {Tecza}, Matthias and {Arribas}, Santiago and {Crespo}, Alejandro and {Estrada Piqueras}, Alberto and {Garc{\'\i}a Garc{\'\i}a}, Miriam and {Pereira Santaella}, Miguel and {Piqueras L{\'o}pez}, Javier and {Blaizot}, Jeremy and {Bouch{\'e}}, Nicholas and {Boudon}, Didier and {Chapuis}, Diane and {Daguise}, Eric and {Disseau}, Karen and {Guibert}, Matthieu and {Jarno}, Aurelien and {Jeanneau}, Alexandre and {Laurent}, Florence and {Loupias}, Magali and {Migniau}, Jean-Emmanuel and {Piqueras}, Laure and {Remillieux}, Alban and {Richard}, Johan and {P{\'e}contal-Rousset}, Arlette and {Bardou}, Lisa and {Close}, Madeline M. and {Deshmukh}, Rishi and {Dimoudi}, Sofia and {Dubbledam}, Marc and {King}, David and {Morris}, Simon and {Morris}, Timothy J. and {O'Brien}, Kieran S. and {Staykov}, Lazar and {Swinbank}, Mark and {Townson}, Matthew and {Younger}, Eddy and {Accardo}, Matteo and {Avarez Mendez}, Domingo and {Conzelmann}, Ralf and {Egner}, Sebastian and {George}, Elizabeth M. and {Gont{\'e}}, Frederic and {Hopgood}, Joshua and {Ives}, Derek and {Mehrgan}, Leander and {Mueller}, Eric and {Peroux}, Celine and {Vernet}, Joel and {Alonso Sanchez}, Angel and {Giuseppina}, Battaglia and {Cagigas}, Miguel and {Delgado}, Jos{\'e} Miguel and {Fernandez Izquierdo}, Patricia and {Fragoso L{\'o}pez}, Ana Bel{\'e}n. and {Garcia-Lorenzo}, Maria Bego{\~n}a. and {Hernandez Suarez}, Elvio and {Herreros Linares}, Jos{\'e} Miguel and {Joven}, Enrique and {L{\'o}pez}, Roberto and {Mart{\'\i}n Hernando}, Yolanda and {Mediavilla}, Evencio and {Monreal}, Ana and {Pe{\~n}ate Castro}, Jos{\'e} and {Rasilla}, Jose Luis and {Rebolo}, Rafael and {Rodr{\'\i}guez-Ramos}, Luis Fernando and {Vega Moreno}, Afrodisio and {Viera}, Teodora and {Carlotti}, Alexis and {Correia}, Jean-Jacques and {Delboulbe}, Alain and {Guieu}, Sylvain and {Hours}, Adrien and {Hubert}, Zoltan and {Jocou}, Laurent and {Magnard}, Yves and {Moulin}, Thibaut and {Pancher}, Fabrice and {Rabou}, Patrick and {Stadler}, Eric and {Contini}, Thierry and {Larrieu}, Marie and {Fantei-Caujolle}, Yan and {Lecron}, Daniel and {Rousseau}, Sylvain and {Beltramo-Martin}, Olivier and {Bon}, William and {Bonnefoi}, Anne and {Ceria}, William and {Choquet}, Elodie and {Correia}, Carlos and {Costille}, Anne and {Dohlen}, Kjetil and {Ducret}, Franck and {El-Hadi}, Kacem and {Epinat}, Benoit and {Fetick}, Romain and {Gach}, Jean-Luc and {Groussin}, Olivier and {Jaafar}, Issa and {Le Merrer}, Joel and {Llored}, Marc and {Pedreros}, Felipe and {Renault}, Edgard and {Sanchez}, Patrice and {Vigan}, Arthur and {Vola}, Pascal and {Lim}, Caroline and {Vedrenne}, Nicola and {Petit}, Cyril and {Sauvage}, Jean-Francois and {Bagci}, Taha and {Cann}, Nick and {Chao Ortiz}, Jorge and {Elliott}, Ellis and {Seitis}, Tea and {Tosh}, Ian and {Anderson}, Josh and {Black}, Martin and {Bond}, Charlotte and {Born}, Andy J. and {Campbell}, Kenny and {Campbell}, Neil and {Carruthers}, James and {Cochrane}, William and {Dobson}, Naomi and {Evans}, Chris J. and {Gallie}, Angus and {Gonzalez}, Oscar and {Harman}, Joel and {Henry}, David M. and {Humphreys}, William and {Louth}, Tom and {Miller}, Chris and {Montgomery}, David M. and {Murray}, John and {O'Malley}, Norman and {Ritchie}, Lynn and {Sanchez-Janssen}, Ruben and {Schwartz}, Noah and {Smith}, Patrick and {Watt}, Stuart and {Wells}, Martyn and {Wilson}, Sandi and {Gultekin}, Kayhan and {Mateo}, Mario and {Meyer}, Michael and {Valluri}, Monica and {Ahmad}, Munadi and {Booth}, Michael and {Capone}, John I. and {Cappellari}, Michele and {Gooding}, David and {Grisdale}, Kearn and {Hidalgo}, Andrea and {Kariuki}, James and {Lewis}, Ian J. and {Lowe}, Adam and {Lynn}, Jim and {Menduina}, Alvaro and {Ozer}, Zeynep and {Preece}, Roy and {Rigopoulou}, Dimitra and {Rodrigues}, Myriam and {Routledge}, Laurence},
        title = "{HARMONI: first light spectroscopy for the ELT: instrument final design and quantitative performance predictions}",
    booktitle = {Ground-based and Airborne Instrumentation for Astronomy VIII},
         year = 2020,
       editor = {{Evans}, Christopher J. and {Bryant}, Julia J. and {Motohara}, Kentaro},
       series = {Society of Photo-Optical Instrumentation Engineers (SPIE) Conference Series},
       volume = {11447},
        month = dec,
          eid = {114471W},
        pages = {114471W},
          doi = {10.1117/12.2562144},
       adsurl = {https://ui.adsabs.harvard.edu/abs/2020SPIE11447E..1WT}
}

@ARTICLE{Cappellari13a,
       author = {{Cappellari}, Michele and {Scott}, Nicholas and {Alatalo}, Katherine and {Blitz}, Leo and {Bois}, Maxime and {Bournaud}, Fr{\'e}d{\'e}ric and {Bureau}, M. and {Crocker}, Alison F. and {Davies}, Roger L. and {Davis}, Timothy A. and {de Zeeuw}, P.~T. and {Duc}, Pierre-Alain and {Emsellem}, Eric and {Khochfar}, Sadegh and {Krajnovi{\'c}}, Davor and {Kuntschner}, Harald and {McDermid}, Richard M. and {Morganti}, Raffaella and {Naab}, Thorsten and {Oosterloo}, Tom and {Sarzi}, Marc and {Serra}, Paolo and {Weijmans}, Anne-Marie and {Young}, Lisa M.},
        title = "{The ATLAS$^{3D}$ project - XV. Benchmark for early-type galaxies scaling relations from 260 dynamical models: mass-to-light ratio, dark matter, Fundamental Plane and Mass Plane}",
      journal = {\mnras},
         year = 2013,
        month = jul,
       volume = {432},
       number = {3},
        pages = {1709-1741},
          doi = {10.1093/mnras/stt562},
archivePrefix = {arXiv},
       eprint = {1208.3522},
 primaryClass = {astro-ph.CO},
       adsurl = {https://ui.adsabs.harvard.edu/abs/2013MNRAS.432.1709C}
}

@INPROCEEDINGS{deZeeuw01,
       author = {{de Zeeuw}, Tim},
        title = "{Evidence for Massive Black Holes in Nearby Galactic Nuclei}",
    booktitle = {Black Holes in Binaries and Galactic Nuclei},
         year = 2001,
       editor = {{Kaper}, Lex and {Heuvel}, Edward P.~J. Van Den and {Woudt}, Patrick A.},
        month = jan,
        pages = {78},
          doi = {10.1007/10720995_12},
archivePrefix = {arXiv},
       eprint = {astro-ph/0009249},
 primaryClass = {astro-ph},
       adsurl = {https://ui.adsabs.harvard.edu/abs/2001bhbg.conf...78D}
}

@ARTICLE{Krajnovic18b,
       author = {{Krajnovi{\'c}}, Davor and {Cappellari}, Michele and {McDermid}, Richard M. and {Thater}, Sabine and {Nyland}, Kristina and {de Zeeuw}, P.~T. and {Falc{\'o}n-Barroso}, Jes{\'u}s and {Khochfar}, Sadegh and {Kuntschner}, Harald and {Sarzi}, Marc and {Young}, Lisa M.},
        title = "{A quartet of black holes and a missing duo: probing the low end of the M$_{BH}$-{\ensuremath{\sigma}} relation with the adaptive optics assisted integral-field spectroscopy}",
      journal = {\mnras},
         year = 2018,
        month = jul,
       volume = {477},
       number = {3},
        pages = {3030-3064},
          doi = {10.1093/mnras/sty778},
archivePrefix = {arXiv},
       eprint = {1803.08055},
 primaryClass = {astro-ph.GA},
       adsurl = {https://ui.adsabs.harvard.edu/abs/2018MNRAS.477.3030K}
}

@INPROCEEDINGS{Thatte16,
       author = {{Thatte}, Niranjan A. and {Clarke}, Fraser and {Bryson}, Ian and {Shnetler}, Hermine and {Tecza}, Matthias and {Fusco}, Thierry and {Bacon}, Roland M. and {Richard}, Johan and {Mediavilla}, Evencio and {Neichel}, Beno{\^\i}t and {Arribas}, Santiago and {Garcia-Lorenzo}, Bego{\~n}a. and {Evans}, Christopher J. and {Remillieux}, Alban and {El Madi}, Kacem and {Herreros}, Jose Miguel and {Melotte}, Dave and {O'Brien}, Kieran and {Tosh}, Ian A. and {Vernet}, Jo{\"e}l. and {Hammersley}, Peter and {Ives}, Derek J. and {Finger}, Gert and {Houghton}, Ryan and {Rigopoulou}, Dimitra and {Lynn}, James D. and {Allen}, Jamie R. and {Zieleniewski}, Simon D. and {Kendrew}, Sarah and {Ferraro-Wood}, Vanessa and {P{\'e}contal-Rousset}, Arlette and {Kosmalski}, Johan and {Laurent}, Florence and {Loupias}, Magali and {Piqueras}, Laure and {Renault}, Edgar and {Blaizot}, Jeremy and {Daguis{\'e}}, Eric and {Migniau}, Jean-Emmanuel and {Jarno}, Aur{\'e}lien and {Born}, Andy and {Gallie}, Angus M. and {Montgomery}, David M. and {Henry}, David and {Schwartz}, Noah and {Taylor}, William and {Zins}, G{\'e}rard and {Rodr{\'\i}guez-Ramos}, Luis Fernando and {Cagigas}, Miguel and {Battaglia}, Giuseppina and {Rebolo L{\'o}pez}, Refael and {Hern{\'a}ndez Su{\'a}rez}, Elvio and {Gigante-Ripoll}, Jos{\'e} Vicente and {Piqueras L{\'o}pez}, Javier and {Villar Martin}, Montserrat and {Correia}, Carlos and {Pascal}, Sandrine and {Blanco}, Leonardo and {Vola}, Pascal and {Epinat}, Benoit and {Peroux}, Celine and {Vigan}, Arthur and {Dohlen}, Kjetil and {Sauvage}, Jean-Francois and {Lee}, Martin and {Carlotti}, Alexis and {Verinaud}, Christophe and {Morris}, Tim and {Myers}, Richard and {Reeves}, Andrew and {Swinbank}, Mark and {Calcines}, Ariadna and {Larrieu}, Marrie},
        title = "{The E-ELT first light spectrograph HARMONI: capabilities and modes}",
    booktitle = {Ground-based and Airborne Instrumentation for Astronomy VI},
         year = 2016,
       editor = {{Evans}, Christopher J. and {Simard}, Luc and {Takami}, Hideki},
       series = {Society of Photo-Optical Instrumentation Engineers (SPIE) Conference Series},
       volume = {9908},
        month = aug,
          eid = {99081X},
        pages = {99081X},
          doi = {10.1117/12.2230629},
       adsurl = {https://ui.adsabs.harvard.edu/abs/2016SPIE.9908E..1XT}
}

@ARTICLE{Davies21,
       author = {{Davies}, R. and {H{\"o}rmann}, V. and {Rabien}, S. and {Sturm}, E. and {Alves}, J. and {Cl{\'e}net}, Y. and {Kotilainen}, J. and {Lang-Bardl}, F. and {Nicklas}, H. and {Pott}, J. -U. and {Tolstoy}, E. and {Vulcani}, B. and {MICADO Consortium}},
        title = "{MICADO: The Multi-Adaptive Optics Camera for Deep Observations}",
      journal = {The Messenger},
         year = 2021,
        month = mar,
       volume = {182},
        pages = {17-21},
          doi = {10.18727/0722-6691/5217},
archivePrefix = {arXiv},
       eprint = {2103.11631},
 primaryClass = {astro-ph.IM},
       adsurl = {https://ui.adsabs.harvard.edu/abs/2021Msngr.182...17D}
}

@ARTICLE{Zieleniewski15,
       author = {{Zieleniewski}, S. and {Thatte}, N. and {Kendrew}, S. and {Houghton}, R.~C.~W. and {Swinbank}, A.~M. and {Tecza}, M. and {Clarke}, F. and {Fusco}, T.},
        title = "{HSIM: a simulation pipeline for the HARMONI integral field spectrograph on the European ELT}",
      journal = {\mnras},
         year = 2015,
        month = nov,
       volume = {453},
       number = {4},
        pages = {3754-3765},
          doi = {10.1093/mnras/stv1860},
archivePrefix = {arXiv},
       eprint = {1508.04441},
 primaryClass = {astro-ph.IM},
       adsurl = {https://ui.adsabs.harvard.edu/abs/2015MNRAS.453.3754Z}
}

@ARTICLE{Oke74,
       author = {{Oke}, J.~B.},
        title = "{Absolute Spectral Energy Distributions for White Dwarfs}",
      journal = {\apjs},
         year = 1974,
        month = feb,
       volume = {27},
        pages = {21},
          doi = {10.1086/190287},
       adsurl = {https://ui.adsabs.harvard.edu/abs/1974ApJS...27...21O}
}

@ARTICLE{Cardelli89,
       author = {{Cardelli}, Jason A. and {Clayton}, Geoffrey C. and {Mathis}, John S.},
        title = "{The Relationship between Infrared, Optical, and Ultraviolet Extinction}",
      journal = {\apj},
         year = 1989,
        month = oct,
       volume = {345},
        pages = {245},
          doi = {10.1086/167900},
       adsurl = {https://ui.adsabs.harvard.edu/abs/1989ApJ...345..245C}
}

@ARTICLE{Schlafly11,
       author = {{Schlafly}, Edward F. and {Finkbeiner}, Douglas P.},
        title = "{Measuring Reddening with Sloan Digital Sky Survey Stellar Spectra and Recalibrating SFD}",
      journal = {\apj},
         year = 2011,
        month = aug,
       volume = {737},
       number = {2},
          eid = {103},
        pages = {103},
          doi = {10.1088/0004-637X/737/2/103},
archivePrefix = {arXiv},
       eprint = {1012.4804},
 primaryClass = {astro-ph.GA},
       adsurl = {https://ui.adsabs.harvard.edu/abs/2011ApJ...737..103S}
}

@INPROCEEDINGS{Davies10,
       author = {{Davies}, Richard and {Ageorges}, N. and {Barl}, L. and {Bedin}, L.~R. and {Bender}, R. and {Bernardi}, P. and {Chapron}, F. and {Clenet}, Y. and {Deep}, A. and {Deul}, E. and {Drost}, M. and {Eisenhauer}, F. and {Falomo}, R. and {Fiorentino}, G. and {F{\"o}rster Schreiber}, N.~M. and {Gendron}, E. and {Genzel}, R. and {Gratadour}, D. and {Greggio}, L. and {Grupp}, F. and {Held}, E. and {Herbst}, T. and {Hess}, H. -J. and {Hubert}, Z. and {Jahnke}, K. and {Kuijken}, K. and {Lutz}, D. and {Magrin}, D. and {Muschielok}, B. and {Navarro}, R. and {Noyola}, E. and {Paumard}, T. and {Piotto}, G. and {Ragazzoni}, R. and {Renzini}, A. and {Rousset}, G. and {Rix}, H. -W. and {Saglia}, R. and {Tacconi}, L. and {Thiel}, M. and {Tolstoy}, E. and {Trippe}, S. and {Tromp}, N. and {Valentijn}, E.~A. and {Verdoes Kleijn}, G. and {Wegner}, M.},
        title = "{MICADO: the E-ELT adaptive optics imaging camera}",
    booktitle = {Ground-based and Airborne Instrumentation for Astronomy III},
         year = 2010,
       editor = {{McLean}, Ian S. and {Ramsay}, Suzanne K. and {Takami}, Hideki},
       series = {Society of Photo-Optical Instrumentation Engineers (SPIE) Conference Series},
       volume = {7735},
        month = jul,
          eid = {77352A},
        pages = {77352A},
          doi = {10.1117/12.856379},
archivePrefix = {arXiv},
       eprint = {1005.5009},
 primaryClass = {astro-ph.IM},
       adsurl = {https://ui.adsabs.harvard.edu/abs/2010SPIE.7735E..2AD}
}

@ARTICLE{Davis20,
       author = {{Davis}, Timothy A. and {Nguyen}, Dieu D. and {Seth}, Anil C. and
         {Greene}, Jenny E. and {Nyland}, Kristina and {Barth}, Aaron J. and
         {Bureau}, Martin and {Cappellari}, Michele and {den Brok}, Mark and
         {Iguchi}, Satoru and {Lelli}, Federico and {Liu}, Lijie and
         {Neumayer}, Nadine and {North}, Eve V. and {Onishi}, Kyoko and
         {Sarzi}, Marc and {Smith}, Mark D. and {Williams}, Thomas G.},
        title = "{Revealing the intermediate-mass black hole at the heart of the dwarf galaxy NGC 404 with sub-parsec resolution ALMA observations}",
      journal = {\mnras},
         year = 2020,
        month = jul,
       volume = {496},
       number = {4},
        pages = {4061-4078},
          doi = {10.1093/mnras/staa1567},
archivePrefix = {arXiv},
       eprint = {2007.05536},
 primaryClass = {astro-ph.GA},
       adsurl = {https://ui.adsabs.harvard.edu/abs/2020MNRAS.496.4061D}
}

@ARTICLE{Smith19,
       author = {{Smith}, Mark D. and {Bureau}, Martin and {Davis}, Timothy A. and
         {Cappellari}, Michele and {Liu}, Lijie and {North}, Eve V. and
         {Onishi}, Kyoko and {Iguchi}, Satoru and {Sarzi}, Marc},
        title = "{WISDOM project - IV. A molecular gas dynamical measurement of the supermassive black hole mass in NGC 524}",
      journal = {\mnras},
         year = 2019,
        month = may,
       volume = {485},
       number = {3},
        pages = {4359-4374},
          doi = {10.1093/mnras/stz625},
archivePrefix = {arXiv},
       eprint = {1903.03124},
 primaryClass = {astro-ph.GA},
       adsurl = {https://ui.adsabs.harvard.edu/abs/2019MNRAS.485.4359S}
}

@ARTICLE{Nguyen19,
       author = {{Nguyen}, Dieu D. and {Seth}, Anil C. and {Neumayer}, Nadine and {Iguchi}, Satoru and {Cappellari}, Michelle and {Strader}, Jay and {Chomiuk}, Laura and {Tremou}, Evangelia and {Pacucci}, Fabio and {Nakanishi}, Kouichiro and {Bahramian}, Arash and {Nguyen}, Phuong M. and {den Brok}, Mark and {Ahn}, Christopher C. and {Voggel}, Karina T. and {Kacharov}, Nikolay and {Tsukui}, Takafumi and {Ly}, Cuc K. and {Dumont}, Antoine and {Pechetti}, Renuka},
        title = "{Improved Dynamical Constraints on the Masses of the Central Black Holes in Nearby Low-mass Early-type Galactic Nuclei and the First Black Hole Determination for NGC 205}",
      journal = {\apj},
         year = 2019,
        month = feb,
       volume = {872},
       number = {1},
          eid = {104},
        pages = {104},
          doi = {10.3847/1538-4357/aafe7a},
archivePrefix = {arXiv},
       eprint = {1901.05496},
 primaryClass = {astro-ph.GA},
       adsurl = {https://ui.adsabs.harvard.edu/abs/2019ApJ...872..104N}
}

@ARTICLE{Nguyen18,
       author = {{Nguyen}, Dieu D. and {Seth}, Anil C. and {Neumayer}, Nadine and {Kamann}, Sebastian and {Voggel}, Karina T. and {Cappellari}, Michele and {Picotti}, Arianna and {Nguyen}, Phuong M. and {B{\"o}ker}, Torsten and {Debattista}, Victor and {Caldwell}, Nelson and {McDermid}, Richard and {Bastian}, Nathan and {Ahn}, Christopher C. and {Pechetti}, Renuka},
        title = "{Nearby Early-type Galactic Nuclei at High Resolution: Dynamical Black Hole and Nuclear Star Cluster Mass Measurements}",
      journal = {\apj},
         year = 2018,
        month = may,
       volume = {858},
       number = {2},
          eid = {118},
        pages = {118},
          doi = {10.3847/1538-4357/aabe28},
archivePrefix = {arXiv},
       eprint = {1711.04314},
 primaryClass = {astro-ph.GA},
       adsurl = {https://ui.adsabs.harvard.edu/abs/2018ApJ...858..118N}
}

@ARTICLE{Nguyen17conf,
       author = {{Nguyen}, Dieu D.},
        title = "{Improved dynamical constraints on the mass of the central black hole in NGC 404}",
      journal = {arXiv e-prints},
         year = 2017,
        month = dec,
          eid = {arXiv:1712.02470},
        pages = {arXiv:1712.02470},
          doi = {10.48550/arXiv.1712.02470},
archivePrefix = {arXiv},
       eprint = {1712.02470},
 primaryClass = {astro-ph.GA},
       adsurl = {https://ui.adsabs.harvard.edu/abs/2017arXiv171202470N}
}

@ARTICLE{Nguyen17,
       author = {{Nguyen}, Dieu D. and {Seth}, Anil C. and {den Brok}, Mark and {Neumayer}, Nadine and {Cappellari}, Michele and {Barth}, Aaron J. and {Caldwell}, Nelson and {Williams}, Benjamin F. and {Binder}, Breanna},
        title = "{Improved Dynamical Constraints on the Mass of the Central Black Hole in NGC 404}",
      journal = {\apj},
         year = 2017,
        month = feb,
       volume = {836},
       number = {2},
          eid = {237},
        pages = {237},
          doi = {10.3847/1538-4357/aa5cb4},
archivePrefix = {arXiv},
       eprint = {1610.09385},
 primaryClass = {astro-ph.GA},
       adsurl = {https://ui.adsabs.harvard.edu/abs/2017ApJ...836..237N}
}

@ARTICLE{Nguyen23,
       author = {{Nguyen}, Dieu D. and {Cappellari}, Michele and {Pereira-Santaella}, Miguel},
        title = "{Simulating supermassive black hole mass measurements for a sample of ultramassive galaxies using ELT/HARMONI high-spatial-resolution integral-field stellar kinematics}",
      journal = {\mnras},
         year = 2023,
        month = dec,
       volume = {526},
       number = {3},
        pages = {3548-3569},
          doi = {10.1093/mnras/stad2860},
archivePrefix = {arXiv},
       eprint = {2302.10012},
 primaryClass = {astro-ph.GA},
       adsurl = {https://ui.adsabs.harvard.edu/abs/2023MNRAS.526.3548N}
}

@ARTICLE{Walsh13,
       author = {{Walsh}, Jonelle L. and {Barth}, Aaron J. and {Ho}, Luis C. and {Sarzi}, Marc},
        title = "{The M87 Black Hole Mass from Gas-dynamical Models of Space Telescope Imaging Spectrograph Observations}",
      journal = {\apj},
         year = 2013,
        month = jun,
       volume = {770},
       number = {2},
          eid = {86},
        pages = {86},
          doi = {10.1088/0004-637X/770/2/86},
archivePrefix = {arXiv},
       eprint = {1304.7273},
 primaryClass = {astro-ph.CO},
       adsurl = {https://ui.adsabs.harvard.edu/abs/2013ApJ...770...86W}
}

@ARTICLE{Barth16,
   author = {{Barth}, A.~J. and {Boizelle}, B.~D. and {Darling}, J. and {Baker}, A.~J. and 
  {Buote}, D.~A. and {Ho}, L.~C. and {Walsh}, J.~L.},
    title = "{Measurement of the Black Hole Mass in NGC 1332 from ALMA Observations at 0.044 arcsecond Resolution}",
  journal = {\apjl},
archivePrefix = "arXiv",
   eprint = {1605.01346},
     year = 2016,
    month = may,
   volume = 822,
      eid = {L28},
    pages = {L28},
      doi = {10.3847/2041-8205/822/2/L28},
   adsurl = {http://adsabs.harvard.edu/abs/2016ApJ...822L..28B}
}

@ARTICLE{Nguyen14,
   author = {{Nguyen}, D.~D. and {Seth}, A.~C. and {Reines}, A.~E. and {den Brok}, M. and 
  {Sand}, D. and {McLeod}, B.},
    title = "{Extended Structure and Fate of the Nucleus in Henize 2-10}",
  journal = {\apj},
archivePrefix = "arXiv",
   eprint = {1408.4446},
     year = 2014,
    month = oct,
   volume = 794,
      eid = {34},
    pages = {34},
      doi = {10.1088/0004-637X/794/1/34},
   adsurl = {http://adsabs.harvard.edu/abs/2014ApJ...794...34N}
}

@ARTICLE{Krajnovic18a,
       author = {{Krajnovi{\'c}}, Davor and {Cappellari}, Michele and
         {McDermid}, Richard M.},
        title = "{Two channels of supermassive black hole growth as seen on the galaxies mass-size plane}",
      journal = {\mnras},
         year = 2018,
        month = feb,
       volume = {473},
       number = {4},
        pages = {5237-5247},
          doi = {10.1093/mnras/stx2704},
archivePrefix = {arXiv},
       eprint = {1707.04274},
 primaryClass = {astro-ph.GA},
       adsurl = {https://ui.adsabs.harvard.edu/abs/2018MNRAS.473.5237K}
}

@ARTICLE{Thater23,
       author = {{Thater}, Sabine and {Lyubenova}, Mariya and {Fahrion}, Katja and {Mart{\'\i}n-Navarro}, Ignacio and {Jethwa}, Prashin and {Nguyen}, Dieu D. and {van de Ven}, Glenn},
        title = "{Effect of the initial mass function on the dynamical SMBH mass estimate in the nucleated early-type galaxy FCC 47}",
      journal = {\aap},
         year = 2023,
        month = jul,
       volume = {675},
          eid = {A18},
        pages = {A18},
          doi = {10.1051/0004-6361/202245362},
archivePrefix = {arXiv},
       eprint = {2304.13310},
 primaryClass = {astro-ph.GA},
       adsurl = {https://ui.adsabs.harvard.edu/abs/2023A&A...675A..18T}
}

@INPROCEEDINGS{Nguyen19conf,
       author = {{Nguyen}, Dieu D.},
        title = "{Uncovering the Census of Black Holes in sub-Milky Way Mass Galaxies}",
    booktitle = {ALMA2019: Science Results and Cross-Facility Synergies},
         year = 2019,
        month = dec,
          eid = {106},
        pages = {106},
          doi = {10.5281/zenodo.3585410},
       adsurl = {https://ui.adsabs.harvard.edu/abs/2019asrc.confE.106N}
}

@ARTICLE{Magorrian98,
       author = {{Magorrian}, John and {Tremaine}, Scott and {Richstone}, Douglas and {Bender}, Ralf and {Bower}, Gary and {Dressler}, Alan and {Faber}, S.~M. and {Gebhardt}, Karl and {Green}, Richard and {Grillmair}, Carl and {Kormendy}, John and {Lauer}, Tod},
        title = "{The Demography of Massive Dark Objects in Galaxy Centers}",
      journal = {\aj},
         year = 1998,
        month = jun,
       volume = {115},
       number = {6},
        pages = {2285-2305},
          doi = {10.1086/300353},
archivePrefix = {arXiv},
       eprint = {astro-ph/9708072},
 primaryClass = {astro-ph},
       adsurl = {https://ui.adsabs.harvard.edu/abs/1998AJ....115.2285M}
}

@ARTICLE{Kormendy95,
       author = {{Kormendy}, John and {Richstone}, Douglas},
        title = "{Inward Bound---The Search For Supermassive Black Holes In Galactic Nuclei}",
      journal = {\araa},
         year = 1995,
        month = jan,
       volume = {33},
        pages = {581},
          doi = {10.1146/annurev.aa.33.090195.003053},
       adsurl = {https://ui.adsabs.harvard.edu/abs/1995ARA&A..33..581K}
}

@ARTICLE{Fabian12,
       author = {{Fabian}, A.~C.},
        title = "{Observational Evidence of Active Galactic Nuclei Feedback}",
      journal = {\araa},
         year = 2012,
        month = sep,
       volume = {50},
        pages = {455-489},
          doi = {10.1146/annurev-astro-081811-125521},
archivePrefix = {arXiv},
       eprint = {1204.4114},
 primaryClass = {astro-ph.CO},
       adsurl = {https://ui.adsabs.harvard.edu/abs/2012ARA&A..50..455F}
}

@ARTICLE{Ferrarese00,
       author = {{Ferrarese}, Laura and {Merritt}, David},
        title = "{A Fundamental Relation between Supermassive Black Holes and Their Host Galaxies}",
      journal = {\apjl},
         year = 2000,
        month = aug,
       volume = {539},
       number = {1},
        pages = {L9-L12},
          doi = {10.1086/312838},
archivePrefix = {arXiv},
       eprint = {astro-ph/0006053},
 primaryClass = {astro-ph},
       adsurl = {https://ui.adsabs.harvard.edu/abs/2000ApJ...539L...9F}
}

@ARTICLE{Gebhardt00,
       author = {{Gebhardt}, Karl and {Bender}, Ralf and {Bower}, Gary and {Dressler}, Alan and {Faber}, S.~M. and {Filippenko}, Alexei V. and {Green}, Richard and {Grillmair}, Carl and {Ho}, Luis C. and {Kormendy}, John and {Lauer}, Tod R. and {Magorrian}, John and {Pinkney}, Jason and {Richstone}, Douglas and {Tremaine}, Scott},
        title = "{A Relationship between Nuclear Black Hole Mass and Galaxy Velocity Dispersion}",
      journal = {\apjl},
         year = 2000,
        month = aug,
       volume = {539},
       number = {1},
        pages = {L13-L16},
          doi = {10.1086/312840},
archivePrefix = {arXiv},
       eprint = {astro-ph/0006289},
 primaryClass = {astro-ph},
       adsurl = {https://ui.adsabs.harvard.edu/abs/2000ApJ...539L..13G}
}

@ARTICLE{Greene20,
       author = {{Greene}, Jenny E. and {Strader}, Jay and {Ho}, Luis C.},
        title = "{Intermediate-Mass Black Holes}",
      journal = {\araa},
         year = 2020,
        month = aug,
       volume = {58},
        pages = {257-312},
          doi = {10.1146/annurev-astro-032620-021835},
archivePrefix = {arXiv},
       eprint = {1911.09678},
 primaryClass = {astro-ph.GA},
       adsurl = {https://ui.adsabs.harvard.edu/abs/2020ARA&A..58..257G}
}

@ARTICLE{Williams09,
       author = {{Williams}, Rik J. and {Quadri}, Ryan F. and {Franx}, Marijn and {van Dokkum}, Pieter and {Labb{\'e}}, Ivo},
        title = "{Detection of Quiescent Galaxies in a Bicolor Sequence from Z = 0-2}",
      journal = {\apj},
         year = 2009,
        month = feb,
       volume = {691},
       number = {2},
        pages = {1879-1895},
          doi = {10.1088/0004-637X/691/2/1879},
archivePrefix = {arXiv},
       eprint = {0806.0625},
 primaryClass = {astro-ph},
       adsurl = {https://ui.adsabs.harvard.edu/abs/2009ApJ...691.1879W}
}

@ARTICLE{Verro22,
       author = {{Verro}, K. and {Trager}, S.~C. and {Peletier}, R.~F. and {Lan{\c{c}}on}, A. and {Gonneau}, A. and {Vazdekis}, A. and {Prugniel}, P. and {Chen}, Y. -P. and {Coelho}, P.~R.~T. and {S{\'a}nchez-Bl{\'a}zquez}, P. and {Martins}, L. and {Arentsen}, A. and {Lyubenova}, M. and {Falc{\'o}n-Barroso}, J. and {Dries}, M.},
        title = "{The X-shooter Spectral Library (XSL): Data Release 3}",
      journal = {\aap},
         year = 2022,
        month = apr,
       volume = {660},
          eid = {A34},
        pages = {A34},
          doi = {10.1051/0004-6361/202142388},
archivePrefix = {arXiv},
       eprint = {2110.10188},
 primaryClass = {astro-ph.SR},
       adsurl = {https://ui.adsabs.harvard.edu/abs/2022A&A...660A..34V}
}

@ARTICLE{Voggel18,
       author = {{Voggel}, Karina T. and {Seth}, Anil C. and {Neumayer}, Nadine and {Mieske}, Steffen and {Chilingarian}, Igor and {Ahn}, Christopher and {Baumgardt}, Holger and {Hilker}, Michael and {Nguyen}, Dieu D. and {Romanowsky}, Aaron J. and {Walsh}, Jonelle L. and {den Brok}, Mark and {Strader}, Jay},
        title = "{Upper Limits on the Presence of Central Massive Black Holes in Two Ultra-compact Dwarf Galaxies in Centaurus A}",
      journal = {\apj},
         year = 2018,
        month = may,
       volume = {858},
       number = {1},
          eid = {20},
        pages = {20},
          doi = {10.3847/1538-4357/aabae5},
archivePrefix = {arXiv},
       eprint = {1803.09750},
 primaryClass = {astro-ph.GA},
       adsurl = {https://ui.adsabs.harvard.edu/abs/2018ApJ...858...20V}
}

@ARTICLE{vandeSande13,
       author = {{van de Sande}, Jesse and {Kriek}, Mariska and {Franx}, Marijn and {van Dokkum}, Pieter G. and {Bezanson}, Rachel and {Bouwens}, Rychard J. and {Quadri}, Ryan F. and {Rix}, Hans-Walter and {Skelton}, Rosalind E.},
        title = "{Stellar Kinematics of z \raisebox{-0.5ex}\textasciitilde 2 Galaxies and the Inside-out Growth of Quiescent Galaxies}",
      journal = {\apj},
         year = 2013,
        month = jul,
       volume = {771},
       number = {2},
          eid = {85},
        pages = {85},
          doi = {10.1088/0004-637X/771/2/85},
       adsurl = {https://ui.adsabs.harvard.edu/abs/2013ApJ...771...85V}
}

@ARTICLE{vanDokkum09,
       author = {{van Dokkum}, Pieter G. and {Labb{\'e}}, Ivo and {Marchesini}, Danilo and {Quadri}, Ryan and {Brammer}, Gabriel and {Whitaker}, Katherine E. and {Kriek}, Mariska and {Franx}, Marijn and {Rudnick}, Gregory and {Illingworth}, Garth and {Lee}, Kyoung-Soo and {Muzzin}, Adam},
        title = "{The NEWFIRM Medium-Band Survey: Filter Definitions and First Results}",
      journal = {\pasp},
         year = 2009,
        month = jan,
       volume = {121},
       number = {875},
        pages = {2},
          doi = {10.1086/597138},
       adsurl = {https://ui.adsabs.harvard.edu/abs/2009PASP..121....2V}
}

@ARTICLE{Lawrence07,
       author = {{Lawrence}, A. and {Warren}, S.~J. and {Almaini}, O. and {Edge}, A.~C. and {Hambly}, N.~C. and {Jameson}, R.~F. and {Lucas}, P. and {Casali}, M. and {Adamson}, A. and {Dye}, S. and {Emerson}, J.~P. and {Foucaud}, S. and {Hewett}, P. and {Hirst}, P. and {Hodgkin}, S.~T. and {Irwin}, M.~J. and {Lodieu}, N. and {McMahon}, R.~G. and {Simpson}, C. and {Smail}, I. and {Mortlock}, D. and {Folger}, M.},
        title = "{The UKIRT Infrared Deep Sky Survey (UKIDSS)}",
      journal = {\mnras},
         year = 2007,
        month = aug,
       volume = {379},
       number = {4},
        pages = {1599-1617},
          doi = {10.1111/j.1365-2966.2007.12040.x},
archivePrefix = {arXiv},
       eprint = {astro-ph/0604426},
 primaryClass = {astro-ph},
       adsurl = {https://ui.adsabs.harvard.edu/abs/2007MNRAS.379.1599L}
}

@ARTICLE{vanderWel16,
       author = {{van der Wel}, A. and {Noeske}, K. and {Bezanson}, R. and {Pacifici}, C. and {Gallazzi}, A. and {Franx}, M. and {Mu{\~n}oz-Mateos}, J.~C. and {Bell}, E.~F. and {Brammer}, G. and {Charlot}, S. and {Chauk{\'e}}, P. and {Labb{\'e}}, I. and {Maseda}, M.~V. and {Muzzin}, A. and {Rix}, H. -W. and {Sobral}, D. and {van de Sande}, J. and {van Dokkum}, P.~G. and {Wild}, V. and {Wolf}, C.},
        title = "{The VLT LEGA-C Spectroscopic Survey: The Physics of Galaxies at a Lookback Time of 7 Gyr}",
      journal = {\apjs},
         year = 2016,
        month = apr,
       volume = {223},
       number = {2},
          eid = {29},
        pages = {29},
          doi = {10.3847/0067-0049/223/2/29},
archivePrefix = {arXiv},
       eprint = {1603.05479},
 primaryClass = {astro-ph.GA},
       adsurl = {https://ui.adsabs.harvard.edu/abs/2016ApJS..223...29V}
}

@ARTICLE{vanderWel21,
       author = {{van der Wel}, Arjen and {Bezanson}, Rachel and {D'Eugenio}, Francesco and {Straatman}, Caroline and {Franx}, Marijn and {van Houdt}, Josha and {Maseda}, Michael V. and {Gallazzi}, Anna and {Wu}, Po-Feng and {Pacifici}, Camilla and {Barisic}, Ivana and {Brammer}, Gabriel B. and {Munoz-Mateos}, Juan Carlos and {Vervalcke}, Sarah and {Zibetti}, Stefano and {Sobral}, David and {de Graaff}, Anna and {Calhau}, Joao and {Kaushal}, Yasha and {Muzzin}, Adam and {Bell}, Eric F. and {van Dokkum}, Pieter G.},
        title = "{The Large Early Galaxy Astrophysics Census (LEGA-C) Data Release 3: 3000 High-quality Spectra of K$_{s}$-selected Galaxies at z > 0.6}",
      journal = {\apjs},
         year = 2021,
        month = oct,
       volume = {256},
       number = {2},
          eid = {44},
        pages = {44},
          doi = {10.3847/1538-4365/ac1356},
archivePrefix = {arXiv},
       eprint = {2108.00744},
 primaryClass = {astro-ph.GA},
       adsurl = {https://ui.adsabs.harvard.edu/abs/2021ApJS..256...44V}
}

@ARTICLE{Whitaker10,
       author = {{Whitaker}, Katherine E. and {van Dokkum}, Pieter G. and {Brammer}, Gabriel and {Kriek}, Mariska and {Franx}, Marijn and {Labb{\'e}}, Ivo and {Marchesini}, Danilo and {Quadri}, Ryan F. and {Bezanson}, Rachel and {Illingworth}, Garth D. and {Lee}, Kyoung-Soo and {Muzzin}, Adam and {Rudnick}, Gregory and {Wake}, David A.},
        title = "{The Age Spread of Quiescent Galaxies with the NEWFIRM Medium-band Survey: Identification of the Oldest Galaxies Out to z \raisebox{-0.5ex}\textasciitilde 2}",
      journal = {\apj},
         year = 2010,
        month = aug,
       volume = {719},
       number = {2},
        pages = {1715-1732},
          doi = {10.1088/0004-637X/719/2/1715},
archivePrefix = {arXiv},
       eprint = {1006.5453},
 primaryClass = {astro-ph.CO},
       adsurl = {https://ui.adsabs.harvard.edu/abs/2010ApJ...719.1715W}
}

@ARTICLE{Kendrew16,
       author = {{Kendrew}, S. and {Zieleniewski}, S. and {Houghton}, R.~C.~W. and {Thatte}, N. and {Devriendt}, J. and {Tecza}, M. and {Clarke}, F. and {O'Brien}, K. and {H{\"a}u{\ss}ler}, B.},
        title = "{Simulated stellar kinematics studies of high-redshift galaxies with the HARMONI Integral Field Spectrograph}",
      journal = {\mnras},
         year = 2016,
        month = may,
       volume = {458},
       number = {3},
        pages = {2405-2422},
          doi = {10.1093/mnras/stw438},
archivePrefix = {arXiv},
       eprint = {1602.06983},
 primaryClass = {astro-ph.GA},
       adsurl = {https://ui.adsabs.harvard.edu/abs/2016MNRAS.458.2405K}
}

@INPROCEEDINGS{Leschinski16,
       author = {{Leschinski}, K. and {Czoske}, O. and {K{\"o}hler}, R. and {Mach}, M. and {Zeilinger}, W. and {Verdoes Kleijn}, G. and {Alves}, J. and {Kausch}, W. and {Przybilla}, N.},
        title = "{SimCADO: an instrument data simulator package for MICADO at the E-ELT}",
    booktitle = {Modeling, Systems Engineering, and Project Management for Astronomy VI},
         year = 2016,
       editor = {{Angeli}, George Z. and {Dierickx}, Philippe},
       series = {Society of Photo-Optical Instrumentation Engineers (SPIE) Conference Series},
       volume = {9911},
        month = aug,
          eid = {991124},
        pages = {991124},
          doi = {10.1117/12.2232483},
archivePrefix = {arXiv},
       eprint = {1609.01480},
 primaryClass = {astro-ph.IM},
       adsurl = {https://ui.adsabs.harvard.edu/abs/2016SPIE.9911E..24L}
}

@ARTICLE{Nguyen2025b,
       author = {{Nguyen}, Dieu D. and {Cappellari}, Michele and {Ngo}, Hai N. and {Le}, Tinh Q.~T. and {Le}, Tuan N. and {Ho}, Khue N.~H. and {Nguyen}, An K. and {On}, Phong T. and {Tong}, Huy G. and {Thatte}, Niranjan and {Pereira-Santaella}, Miguel},
        title = "{Simulating Intermediate Black Hole Mass Measurements for a Sample of Galaxies with Nuclear Star Clusters Using ELT/HARMONI High Spatial Resolution Integral-field Stellar Kinematics}",
      journal = {\aj},
         year = 2025,
        month = aug,
       volume = {170},
       number = {2},
          eid = {124},
        pages = {124},
          doi = {10.3847/1538-3881/ade9ba},
       adsurl = {https://ui.adsabs.harvard.edu/abs/2025AJ....170..124N}
}

@ARTICLE{Nguyen2025a,
       author = {{Nguyen}, Dieu D. and {Ngo}, Hai N. and {Le}, Tinh Q.~T. and {Graham}, Alister W. and {Soria}, Roberto and {Chilingarian}, Igor V. and {Thatte}, Niranjan and {Phuong}, N.~T. and {Hoang}, Thiem and {Pereira-Santaella}, Miguel and {Durre}, Mark and {Pham}, Diep N. and {Ngoc Tram}, Le and {Ngoc}, Nguyen B. and {L{\^e}}, Ng{\^a}n},
        title = "{Supermassive black hole mass measurement in the spiral galaxy NGC 4736 using JWST/NIRSpec stellar kinematics}",
      journal = {\aap},
         year = 2025,
        month = jun,
       volume = {698},
          eid = {L9},
        pages = {L9},
          doi = {10.1051/0004-6361/202554672},
archivePrefix = {arXiv},
       eprint = {2505.09941},
 primaryClass = {astro-ph.GA},
       adsurl = {https://ui.adsabs.harvard.edu/abs/2025A&A...698L...9N}
}

@ARTICLE{McConnell11,
       author = {{McConnell}, Nicholas J. and {Ma}, Chung-Pei and {Gebhardt}, Karl and {Wright}, Shelley A. and {Murphy}, Jeremy D. and {Lauer}, Tod R. and {Graham}, James R. and {Richstone}, Douglas O.},
        title = "{Two ten-billion-solar-mass black holes at the centres of giant elliptical galaxies}",
      journal = {\nat},
         year = 2011,
        month = dec,
       volume = {480},
       number = {7376},
        pages = {215-218},
          doi = {10.1038/nature10636},
archivePrefix = {arXiv},
       eprint = {1112.1078},
 primaryClass = {astro-ph.CO},
       adsurl = {https://ui.adsabs.harvard.edu/abs/2011Natur.480..215M}
}

@ARTICLE{Saglia16,
       author = {{Saglia}, R.~P. and {Opitsch}, M. and {Erwin}, P. and {Thomas}, J. and {Beifiori}, A. and {Fabricius}, M. and {Mazzalay}, X. and {Nowak}, N. and {Rusli}, S.~P. and {Bender}, R.},
        title = "{The SINFONI Black Hole Survey: The Black Hole Fundamental Plane Revisited and the Paths of (Co)evolution of Supermassive Black Holes and Bulges}",
      journal = {\apj},
         year = 2016,
        month = feb,
       volume = {818},
       number = {1},
          eid = {47},
        pages = {47},
          doi = {10.3847/0004-637X/818/1/47},
archivePrefix = {arXiv},
       eprint = {1601.00974},
 primaryClass = {astro-ph.GA},
       adsurl = {https://ui.adsabs.harvard.edu/abs/2016ApJ...818...47S}
}

@ARTICLE{Ahn18,
       author = {{Ahn}, Christopher P. and {Seth}, Anil C. and {Cappellari}, Michele and {Krajnovi{\'c}}, Davor and {Strader}, Jay and {Voggel}, Karina T. and {Walsh}, Jonelle L. and {Bahramian}, Arash and {Baumgardt}, Holger and {Brodie}, Jean and {Chilingarian}, Igor and {Chomiuk}, Laura and {den Brok}, Mark and {Frank}, Matthias and {Hilker}, Michael and {McDermid}, Richard M. and {Mieske}, Steffen and {Neumayer}, Nadine and {Nguyen}, Dieu D. and {Pechetti}, Renuka and {Romanowsky}, Aaron J. and {Spitler}, Lee},
        title = "{The Black Hole in the Most Massive Ultracompact Dwarf Galaxy M59-UCD3}",
      journal = {\apj},
         year = 2018,
        month = may,
       volume = {858},
       number = {2},
          eid = {102},
        pages = {102},
          doi = {10.3847/1538-4357/aabc57},
archivePrefix = {arXiv},
       eprint = {1804.02399},
 primaryClass = {astro-ph.GA},
       adsurl = {https://ui.adsabs.harvard.edu/abs/2018ApJ...858..102A}
}

@ARTICLE{Gustafsson08,
       author = {{Gustafsson}, B. and {Edvardsson}, B. and {Eriksson}, K. and {J{\o}rgensen}, U.~G. and {Nordlund}, {\r{A}}. and {Plez}, B.},
        title = "{A grid of MARCS model atmospheres for late-type stars. I. Methods and general properties}",
      journal = {\aap},
         year = 2008,
        month = aug,
       volume = {486},
       number = {3},
        pages = {951-970},
          doi = {10.1051/0004-6361:200809724},
archivePrefix = {arXiv},
       eprint = {0805.0554},
 primaryClass = {astro-ph},
       adsurl = {https://ui.adsabs.harvard.edu/abs/2008A&A...486..951G}
}

@ARTICLE{Maraston11,
       author = {{Maraston}, C. and {Str{\"o}mb{\"a}ck}, G.},
        title = "{Stellar population models at high spectral resolution}",
      journal = {\mnras},
         year = 2011,
        month = dec,
       volume = {418},
       number = {4},
        pages = {2785-2811},
          doi = {10.1111/j.1365-2966.2011.19738.x},
archivePrefix = {arXiv},
       eprint = {1109.0543},
 primaryClass = {astro-ph.CO},
       adsurl = {https://ui.adsabs.harvard.edu/abs/2011MNRAS.418.2785M}
}

@ARTICLE{Cappellari23,
       author = {{Cappellari}, Michele},
        title = "{Full spectrum fitting with photometry in PPXF: stellar population versus dynamical masses, non-parametric star formation history and metallicity for 3200 LEGA-C galaxies at redshift z {\ensuremath{\approx}} 0.8}",
      journal = {\mnras},
         year = 2023,
        month = dec,
       volume = {526},
       number = {3},
        pages = {3273-3300},
          doi = {10.1093/mnras/stad2597},
archivePrefix = {arXiv},
       eprint = {2208.14974},
 primaryClass = {astro-ph.GA},
       adsurl = {https://ui.adsabs.harvard.edu/abs/2023MNRAS.526.3273C}
}

@ARTICLE{Cappellari03,
       author = {{Cappellari}, Michele and {Copin}, Yannick},
        title = "{Adaptive spatial binning of integral-field spectroscopic data using Voronoi tessellations}",
      journal = {\mnras},
         year = 2003,
        month = jun,
       volume = {342},
       number = {2},
        pages = {345-354},
          doi = {10.1046/j.1365-8711.2003.06541.x},
archivePrefix = {arXiv},
       eprint = {astro-ph/0302262},
 primaryClass = {astro-ph},
       adsurl = {https://ui.adsabs.harvard.edu/abs/2003MNRAS.342..345C}
}

@ARTICLE{AstropyCollaboration22,
       author = {{Astropy Collaboration} and {Price-Whelan}, Adrian M. and {Lim}, Pey Lian and {Earl}, Nicholas and {Starkman}, Nathaniel and {Bradley}, Larry and {Shupe}, David L. and {Patil}, Aarya A. and {Corrales}, Lia and {Brasseur}, C.~E. and {N{\"o}the}, Maximilian and {Donath}, Axel and {Tollerud}, Erik and {Morris}, Brett M. and {Ginsburg}, Adam and {Vaher}, Eero and {Weaver}, Benjamin A. and {Tocknell}, James and {Jamieson}, William and {van Kerkwijk}, Marten H. and {Robitaille}, Thomas P. and {Merry}, Bruce and {Bachetti}, Matteo and {G{\"u}nther}, H. Moritz and {Aldcroft}, Thomas L. and {Alvarado-Montes}, Jaime A. and {Archibald}, Anne M. and {B{\'o}di}, Attila and {Bapat}, Shreyas and {Barentsen}, Geert and {Baz{\'a}n}, Juanjo and {Biswas}, Manish and {Boquien}, M{\'e}d{\'e}ric and {Burke}, D.~J. and {Cara}, Daria and {Cara}, Mihai and {Conroy}, Kyle E. and {Conseil}, Simon and {Craig}, Matthew W. and {Cross}, Robert M. and {Cruz}, Kelle L. and {D'Eugenio}, Francesco and {Dencheva}, Nadia and {Devillepoix}, Hadrien A.~R. and {Dietrich}, J{\"o}rg P. and {Eigenbrot}, Arthur Davis and {Erben}, Thomas and {Ferreira}, Leonardo and {Foreman-Mackey}, Daniel and {Fox}, Ryan and {Freij}, Nabil and {Garg}, Suyog and {Geda}, Robel and {Glattly}, Lauren and {Gondhalekar}, Yash and {Gordon}, Karl D. and {Grant}, David and {Greenfield}, Perry and {Groener}, Austen M. and {Guest}, Steve and {Gurovich}, Sebastian and {Handberg}, Rasmus and {Hart}, Akeem and {Hatfield-Dodds}, Zac and {Homeier}, Derek and {Hosseinzadeh}, Griffin and {Jenness}, Tim and {Jones}, Craig K. and {Joseph}, Prajwel and {Kalmbach}, J. Bryce and {Karamehmetoglu}, Emir and {Ka{\l}uszy{\'n}ski}, Miko{\l}aj and {Kelley}, Michael S.~P. and {Kern}, Nicholas and {Kerzendorf}, Wolfgang E. and {Koch}, Eric W. and {Kulumani}, Shankar and {Lee}, Antony and {Ly}, Chun and {Ma}, Zhiyuan and {MacBride}, Conor and {Maljaars}, Jakob M. and {Muna}, Demitri and {Murphy}, N.~A. and {Norman}, Henrik and {O'Steen}, Richard and {Oman}, Kyle A. and {Pacifici}, Camilla and {Pascual}, Sergio and {Pascual-Granado}, J. and {Patil}, Rohit R. and {Perren}, Gabriel I. and {Pickering}, Timothy E. and {Rastogi}, Tanuj and {Roulston}, Benjamin R. and {Ryan}, Daniel F. and {Rykoff}, Eli S. and {Sabater}, Jose and {Sakurikar}, Parikshit and {Salgado}, Jes{\'u}s and {Sanghi}, Aniket and {Saunders}, Nicholas and {Savchenko}, Volodymyr and {Schwardt}, Ludwig and {Seifert-Eckert}, Michael and {Shih}, Albert Y. and {Jain}, Anany Shrey and {Shukla}, Gyanendra and {Sick}, Jonathan and {Simpson}, Chris and {Singanamalla}, Sudheesh and {Singer}, Leo P. and {Singhal}, Jaladh and {Sinha}, Manodeep and {Sip{\H{o}}cz}, Brigitta M. and {Spitler}, Lee R. and {Stansby}, David and {Streicher}, Ole and {{\v{S}}umak}, Jani and {Swinbank}, John D. and {Taranu}, Dan S. and {Tewary}, Nikita and {Tremblay}, Grant R. and {de Val-Borro}, Miguel and {Van Kooten}, Samuel J. and {Vasovi{\'c}}, Zlatan and {Verma}, Shresth and {de Miranda Cardoso}, Jos{\'e} Vin{\'\i}cius and {Williams}, Peter K.~G. and {Wilson}, Tom J. and {Winkel}, Benjamin and {Wood-Vasey}, W.~M. and {Xue}, Rui and {Yoachim}, Peter and {Zhang}, Chen and {Zonca}, Andrea and {Astropy Project Contributors}},
        title = "{The Astropy Project: Sustaining and Growing a Community-oriented Open-source Project and the Latest Major Release (v5.0) of the Core Package}",
      journal = {\apj},
         year = 2022,
        month = aug,
       volume = {935},
       number = {2},
          eid = {167},
        pages = {167},
          doi = {10.3847/1538-4357/ac7c74},
archivePrefix = {arXiv},
       eprint = {2206.14220},
 primaryClass = {astro-ph.IM},
       adsurl = {https://ui.adsabs.harvard.edu/abs/2022ApJ...935..167A}
}

@Article{Gebhardt2003,
  author   = {Gebhardt, K. and Richstone, D. and Tremaine, S. and Lauer, T. R. and Bender, R. and Bower, G. and Dressler, A. and Faber, S. M. and Filippenko, A. V. and Green, R. and Grillmair, C. and Ho, L. C. and Kormendy, J. and Magorrian, J. and Pinkney, J.},
  journal  = {\apj},
  title    = {{Axisymmetric Dynamical Models of the Central Regions of Galaxies}},
  year     = {2003},
  month    = jan,
  pages    = {92--115},
  volume   = {583},
  adsurl   = {https://ui.adsabs.harvard.edu/abs/2003ApJ...583...92G},
  doi      = {10.1086/345081},
  eprint   = {astro-ph/0209483},
}

@Article{Cappellari2009,
  author        = {Cappellari, M. and Neumayer, N. and Reunanen, J. and van der Werf, P. P. and de Zeeuw, P. T. and Rix, H.-W.},
  journal       = {\mnras},
  title         = {{The mass of the black hole in Centaurus A from SINFONI AO-assisted integral-field observations of stellar kinematics}},
  year          = {2009},
  month         = apr,
  pages         = {660--674},
  volume        = {394},
  adsurl        = {https://ui.adsabs.harvard.edu/abs/2009MNRAS.394..660C},
  archiveprefix = {arXiv},
  doi           = {10.1111/j.1365-2966.2008.14377.x},
  eprint        = {0812.1000},
}

@Article{Rusli2013,
  author        = {Rusli, S. P. and Thomas, J. and Saglia, R. P. and Fabricius, M. and Erwin, P. and Bender, R. and Nowak, N. and Lee, C. H. and Riffeser, A. and Sharp, R.},
  journal       = {\aj},
  title         = {{The Influence of Dark Matter Halos on Dynamical Estimates of Black Hole Mass: 10 New Measurements for High-{$\sigma$} Early-type Galaxies}},
  year          = {2013},
  month         = sep,
  pages         = {45},
  volume        = {146},
  adsurl        = {https://ui.adsabs.harvard.edu/abs/2013AJ....146...45R},
  archiveprefix = {arXiv},
  doi           = {10.1088/0004-6256/146/3/45},
  eid           = {45},
  eprint        = {1306.1124},
  primaryclass  = {astro-ph.CO},
}

@Article{Walsh2016,
  author        = {Walsh, J. L. and van den Bosch, R. C. E. and Gebhardt, K. and Y{\i}ld{\i}r{\i}m, A. and Richstone, D. O. and G{\"u}ltekin, K. and Husemann, B.},
  journal       = {\apj},
  title         = {{A 5 x 10$^{9}$ Msun Black Hole in NGC 1277 from Adaptive Optics Spectroscopy}},
  year          = {2016},
  month         = jan,
  pages         = {2},
  volume        = {817},
  adsurl        = {https://ui.adsabs.harvard.edu/abs/2016ApJ...817....2W},
  archiveprefix = {arXiv},
  doi           = {10.3847/0004-637X/817/1/2},
  eid           = {2},
  eprint        = {1511.04455},
}

@Article{Shapiro2006,
  author  = {Shapiro, K. L. and Cappellari, M. and de Zeeuw, T. and McDermid, R. M. and Gebhardt, K. and van den Bosch, R. C. E. and Statler, T. S.},
  journal = {\mnras},
  title   = {The black hole in NGC 3379: a comparison of gas and stellar dynamical mass measurements with HST and integral-field data},
  year    = {2006},
  month   = aug,
  pages   = {559--579},
  volume  = {370},
  adsurl  = {https://ui.adsabs.harvard.edu/abs/2006MNRAS.370..559S},
  doi     = {10.1111/j.1365-2966.2006.10537.x},
  eprint  = {arXiv:astro-ph/0605479},
}

@Article{Barth2001,
  author        = {Barth, Aaron J. and Sarzi, Marc and Rix, Hans-Walter and Ho, Luis C. and Filippenko, Alexei V. and Sargent, Wallace L.~W.},
  journal       = {\apj},
  title         = {Evidence for a Supermassive Black Hole in the S0 Galaxy NGC 3245},
  year          = {2001},
  month         = jul,
  number        = {2},
  pages         = {685--708},
  volume        = {555},
  archiveprefix = {arXiv},
  doi           = {10.1086/321523},
  eprint        = {astro-ph/0012213},
  primaryclass  = {astro-ph},
  url           = {https://ui.adsabs.harvard.edu/abs/2001ApJ...555..685B},
}

@Article{Cappellari2006,
  author   = {Cappellari, M. and Bacon, R. and Bureau, M. and Damen, M. C. and Davies, R. L. and de Zeeuw, P. T. and Emsellem, E. and Falc{\'o}n-Barroso, J. and Krajnovi{\'c}, D. and Kuntschner, H. and McDermid, R. M. and Peletier, R. F. and Sarzi, M. and van den Bosch, R. C. E. and van de Ven, G.},
  journal  = {\mnras},
  title    = {The SAURON project - IV. The mass-to-light ratio, the virial mass estimator and the Fundamental Plane of elliptical and lenticular galaxies},
  year     = {2006},
  month    = mar,
  pages    = {1126--1150},
  volume   = {366},
  adsurl   = {https://ui.adsabs.harvard.edu/abs/2006MNRAS.366.1126C},
  doi      = {10.1111/j.1365-2966.2005.09981.x},
  eprint   = {arXiv:astro-ph/0505042},
}

@ARTICLE{Zhu2025,
       author = {{Zhu}, Kai and {Cappellari}, Michele and {Mao}, Shude and {Lu}, Shengdong and {Li}, Ran and {Shi}, Yong and {Simon}, David A. and {Fu}, Youquan and {Wang}, Xiaohan},
        title = "{MaNGA DynPop. VII. A Unified Bulge{\textendash}Disk{\textendash}Halo Model for Explaining Diversity in Circular Velocity Curves of 6000 Spiral and Early-type Galaxies}",
      journal = {\apjs},
         year = 2025,
        month = oct,
       volume = {280},
       number = {2},
          eid = {55},
        pages = {55},
          doi = {10.3847/1538-4365/adfb7c},
archivePrefix = {arXiv},
       eprint = {2503.06968},
 primaryClass = {astro-ph.GA},
       adsurl = {https://ui.adsabs.harvard.edu/abs/2025ApJS..280...55Z}
}

@Article{Wright2006,
  author        = {Wright, E.~L.},
  journal       = {\pasp},
  title         = {A Cosmology Calculator for the World Wide Web},
  year          = {2006},
  month         = dec,
  number        = {850},
  pages         = {1711-1715},
  volume        = {118},
  archiveprefix = {arXiv},
  doi           = {10.1086/510102},
  eprint        = {astro-ph/0609593},
  primaryclass  = {astro-ph},
  url           = {https://ui.adsabs.harvard.edu/abs/2006PASP..118.1711W},
}

@InProceedings{Davies2018,
  author        = {Davies, R. and Alves, J. and Cl{\'e}net, Y. and Lang-Bardl, F. and Nicklas, H. and Pott, J. U. and Ragazzoni, R. and Tolstoy, E. and Amico, P. and Anwand-Heerwart, H. and Barboza, S. and Barl, L. and Baudoz, P. and Bender, R. and Bezawada, N. and Bizenberger, P. and Boland, W. and Bonifacio, P. and Borgo, B. and Buey, T. and Chapron, F. and Chemla, F. and Cohen, M. and Czoske, O. and D{\'e}o, V. and Disseau, K. and Dreizler, S. and Dupuis, O. and Fabricius, M. and Falomo, R. and Fedou, P. and F{\"o}rster Schreiber, N. and Garrel, V. and Geis, N. and Gemperlein, H. and Gendron, E. and Genzel, R. and Gillessen, S. and Gl{\"u}ck, M. and Grupp, F. and Hartl, M. and H{\"a}user, M. and Hess, H. J. and Hofferbert, R. and Hopp, U. and H{\"o}rmann, V. and Hubert, Z. and Huby, E. and Huet, J. M. and Hutterer, V. and Ives, D. and Janssen, A. and Jellema, W. and Kausch, W. and Kerber, F. and Kravcar, H. and Le Ruyet, B. and Leschinski, K. and Mandla, C. and Manhart, M. and Massari, D. and Mei, S. and Merlin, F. and Mohr, L. and Monna, A. and Muench, N. and M{\"u}ller, F. and Musters, G. and Navarro, R. and Neumann, U. and Neumayer, N. and Niebsch, J. and Plattner, M. and Przybilla, N. and Rabien, S. and Ramlau, R. and Ramos, J. and Ramsay, S. and Rhode, P. and Richter, A. and Richter, J. and Rix, H. W. and Rodeghiero, G. and Rohloff, R. R. and Rosensteiner, M. and Rousset, G. and Schlichter, J. and Schubert, J. and Sevin, A. and Stuik, R. and Sturm, E. and Thomas, J. and Tromp, N. and Verdoes-Kleijn, G. and Vidal, F. and Wagner, R. and Wegner, M. and Zeilinger, W. and Ziegleder, J. and Ziegler, B. and Zins, G.},
  booktitle     = {Ground-based and Airborne Instrumentation for Astronomy VII},
  title         = {The MICADO first light imager for the ELT: overview, operation, simulation},
  year          = {2018},
  editor        = {{Evans}, Christopher J. and {Simard}, Luc and {Takami}, Hideki},
  month         = jul,
  pages         = {107021S},
  series        = {Society of Photo-Optical Instrumentation Engineers (SPIE) Conference Series},
  volume        = {10702},
  archiveprefix = {arXiv},
  doi           = {10.1117/12.2311483},
  eid           = {107021S},
  eprint        = {1807.10003},
  primaryclass  = {astro-ph.IM},
  url           = {https://ui.adsabs.harvard.edu/abs/2018SPIE10702E..1SD},
}

@Book{VanRossum2009,
  author    = {Van Rossum, Guido and Drake, Fred L.},
  publisher = {CreateSpace},
  title     = {Python 3 Reference Manual},
  year      = {2009},
  address   = {Scotts Valley, CA},
  isbn      = {1441412697},
}

@Article{Hunter2007,
  author    = {Hunter, J. D.},
  journal   = {Computing In Science \& Engineering},
  title     = {Matplotlib: A 2D graphics environment},
  year      = {2007},
  number    = {3},
  pages     = {90--95},
  volume    = {9},
  doi       = {10.1109/MCSE.2007.55},
  publisher = {IEEE COMPUTER SOC},
}

@Article{Harris2020,
  author    = {Harris, Charles R. and Millman, K. Jarrod and van der Walt, St{\'{e}}fan J. and Gommers, Ralf and Virtanen, Pauli and Cournapeau, David and Wieser, Eric and Taylor, Julian and Berg, Sebastian and Smith, Nathaniel J. and Kern, Robert and Picus, Matti and Hoyer, Stephan and van Kerkwijk, Marten H. and Brett, Matthew and Haldane, Allan and del R{\'{\i}}o, Jaime Fern{\'{a}}ndez and Wiebe, Mark and Peterson, Pearu and G{\'{e}}rard-Marchant, Pierre and Sheppard, Kevin and Reddy, Tyler and Weckesser, Warren and Abbasi, Hameer and Gohlke, Christoph and Oliphant, Travis E.},
  journal   = {Nature},
  title     = {Array programming with {NumPy}},
  year      = {2020},
  month     = sep,
  number    = {7825},
  pages     = {357--362},
  volume    = {585},
  doi       = {10.1038/s41586-020-2649-2},
  publisher = {Springer Science and Business Media {LLC}},
}

@Article{Virtanen2020,
  author    = {Virtanen, Pauli and Gommers, Ralf and Oliphant, Travis E. and Haberland, Matt and Reddy, Tyler and Cournapeau, David and Burovski, Evgeni and Peterson, Pearu and Weckesser, Warren and Bright, Jonathan and van der Walt, St{\'{e}}fan J. and Brett, Matthew and Wilson, Joshua and Millman, K. Jarrod and Mayorov, Nikolay and Nelson, Andrew R. J. and Jones, Eric and Kern, Robert and Larson, Eric and Carey, C. J. and Polat, {\.{I}}lhan and Feng, Yu and Moore, Eric W. and VanderPlas, Jake and Laxalde, Denis and Perktold, Josef and Cimrman, Robert and Henriksen, Ian and Quintero, E. A. and Harris, Charles R. and Archibald, Anne M. and Ribeiro, Ant{\^{o}}nio H. and Pedregosa, Fabian and van Mulbregt, Paul and Vijaykumar, Aditya and Bardelli, Alessandro Pietro and Rothberg, Alex and Hilboll, Andreas and Kloeckner, Andreas and Scopatz, Anthony and Lee, Antony and Rokem, Ariel and Woods, C. Nathan and Fulton, Chad and Masson, Charles and Häggström, Christian and Fitzgerald, Clark and Nicholson, David A. and Hagen, David R. and Pasechnik, Dmitrii V. and Olivetti, Emanuele and Martin, Eric and Wieser, Eric and Silva, Fabrice and Lenders, Felix and Wilhelm, Florian and Young, G. and Price, Gavin A. and Ingold, Gert-Ludwig and Allen, Gregory E. and Lee, Gregory R. and Audren, Herv{\'{e}} and Probst, Irvin and Dietrich, Jörg P. and Silterra, Jacob and Webber, James T. and Slavi{\v{c}}, Janko and Nothman, Joel and Buchner, Johannes and Kulick, Johannes and Schönberger, Johannes L. and de Miranda Cardoso, Jos{\'{e}} Vin{\'{\i}}cius and Reimer, Joscha and Harrington, Joseph and Rodr{\'{\i}}guez, Juan Luis Cano and Nunez-Iglesias, Juan and Kuczynski, Justin and Tritz, Kevin and Thoma, Martin and Newville, Matthew and Kümmerer, Matthias and Bolingbroke, Maximilian and Tartre, Michael and Pak, Mikhail and Smith, Nathaniel J. and Nowaczyk, Nikolai and Shebanov, Nikolay and Pavlyk, Oleksandr and Brodtkorb, Per A. and Lee, Perry and McGibbon, Robert T. and Feldbauer, Roman and Lewis, Sam and Tygier, Sam and Sievert, Scott and Vigna, Sebastiano and Peterson, Stefan and More, Surhud and Pudlik, Tadeusz and Oshima, Takuya and Pingel, Thomas J. and Robitaille, Thomas P. and Spura, Thomas and Jones, Thouis R. and Cera, Tim and Leslie, Tim and Zito, Tiziano and Krauss, Tom and Upadhyay, Utkarsh and Halchenko, Yaroslav O. and V{\'{a}}zquez-Baeza, Yoshiki},
  journal   = {Nature Methods},
  title     = {{SciPy} 1.0: fundamental algorithms for scientific computing in Python},
  year      = {2020},
  month     = feb,
  number    = {3},
  pages     = {261--272},
  volume    = {17},
  doi       = {10.1038/s41592-019-0686-2},
  publisher = {Springer Science and Business Media {LLC}},
}

@Misc{bradley2024,
  author    = {Larry Bradley and Brigitta Sip{\H o}cz and Thomas Robitaille and Erik Tollerud and Z\`e Vin{\'{\i}}cius and Christoph Deil and Kyle Barbary and Tom J Wilson and Ivo Busko and Axel Donath and Hans Moritz G{\"u}nther and Mihai Cara and P. L. Lim and Sebastian Me{\ss}linger and Simon Conseil and Zach Burnett and Azalee Bostroem and Michael Droettboom and E. M. Bray and Lars Andersen Bratholm and Adam Ginsburg and William Jamieson and Geert Barentsen and Matt Craig and Brett M. Morris and Marshall Perrin and Shivangee Rathi and Sergio Pascual and Iskren Y. Georgiev},
  month     = oct,
  title     = {astropy/photutils: 2.0.2},
  year      = {2024},
  doi       = {10.5281/zenodo.13989456},
  publisher = {Zenodo},
  url       = {https://doi.org/10.5281/zenodo.13989456},
  version   = {2.0.2},
}

@Article{Maiolino2024,
  author        = {Maiolino, Roberto and Scholtz, Jan and Curtis-Lake, Emma and Carniani, Stefano and Baker, William and de Graaff, Anna and Tacchella, Sandro and {\"U}bler, Hannah and D'Eugenio, Francesco and Witstok, Joris and Curti, Mirko and Arribas, Santiago and Bunker, Andrew J. and Charlot, St{\'e}phane and Chevallard, Jacopo and Eisenstein, Daniel J. and Egami, Eiichi and Ji, Zhiyuan and Jones, Gareth C. and Lyu, Jianwei and Rawle, Tim and Robertson, Brant and Rujopakarn, Wiphu and Perna, Michele and Sun, Fengwu and Venturi, Giacomo and Williams, Christina C. and Willott, Chris},
  journal       = {\aap},
  title         = {JADES: The diverse population of infant black holes at $4 < z < 11$: Merging, tiny, poor, but mighty},
  year          = {2024},
  month         = nov,
  pages         = {A145},
  volume        = {691},
  adsurl        = {https://ui.adsabs.harvard.edu/abs/2024A%26A...691A.145M},
  archiveprefix = {arXiv},
  doi           = {10.1051/0004-6361/202347640},
  eid           = {A145},
  eprint        = {2308.01230},
  primaryclass  = {astro-ph.GA},
}


\bsp	
\label{lastpage}
\end{document}